\documentclass[aps,prd,nofootinbib]{revtex4}
\usepackage{graphicx}
\usepackage{amssymb,amsmath}
\def\ghz{\rm GHz}
\def\Ghz{\rm GHz}
\def\nside{N_{\rm side}}
\def\lmax{l_{\rm max}}
\def\rmf{{\rm f}}
\def\rmn{{\rm n}}
\def\rmc{{\rm c}}
\def\rmD{{\rm D}}
\def\rmCC{{\rm C}}
\def\lsim{\mathrel{\rlap{\lower4pt\hbox{\hskip1pt$\sim$}}
    \raise1pt\hbox{$<$}}}                
\def\gsim{\mathrel{\rlap{\lower4pt\hbox{\hskip1pt$\sim$}}
    \raise1pt\hbox{$>$}}}         

\begin{document}

\title{CMB anisotropy power spectrum using linear combinations of WMAP maps}
\author{Rajib Saha$^{1,2,3}$, Simon Prunet$^{3}$, Pankaj Jain$^{1}$ \& 
Tarun Souradeep$^2$}
\affiliation{$^{1}$Department of Physics, Indian Institute of Technology, 
Kanpur, U.P, 208016, India}
\affiliation{$^{2}$Inter-University Centre for
  Astronomy and Astrophysics, Post Bag 4, Ganeshkhind, Pune 411007,
  India\\}
\affiliation{$^{3}$Institut d'Astrophysique de Paris, 98 bis Boulevard Arago, F-75014 Paris, France.}

\begin{abstract}
In recent years the goal of estimating different cosmological
parameters precisely has set new challenges in the effort to
accurately measure the angular power spectrum of CMB. This has
required removal of foreground contamination as well as detector noise
bias with reliability and precision. Recently, a novel {\em
model-independent} method for the estimation of CMB angular power
spectrum solely from multi-frequency observations has been proposed
and implemented on the first year WMAP
data by Saha et al.~2006. All previous estimates of power
spectrum of CMB are based upon foreground templates using data sets
from different experiments. However our methodology demonstrates that
{\em CMB angular spectrum can be reliably estimated with precision
from a self contained analysis of the WMAP data}. In this work we
provide a detailed description of this method. We also study and 
identify the biases present in our power spectrum estimate. 
We apply our methodoly to extract
the power spectrum from the WMAP 1 year and 3 year data.
\end{abstract}
\maketitle
\section{Introduction}
Starting from the end of the last millenium remarkable progress in cosmology has been made by the
precise measurements of the anisotropies in CMB from different ground
based as well as satellite observations~\cite{arch1,cob1,boomerang,cbi1}. 
The Wilkinson Microwave Anisotropy Probe (WMAP) 
measures  the CMB  anisotropy over the $5$ frequency bands at $23~\ghz$~(K),
$33~\ghz$~(Ka), $41~\ghz$~(Q), $61~\ghz$~(V) and $94~\ghz$~(W). 
The observation system of the WMAP satellite consists of 10
differencing assemblies (DA), \cite{Bennett, Bennett2, WMAP,
Hinshaw1}, one each for K and Ka bands, two for Q band, two for V band
and four for W band. They are labeled as K, Ka, Q1, Q2, V1, V2, W1,
W2, W3, and W4 DA respectively. In the 1 year and 3 year data release
the WMAP science team has provided the science community with large
amount of high quality data sets measured by these 10 DAs. However,
extracting the primordial signal from these large data set is a non
trivial task. The anisotropies in CMB are weak in comparison to
those originating due to radiation in the local universe, which
inevitably contaminate the observed signal. These dominant foreground 
emissions are from within the milky way as well as from extragalactic
point sources ~\cite{bouc_gisp99}. In the low frequency microwave regime the strongest
contamination comes from the synchrotron and free free emission
\cite{Hinshaw, Bennett}. At higher frequencies, where synchrotron and
free free emissions are low, dust emission dominates. A reliable
extraction of CMB signal from the multicomponent foreground
contaminated data is thus complicated. There exists several methods in
literature to remove foregrounds using foreground tracer templates
\cite{Bennett1,Hinshaw,Hinshaw06} built from observations from other
experiments. However, this requires a prior model of spatio-spectral dependence 
for all the foreground components. The effect of uncertanities 
in the foreground models to estimate CMB anisotropies have
 been discussed in Refs.~\cite{dodelson,Tegmark98}. 
The foreground cleaning is applicable in the
region away from the galactic plane. Detector noise is another
important concern that has to be addressed in order to precisely measure the
angular power spectrum. The angular power spectrum from detector noise
is dominant over the CMB power spectrum at the small angular
scales. The detector noise, being a random quantity, is treated
differently from foreground contamination which are treated here as 
fixed templates on the sky. However each of the DA's of WMAP has uncorrelated noise
property \cite{Hinshaw,Jarosik,Hinshaw06, Jarosik06}. WMAP science
team used this property to remove detector noise bias in a cross
correlated power spectrum obtained from two different DA
\cite{Hinshaw,Hinshaw06} using DA maps with frequencies $41~\ghz$,
$61~\ghz$ and $94~\ghz$.

An interesting model independent method to remove foregrounds from the
multi-frequency observations of CMB without any assumption about
foreground components has been proposed in Ref. \cite{Tegmark96} and
implemented in Ref. \cite{Tegmark} in order to extract the CMB signal
from the WMAP data. The foreground emissions were removed by
exploiting the fact that their contributions in different spectral
bands are considerably different while the CMB anisotropy power
spectrum is same in all the bands in unit of thermodynamic temperature 
due to the Planck blackbody energy spectrum of the CMB, \cite{FIRAS,fix96}. 
The main advantage of this foreground cleaning method is that it is
totally free from any assumption about foreground modeling. Another
advantage is that it is computationally fast. However the auto
power spectrum obtained from a single cleaned map as reported in Ref.~\cite{Tegmark}
is not directly usable for primordial power spectrum estimation at smaller angular scale. This is because the detector noise bias dominates over the CMB power spectrum at smaller angular scale, beyond the beam width of WMAP detectors.

In earlier publications \cite{sah06, sah06_proc,erik06} we extended this
model independent foreground removal method to remove detector noise bias also. 
In this work we describe 
the basic formalism of our previous work in detail. We apply our
method both on the WMAP 1 year and WMAP 3 year data to estimate CMB
angular power spectrum. We form several cleaned maps using different
cross combinations of the DA maps. Finally we form several cross power
spectra where detector noise bias is removed. The results from this analysis
are summarized in figure \ref{1yr3yr}. In this figure we show the power
spectra estimated from the WMAP's 1 year and 3 year data using the
multi-frequency combination of CMB maps. The bottom part of this figure
shows the residual unresolved point source contamination that
were corrected for to obtain these two spectra. These power spectra are 
obtained without making any explicit
model of the foregrounds or the detector noise. In most power spectrum
extraction procedures, only the three highest frequency channels
observed by WMAP have been used to extract CMB power spectrum. We present a
more general procedure where we use observations from all the five
frequency channels of WMAP. The primary merit of the foreground removal method is that 
it avoids any need to remove foregrounds based upon extrapolated flux measurements
at frequencies far away from observational
frequencies of WMAP ~\cite{Hinshaw,Pablo,Patanchon}.

Presence of a bias in the internally cleaned maps has been reported earlier in the
Refs.~\cite{Hinshaw06,Tegmark2000a,erik04}. In this work we also perform a detailed study of the nature of bias 
in the cleaned power spectrum.
 We show that the cleaned power spectrum is not an unbiased estimator 
of the underlying CMB spectrum. Naively one expects that 
there might be some residual foreground contamination 
causing a positive bias in the power spectrum. For a simplified approach,
which cleans the entire sky simultaneously, without sub-dividing into
regions of varying foreground contamination, we are able to analytically 
compute the 
cleaned power spectrum in terms of a CMB signal and the foreground plus 
detector noise covariance matrix. The existence of  bias is easily identified from these results. We also report and quantify an interesting negative bias in the  cleaned 
power spectrum. This negative bias is directly determined by the underlying CMB power
spectrum and is strongest at the lowest multipoles. 
The analytical results for the bias estimations make the model independent foreground removal method
more interesting for use in cosmology. The bias in the cleaned power spectrum can be removed following models of foreground and noise covariance matrices after a model independent foreground removal is performed. This
makes the estimated power spectrum less prone to uncertainties of the foreground modeling compared to a method which tries to minimize foreground using external templates. In this work we debias the CMB anisotropy spectra obtained from WMAP data only at the large multipole regime, $l\ge 400$. It turns out that because of large noise level of WMAP the point source bias is the only issue in this multipole range. We do not attempt to employ a debiasing method at the low multipoles where the negative bias is expected to dominate, because of its complicated nature in the case of the iterative, multiregion cleaning method. A more detailed study on this issue will be reported in a future 
publication.

The plan of this paper is as follows. The basic formalism and methodology to obtain power spectrum is 
described in the section ~\ref{method}.
Here we also discuss the bias present
 in this method. 
The implementation of our methodology on the WMAP 1 year and 
3 year data is discussed in section \ref{wmap_data}. Here we also obtain
an analytic expression for the residual unresolved point source power 
contamination.
The results are described in section \ref{results} and finally 
we conclude in section \ref{conclu}. Some of the notations used in this
report are as follows. We denote matrices and vectors by $\bf bold
 faced$ letters. Any variable (scalar, vector or matrix) which depends
on the stochastic component is denoted by a hat on the top. As an example
we note that the CMB power spectrum is a stochastic variable which is represented as 
$\hat C^c_l$.

\section{Basic Formalism and Methodology}
\label{method}
  In this section we outline the mathematical formalism to 
  obtain the underlying power spectrum. We also quantify the bias
present in the estimated power spectrum using our procedure.

\subsection{Foreground Removal}
\label{fgmethod}
The observed signal at frequency channel $i$ in a
differential telescope like WMAP can be modeled as
\begin{eqnarray}
\widetilde\Delta T^i(\hat n) = \int \left (\Delta T^{\rm c}(\hat n') +
\Delta T^{\rmf i}(\hat n')\right)B^i(\hat n\cdot \hat n')d\hat n' + \Delta
T^{{\rmn}i}(\hat n) \,.
\label{Delta_T}
\end{eqnarray}
Here $\Delta T^{\rmc}(\hat n)$ and $ \Delta T^{\rmf i}(\hat n) $ are
respectively CMB and foreground component of the anisotropy in the
channel,$i$.  The detector noise in the channel, $i$, is $\Delta
T^{\rmn i}$.  The beam function $B^i(\hat n\cdot \hat n')$ represents
the smoothing of the map due to finite resolution of the antenna of
channel, $i$. The beam is assumed to be circularly symmetric
as done in most analysis. We note that the detector noise is not
affected by the beam function.
An experiment such as WMAP provides multi-frequency maps
$\widetilde\Delta T^{ i}(\hat n)$, $i=1,2,...,n_c$ corresponding to
observation of the CMB at $n_c$ different frequency
bands. Equivalently, in the spherical harmonic representation
\begin{equation}
\tilde a^i_{lm} = \left( a_{lm}^{\rmc} + a_{lm}^{\rmf i}\right) B_l^i +
a_{lm}^{\rmn i}\,.
\label{alm}
\end{equation}
where $a_{lm}$ are respective spherical harmonic contributions to the
maps and $B_l$ are Legendre transform coefficients of the beam
$B^i(\hat n\cdot \hat n')$.
The aim is to linearly combine the maps with appropriate weights to
get an optimal estimator of the CMB anisotropy $\Delta T^\rmc(\hat
n')$ that minimizes the contribution from $ \Delta T^\rmf(\hat n')$.
The linear combination of the multi-frequency maps available can be
carried out in pixel space, or, in the equivalent representation in
terms of the spherical harmonic coefficients. The former has been
followed by the WMAP team to produce Internal Linear Combination (ILC)
map and also in a related, but more elaborate approach in Ref.~\cite{erik04}
to produce LILC map. The approach of carrying out a
multi-frequency maps combination in the spherical harmonic space was
proposed in Ref.~\cite{Tegmark96}.  For the
first year of WMAP data this was implemented in 
Ref.~\cite{Tegmark}. This method has the advantage that we
can simultaneously take into account variation of foreground with sky
positions and with different multipoles for a given sky position.

We define a cleaned map as a linearly weighted sum of the maps at
different frequencies,
\begin{equation}
a_{lm}^{\rm Clean}=\sum_{i=1}^{i=n_c}\hat W_l^{i}\frac{a_{lm}^i}{B_l^i} \, .
\label{c_map}
\end{equation}
Here $\hat W_l^i$ is a weight factor which depends upon the multipole
$l$ and the frequency channel, $i$. Since each of the channels has a 
different beam resolution, the maps are deconvolved by the
corresponding circularly symmetric beam transform functions $B_l$ prior to the linear combination.
The total power in the cleaned map at a given multipole $l$ is then
\begin{eqnarray} 
\hat C_l^{Clean}= \frac{1}{2l+1}
\sum_{m=-l}^{m=l}{a_{lm}^{Clean}a_{lm}^{Clean*}} \, .
\label{c_map_p}
\end{eqnarray}
Substituting eq.~(\ref{c_map}) in eq.~(\ref{c_map_p}) we obtain,
\begin{equation}
\hat C_l^{Clean}=\bf \hat W_l\hat {C_l}\bf\hat W_l^T \,,
\label{cleaned_power}
\end{equation}
where the matrix {$\bf \hat C_l$} is given by
\begin{equation}
{\bf \hat {C_l}}=\left(
\begin{array}{cccc}
\frac{\hat C_l^{11}}{B_l^1B_l^1} & .. & ..  & \frac{\hat C_l^{1n_c}}{B_l^1B_l^{n_c}}\\
.. & .. & ..  & ..\\
.. & .. & ..  & ..\\
\frac{\hat C_l^{n_c1}}{B_l^{n_c}B_l^1} & .. & ..  & \frac{\hat C_l^{n_c n_c}}{B_l^{n_c}B_l^{n_c}}
\end{array}
\right) \, ,
\end{equation}
{$\bf \hat W_l$} is a row vector describing the weights for
different channels,
\begin{equation}
{{\bf \hat W_l}=(\hat w_l^1, \hat w_l^2,, \ldots, \hat w_l^{n_c})} \,
\label{weight}
\end{equation}
and $\hat C_l^{i,j}$ is the cross power spectrum between the $i^{th}$
and $j^{th}$ channel,
\begin{equation}
\hat C_l^{i,j}=\sum_{m=-l}^{m=l}\frac{a_{lm}^ia_{lm}^{j*}}{2l+1}=\frac{a_{
l0}^ia_{l0}^j}{2l+1}+2 \Re\sum_{m=1}^{m=l}\frac{a_{lm}^ia_{lm}^{j*}}{2l+1} \, .
\end{equation} 
By construction, the ${\bf \hat C_l}$ matrix is symmetric. The
spherical harmonic coefficients extracted from a map contain CMB
signal and foregrounds, both smoothed by the beam function of the 
optical instrument used in the experiment, as well as detector noise. Using eq.~(\ref{alm}) we obtain,
   \begin{equation}
   a_{lm}= B_la^s_{lm} + \underbrace {B_la^F_{lm} + a^N_{lm}}_{a^{junk}_{lm}} \, ,
   \end{equation}
where $a^{junk}_{lm} $ is used to denote the total non-CMB
contamination in the map. Since the CMB contribution is independent of frequency, the expression for $\frac{\hat C_l^{ij}}{B_l^iB_l^j}$
simplifies to
\begin{eqnarray}
\nonumber\frac{\hat
C_l^{ij}}{B_l^iB_l^j}=\frac{1}{2l+1}\sum_{m=-l}^{m=l}\frac{\left(B_l^i
a_{lm}^{ S}+a_{lm}^{i (junk)
}\right)}{B_l^i}\times\frac{\left(B_l^ja_{lm}^{S* }+a_{lm}^{j (junk)*
}\right)}{B_l^j}= \hat C_l^{S}+\hat C_l^{ij (junk)} \, ,
\end{eqnarray}
for all values of $i$ and $j$.  
Now using eq.~(\ref{cleaned_power}) we obtain 
\begin{equation}
\hat C_l^{Clean}=\hat C_l^{S}\bf \hat W_le_0e^T_0\bf\hat W_l^T+ \bf
\hat W_l\hat C_l^{(junk)}\bf\hat W_l^T \, ,
\label{C_power}
\end{equation}
where { $\bf e_0$ } is a column vector with unit elements
\begin{equation}
\bf e_0 =\left(
\begin{array}{c}
1  \\
..  \\
..  \\
1
\end{array}
\right) \, .
\end{equation}
The CMB signal power in the cleaned map is kept unaltered by
imposing the constraint 
\begin{equation}
{\bf \hat W_le_0=e^T_0\hat W^T}=1 \, ,
\label{constraint}
\end{equation}
on weights $\hat W_l^{i}$. Using eqs.~({\ref{C_power}}) and
({\ref{constraint}}), the total power at a multipole $l$ in the cleaned map
is
\begin{equation}
\hat C_l^{Clean}= \hat C_l^{S}+\bf \hat W_l\hat C_l^{junk}\bf\hat W_l^T \, .
\end{equation}
{\em Thus the CMB signal power is only an additive positive constant in the
expression of the total power of the weighted map.} Hence, choosing
weights that minimize $\hat C_l^{\rm Clean}$ also minimizes the combined
contamination coming from foreground and detector noise. 
It may be shown easily, using the Lagrange's multiplier method, that
minimizing $\hat C_l^{\rm Clean}$ 
subject to the constraint eq.~(\ref{constraint}), gives the following
expression for the weight factors~\cite{Tegmark96,Tegmark2000a},
\begin{equation}
{\bf \hat W_l=\frac{e^T_0 \hat C_l^{-1}}{e^T_0\hat C_l^{-1}e_0}} \, .
\label{Weight}
\end{equation}
Following eqs.~(\ref{cleaned_power}) and (\ref{Weight}) we can express the
power in the cleaned maps neatly as
\begin{equation}
\hat C^{Clean}_l=\frac{1}{\bf e^T_0\hat C_l^{-1}e_0} \, .
\end{equation}

It is important to note that in cases when $\bf \hat C_l$ is singular,
it is possible to generalize the above expressions for weights and cleaned
power spectrum in terms of the {\em Moore-Penrose Generalized inversion} (MPGI) of $\bf \hat
C_l$. The generalized weights and cleaned power spectrum are then
given by
\begin{equation}
{\bf \hat W_l=\frac{e^T_0 \hat C_l^{\dagger}}{e^T_0\hat C_l^{\dagger}e_0}} \, ,
\label{Weight_mpi}
\end{equation}

\begin{equation}
\hat C^{Clean}_l=\frac{1}{\bf e^T_0\hat C_l^{\dagger}e_0} \, .
\label{Clean_mpi}
\end{equation} 
For further details of the analytic derivation of the weights and
cleaned power spectrum we refer to appendix \ref{Weight_eqn}.

\subsection{Biases in the foreground cleaning method}
\label{Bias}

Although the foreground cleaning is performed with the
constraint that CMB power spectrum remains preserved, 
the method biases the final power spectrum. This section is
devoted to a discussion of the full bias in the method.

The existence of some bias is not difficult to anticipate and understand.
The method is intended to perform a minimization of the foreground
power spectrum which is a positive definite quantity. Unless the
foreground cleaning is fully effective at all multipoles, the
minimization would leave some residual foreground which naively would
give rise to a positive bias in the cleaned power spectrum. However,
it is very interesting that there exists an additional negative
bias in the method. This negative bias is strongest at the lowest
multipoles and increases in magnitude with increase in number of
channels that are combined.

Let us consider $n_c$ number of channels in the linear combinations
for the cleaned map.  The maps at each channel consists of the CMB
signal and foreground contamination coming from the different
foreground components (synchrotron, free-free, dust etc). Although,
the number of distinct foreground components could be arbitrary, for
brevity and simplicity, here we club the contributions from all
the components into a single foreground term.  The foreground
contribution can be described by a covariance matrix. Later, after
simplification of the expressions, we can split up the total
foreground covariance matrix in terms of the covariance matrices of
the distinct constituent components. In the entire discussion that
follows we do not treat foregrounds as a stochastic
component. Instead we consider them as fixed templates in all the
realizations. This is entirely justified. We are simply interested in 
computing a foreground free template rather than estimating information
about the distribution of foreground from which they are drawn. We
discuss the bias in few different cases depending upon the rank of the
covariance matrix $\bf \hat C_l$.

\subsubsection{Case: Rank ${\bf \hat C_l} \le n_c-1$}

First we consider an ideal case  wherein detector noise remains absent
and each foreground component follows a rigid frequency scaling on the
entire sky. Let the number 
of foreground components be $n_f$. Then the rank of the 
foreground covariance matrix, $\bf C^f_l$, is also $n_f$.

The $p^{th}$ foreground component for channel $i$ is denoted by
$F^p_0(\theta,\phi) f^i_p$, where the frequency dependence, $f^i_p$
and the spatial (sky) dependence $F^p_0(\theta, \phi)$ are explicitly
separable in the rigid scaling assumption. (Here, $F^p_0(\theta, \phi)$
is the $p^{th}$ foreground template based on frequency $\nu_0$, so
that $f^i_p=1$, for frequency $\nu_0$.) We denote the CMB component by
 $C(\theta, \phi)$. Full signal map at frequency
channel, $i$, is then given by
\begin{eqnarray}
S^i(\theta, \phi) = C(\theta, \phi) +
\sum_{p=1}^{n_f}F^p_0(\theta,\phi)f^i_p \, .
\end{eqnarray}
Alternatively, in the spherical harmonic space,
\begin{eqnarray}
a^i_{lm}= a^c_{lm} + \sum_{p=1}^{n_f}f^i_pa^{p0}_{lm} \, .
\end{eqnarray}
The auto power spectrum of the $i^{th}$ channel
\begin{eqnarray}
\hat C^i_l = \hat C^c_l + 2\sum_{p=1}^{n_f}f^i_p\hat C^{cf(p)0}_l + \sum_{p,p'}^{n_f}f^i_pf^i_{p'} C^{(pp')0}_l \, .
\end{eqnarray}
In the above equation, $C^{(pp')0}_l $ is the correlation between any two
foreground components $p,p'$ and $\hat C^{cf(p)0}_l$ denotes the
chance correlation between CMB signal and $p^{th}$ foreground
component.  The cross power spectrum between two channels $i,j$ is
given by
\begin{eqnarray}
\hat C^{ij}_l = \hat C^c_l + \sum_{p=1}^{n_f}f^i_p\hat C^{cf(p)0}_l + \sum_{p=1}^{n_f}f^j_p\hat C^{cf(p)0}_l+ \sum_{p,p'}^{n_f}f^i_pf^j_{p'} C^{(pp')0}_l \, .
\label{cov}
\end{eqnarray} 
It is convenient to define the vectors,
\begin{eqnarray}
 {\bf  \hat f^{p0}_{l}}= \hat C^{cf(p)0}_l\left(\begin{array}{c}
f^1_p  \\
f^2_p  \\
.     \\
f^{n_c}_p 
\end{array}\right) \, ,
\end{eqnarray}
and 
\begin{equation}
{\bf \hat f^{0}_l}= \sum_{p=1}^{n_f}{\bf \hat f^{p0}_l}  \, .
\end{equation}
A little algebraic manipulation allows us to recast
eq.~(\ref{cov}) as
\begin{eqnarray}
{\bf \hat C_l} = \hat C^c_l{\bf e_0e^T_0} + {\bf \hat f^{0}_l} {\bf e^T_0} + {\bf e_0}{\bf \hat f^{0T}_l} + {\bf C^f_l}  \, .
\end{eqnarray} 
The above equation will be useful to compute the bias in the cleaned power spectrum. On the ensemble average the cleaned power spectrum as given by eq.~(\ref{Clean_mpi}) could be simplified with the help of successive use of a set of theorems reported in Refs.~\cite{CDM,JKB}. An elaborate discussion of these
theorems is given in appendix~\ref{bias_lowl}. Assuming statistically isotropic CMB sky we obtain the following expression for the ensemble average of the cleaned power spectrum
\begin{eqnarray}
 \left<\hat C^{Clean}_l\right> = \left<\hat C^c_l
 \right>-n_f\frac{\left<\hat C^c_l\right>}{2l+1} \, .
 \label{B1}
\end{eqnarray}
We easily note a few interesting aspects of the above equation. First
of all there exists a negative bias.  Secondly, 
because of $\sim {1}/(2l+1)$ decay, this bias is important at the lowest 
multipoles. Another important point to note is
that the bias depends on the underlying CMB power spectrum.  Therefore it
is possible to debias any given statistically isotropic CMB model in this approach by constructing an appropriately scaled estimator.  Lastly, we find that there is no foreground bias. One would naively expect the foregrounds would  contribute a positive bias in the power spectrum. However this does not happen in this case as the rigid scaling assumption along with the condition $n_c\ge n_f+1$ ensure that sufficient amount of spectral information are available to remove all the foregrounds. The negative bias arises as the weights are to be determined from the empirical covariance matrix to take into account information available from the observed data.

\subsubsection{Case: Rank $({\bf C^f_l })= n_c$}
\label{full_rank}
The rigid scaling assumption for the foreground contaminants
considered in the previous section is at best a reasonable
approximation and is known not to be valid in general. As mentioned in Ref.
 ~\cite{bouc_gisp99} a foreground component with varying spectral 
index over the sky could be approximated in terms of two templates, 
provided the variation is small compared to the mean spectral index 
over the sky. A stronger variation will need more 
 than two templates for reasonable modeling. In such a situation, if the number of templates 
required for modeling of all the foreground components exceeds the
number of maps available for linear combination then $\bf C^f_l$ is of full
rank. In this case a positive foreground bias appears along with 
a negative bias. The negative bias is similar to the previous case in that
it remains proportional to $ \frac{\left< \hat C^c_l \right>}{2l+1}$. Keeping aside 
the detector noise for the moment, the ensemble average of the cleaned power 
spectrum is given by
\begin{eqnarray}
\left<\hat C^{Clean}_l\right> = \left < \hat C^{c}_l\right> +{\bf \frac{1}{e^T_0{(C^{f}_l)}^{-1}e_0}}+ (1-n_c)
\frac{\left < \hat C^{c}_l \right>}{2l+1} \, .
\label{an_bias}
\end{eqnarray}
Detailed derivation of the above equation is similar to derivation of eq.~(\ref{B1}).  
As $\bf C^f_l$ is of full rank and a positive definite matrix, the second term 
on the right causes a positive foreground bias.

\subsubsection{Noise and foreground case}
The discussion in the last two sections does not consider any detector 
noise. We now consider the most general case where we have both foreground and
detector noise. Following a method similar to that used in derivation of 
eqs.~(\ref{B1}) and ~(\ref{an_bias})
we can show that, on the ensemble average the cleaned power spectrum in this
case is given by
\begin{eqnarray}
\left < \hat C^{Clean}_l \right >= \left< \hat C^c_l \right>+ {\bf \left< \frac{1}{e^T_0 {(\hat C^{f+N}_l)}^{-1}e_0} \right>} + (1-n_c)\frac{\left <\hat C^{c}_l\right>}{2l+1} \, .
\label{bias_fg_nse}
\end{eqnarray}
We have carried out Monte-Carlo simulations to verify the analytical
result given by eq.~(\ref{B1}). We perform simulations of foreground
cleaning using $n_c=3$ channels, corresponding to $41, 61 $ and $94$
$\Ghz$ frequencies.  The foreground model consists of a synchrotron and
free-free emission. Each of the foreground component is assumed to
follow a rigid frequency scaling all over the sky. First we generate
synchrotron and free-free templates using the Planck Sky 
Model~\footnote {We acknowledge the use of version 1.1 of the Planck
reference sky model, prepared by the members of Working Group 2 and
available at www.planck.fr/heading79.html.} at the $3$ different
frequencies. Since these templates do not follow rigid frequency
scaling all over the sky, ( eg., synchrotron spectral index varies
with the position of the sky), we developed a method to
regenerate rigid scaling synchrotron and free templates at these $3$
different frequency channels~\footnote{For this purpose, 
we first note that for each component the approximate normalization ${\sqrt
{C^{i}_l/C^{23}_l }} $ remains roughly constant for all
$l$. Here $C^{23}_l $ is the synchrotron (or free -free) power
spectrum for frequency $23$ GHz and $C^{i}_l$ is the synchrotron ( or
free-free) power spectrum at any of other $3$ frequencies. We
generate a synchrotron ( or free-free ) template at the $i^{th}$ 
frequency channel following a scaling of the $23$ GHz template by 
the number ${\sqrt {C^{i}_l/C^{23}_l }} $. This
ensures that the emisssion for each foreground component at different
frequencies  follows rigid frequency scaling.}
  
 \begin{figure}[h]
\includegraphics[scale=0.3,angle=-90]{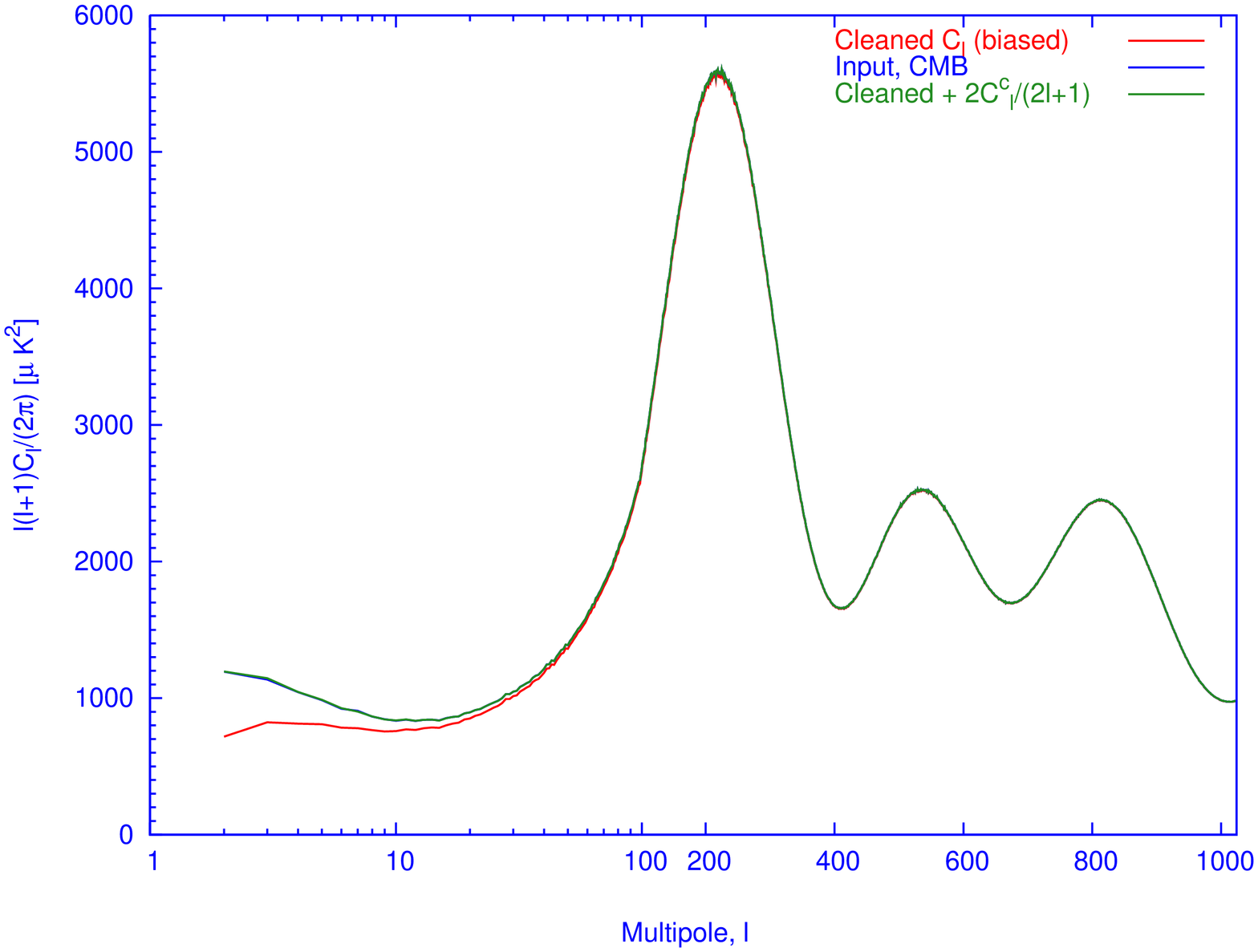}
\includegraphics[scale=0.3,angle=-90]{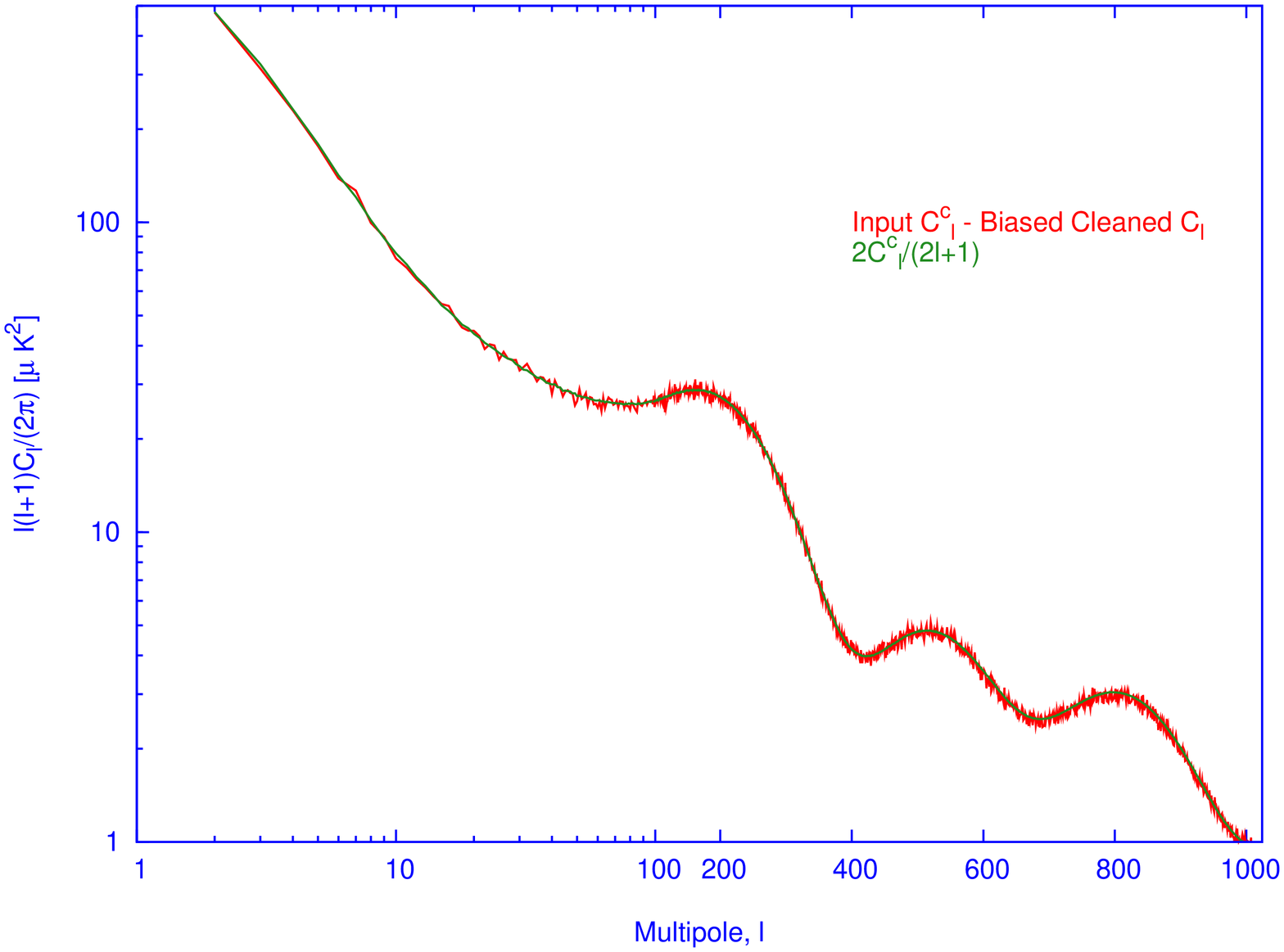}
\caption{The negative bias in the extracted power spectrum at low $l$ is 
shown by the red line in the left panel. The bias corrected spectrum 
is plotted in blue line. However it lies entirely behind the green
line and is not visible.  
 The right panel shows how
well the analytic results for bias match with those obtained from
Monte Carlo simulations. 
 }
\label{bias_fg_A}
\end{figure}  
The left hand panel of the fig.~\ref{bias_fg_A} based on $1000$ Monte Carlo
simulations shows that there exists a negative bias in the cleaning
method. For the assumed model of foreground components with rigid
frequency scaling, the second term on the right in eq.~(\ref{B1}) contributes 
 $-2{\left<\hat C^{c}_l\right>}/{2l+1}$ as a negative bias. In the right hand
panel of the fig.~\ref{bias_fg_A} we explicitly show that the
magnitude of the negative bias is exactly compensated by
$2\frac{\left<\hat C^{c}_l\right>}{2l+1}$. The negative bias is important at the
lowest multipoles and becomes negligible at high $l$, e.g. the bias is only
$\approx 3 \mu K^2$ at multipole $l$ $\approx 800$. The bias corrected
spectrum plotted in green color in the left panel of this
fig.~\ref{bias_fg_A} is completely hidden by the blue curve corresponding
to the input CMB power spectrum which is used to generate random realizations 
of CMB maps. 

The bias in the cleaned power spectrum for different cases described so far
is computed following the assumption that foreground cleaning is simultaneously 
carried out over the entire sky. However for the cleaning purpose of WMAP maps 
we follow a more sophisticated method. The main reason behind opting for a 
sophistication is that the spectrum of foreground emission as well the amplitude 
strongly depend on the location in the sky. The effectiveness of the foreground cleaning in varying foreground 
spectral index has been studied in the existing literature, ~\cite{bouc_gisp99,Tegmark98}.
Following the discussion in section \ref{full_rank} a varying foreground spectra would cause a positive 
foreground bias in the cleaned power spectrum. 
To minimize this bias we followed 
Ref.~\cite{Tegmark} and divide the sky into $9$ regions based on a simple 
estimation of the level of foreground contamination
(see appendix~\ref{mask_making}).
We then carry out foreground cleaning iteratively taking one region at a time
starting with dirtiest region. The advantage of such iterative method is that diffuse 
foreground contamination which is dominant at the low multipoles compared to the detector noise
could be very effectively minimized without a precise model of residual bias. The main principle 
for iterative cleaning is to search for sky regions where foreground emission could be assumed approximately constant or at least nearly constant. Such a philosophy of foreground cleaning has been adopted by WMAP team to produce their Internal Linear Combination Map \cite{Bennett1}. In a future publication we would generalize the bias results reported in this article to the most general method of iterative cleaning. Another advantage of iterative method is that one has the freedom to adjust weights
on different sky positions depending upon the amplitude of the foreground components leading to a 
better cleaning.

\subsection{Power spectrum Estimation}
\label{pow_es}
The power spectrum is estimated by cross correlating multichannel 
foreground cleaned maps
from independent Differencing Assemblies (DA).
The alternative is to use auto correlation and use 
a detector noise model to debias the power
spectrum. In this case the detector noise model uncertainty 
affects the mean of the
cleaned power spectrum. We note that the WMAP team too used
cross-correlation of independent DA to obtain their power spectrum.
Let us consider an hypothetical experiment with $n_c$ number of
frequency channels each of which is denoted by $i$. Each frequency
channel consists of $d$ number of independent detectors denoted by $\rmD^i_j$. 
Thus $\rmD^i_j$ is the map observed by $j^{th}$ detector of
$i^{th}$ channel. We then propose to form several cleaned maps in such
a way that each cleaned map consists of one map from each of the $n_c$
number of channels. By choosing appropriate combinations of DA maps we 
identify two different cleaned maps having totally disjoint set of detectors. 
Then assuming that the
noise properties are uncorrelated in two different detectors,
the noise bias will not affect the cross power spectrum of two such
cleaned maps at the $1^{st}$ order.  The final power spectrum could be a
simple average of all possible such cross power spectra. 
\begin{table}
\begin{tabular}{|c |c |}
\hline
$n_c$-channel combination    & Cleaned Map \\		
\hline
 & \\
$\rmD^1_1+ \rmD^2_1+...+\rmD^{n_c}_1$ & $\rmCC_1$\\
 $\rmD^1_2+ \rmD^2_2+...+\rmD^{n_c}_2$& $\rmCC_2$\\
$\rmD^1_3+ \rmD^2_3+...+\rmD^{n_c}_1$ & $\rmCC_3$\\
$..........................$ & $....$\\
 $\rmD^1_d+ \rmD^2_d+...+\rmD^{n_c}_d$&  $\rmCC_d$\\
  & \\ 
 \hline  
\end{tabular}
 \caption{List of several possible cleaned maps with uncorrelated detector
 noise properties. Here we assume $n_c$ number of frequency bands with 
 each band having $d$ number of independent detectors.}
\label{tab3}
\end{table}
In table \ref{tab3} we show a set of possible cleaned maps 
$\rmCC_j, j=1,2,..,d$ with uncorrelated
detector noise properties. Hence $\frac{d(d-1)}{2}$ number of cross power spectra
can be obtained that do not have any noise bias. In general the
number of possible cleaned maps is more than $d$. If we divide
the combination of cleaned map $1$ in $p$ number of subsets and
the combination of cleaned map $2$ in $q$ number of subsets then one only
needs to make sure that these $p$ and $q$ subsets have
uncorrelated noise properties.

\section{Implementation on WMAP Data}
 \label{wmap_data}
In our method we linearly combine maps corresponding to a set of 4 DAs
 at different frequencies.~\footnote {In 
Ref.~\cite{Tegmark}, a single foreground cleaned map is obtained by
linearly combining 5 maps corresponding to one each for the different
WMAP frequency channels. For the Q, V and W frequency channels, where
more than one maps were available, an averaged map was used. However,
averaging over the DA maps in a given frequency channel precludes any
possibility of removing detector noise bias using cross
correlation.}  We treat the K and Ka maps effectively as
observations of the CMB sky in two different DA at the low
frequencies. Therefore, we use either K or Ka maps in any
combination. In case of the W band, 4 DA maps are available. We simply
form pairwise averaged map taking two of them at a time and form 6
effective DA maps corresponding to W band. W$ij$ represents simply an
averaged map obtained from the $i^{\rm th}$ and $j^{\rm th}$ DA of W
band.  In table \ref{tab2} we list all the $48$ possible linear
combinations of the DA maps that lead to `cleaned' maps, C${\bf i}$
and CA${\bf i}$'s, where {${\bf i}$ = 1, 2, \ldots, 24}. In an
alternative approach, we form combinations to form cleaned maps
excluding the most foreground contaminated K and Ka bands (referred to
as the three-channel $n_c=3$ case). This leads to a total of $24$
cleaned maps. All such combinations are shown in the bottom panel of
the table \ref{tab2}. In this combination the cleaned maps are labeled
as C${\bf i}$ s, where {${\bf i}$ = 1, 2, \ldots, 24}.

\begin{table}
\scriptsize
\begin{tabular}{|c |c|}
\hline
4-channel combinations ($n_c=4$)                          &                            \\		
\hline
 & \\
(K,KA)+Q1+V1+W12=(C1,CA1)&(K,KA)+Q1+V2+W12=(C13,CA13)\\
(K,KA)+Q1+V1+W13=(C2,CA2)&(K,KA)+Q1+V2+W13=(C14,CA14)\\
(K,KA)+Q1+V1+W14=(C3,CA3)&(K,KA)+Q1+V2+W14=(C15,CA15)\\
(K,KA)+Q1+V1+W23=(C4,CA4)&(K,KA)+Q1+V2+W23=(C16,CA16)\\
(K,KA)+Q1+V1+W24=(C5,CA5)&(K,KA)+Q1+V2+W24=(C17,CA17)\\
(K,KA)+Q1+V1+W34=(C6,CA6)&(K,KA)+Q1+V2+W34=(C18,CA18)\\
(K,KA)+Q2+V2+W12=(C7,CA7)&(K,KA)+Q2+V1+W12=(C19,CA19)\\
(K,KA)+Q2+V2+W13=(C8,CA8)&(K,KA)+Q2+V1+W13=(C20,CA20)\\
(K,KA)+Q2+V2+W14=(C9,CA9)&(K,KA)+Q2+V1+W14=(C21,CA21)\\ 
(K,KA)+Q2+V2+W23=(C10,CA10)&(K,KA)+Q2+V1+W23=(C22,CA22)\\ 
(K,KA)+Q2+V2+W24=(C11,CA11)&(K,KA)+Q2+V1+W24=(C23,CA23)\\ 
(K,KA)+Q2+V2+W34=(C12,CA12)&(K,KA)+Q2+V1+W34=(C24,CA24)\\
                           &                           \\
 \hline
 \hline

\hline
3-channel combinations  ($n_c=3$)                        &                            \\
\hline
                   &                            \\
Q1+V1+W12=C1 & Q1+V2+W12=C13\\
Q1+V1+W13=C2 & Q1+V2+W13=C14\\
Q1+V1+W14=C3 & Q1+V2+W14=C15\\
Q1+V1+W23=C4 & Q1+V2+W23=C16\\
Q1+V1+W24=C5 & Q1+V2+W24=C17\\
Q1+V1+W34=C6 & Q1+V2+W34=C18\\
Q2+V2+W12=C7 & Q2+V1+W12=C19\\
Q2+V2+W13=C8 & Q2+V1+W13=C20\\
Q2+V2+W14=C9 & Q2+V1+W14=C21\\
Q2+V2+W23=C10 & Q2+V1+W23=C22\\
Q2+V2+W24=C11 & Q2+V1+W24=C23\\
Q2+V2+W34=C12 & Q2+V1+W34=C24\\
                            &                      \\
\hline
\end{tabular}
 \caption{ 
The table on the top shows 48 different combinations of the DA maps
used in our 4 channel cleaning method. The bottom table shows the 24
possible combinations in the 3 channel cleaning method.}
\label{tab2}
\end{table}

The entire method leading to the power spectrum estimation consists of
three main steps. First we perform foreground cleaning using several
multi-channel combinations of WMAP maps. At the second step we obtain
cross power spectra from these foreground cleaned maps. The foreground
cleaning is similar to Refs.~\cite{Tegmark96,Tegmark}. 
Finally we correct for the estimated residual unresolved 
point source contamination. We note in passing that each of these steps 
are logically modular and each of them could be modified or improved 
independent of the other.

\subsection{Map cleaning}
\label{cleaning}
We use the `raw' DA maps (i.e., that have not undergone any foreground
cleaning process) both for the WMAP 1 year and WMAP 3 year data
release from the LAMBDA website. These maps follow HEALPix~\footnote {For 
comprehensive studies about HEALPix we refer to 
Ref.~\cite{Gorski99a,Gorski99,Calabretta}.} pixelization 
scheme at a resolution level $\nside = 512$ corresponding to approximately
$3$ million sky pixels. All these maps are provided in the
`nested' pixelization scheme which is suitable for nearest neighbor
searches. However for converting maps to spherical harmonic space and
vice versa it is computationally advantageous to convert them to ring
format that facilitates the use of Fast Fourier transformation method
along equal latitudes. Therefore prior to the analysis all the maps
corresponding to 10 DA's were converted to `ring'
pixelization scheme. When converting a map to spherical harmonic
space we restricted ourselves to a maximum multipole, $\lmax = 1024$.

The spectrum of foreground emission has some dependency on the location on the
sky. A better cleaning may be achieved if we partition the entire sky
into certain number of sky parts depending upon the level of
foreground contaminations~\cite{Tegmark}. Then cleaning is done for
each sky-parts iteratively. For each regions we will have then different
weights which are chosen to minimize foreground contamination from that
particular region. Cleaning becomes more efficient 
by allowing the weights to depend on the sky parts
rather than using a single set of weights for the entire sky. 
Following Ref.~\cite{Tegmark} we partition the sky depending
upon the level of foreground contamination.~\footnote{Although this
scheme is found to be effective, it is possible to envisage other
schemes such as those that make foreground contamination in each part
closer to rigid scaling approximation. In an ongoing study we have
investigated and demonstrated improvement in foreground cleaning for
different levels of partitioning~\cite{fgdiagon}.} 

\begin{figure}   
\centering
 \includegraphics[angle=90,width=10.5cm,totalheight=7cm]{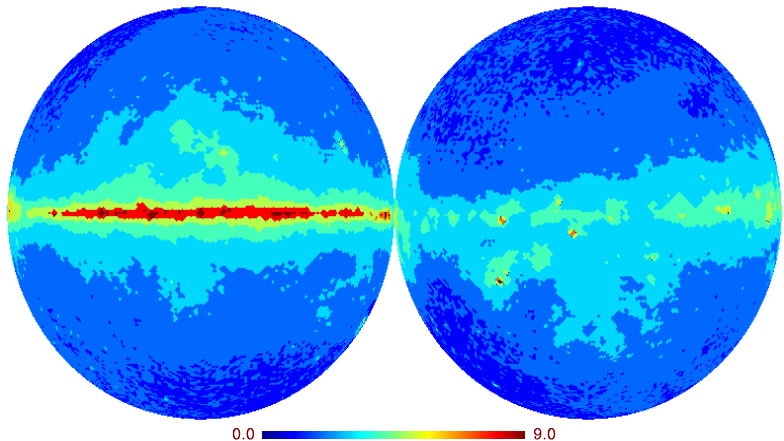}
 \caption{The 9 different masks in
 orthographic projection that correspond to the partitions of the
 sky. Each region is color coded to distinguish them visually. The
 cleanest region is the region farthest away from the galactic
 plane. The dirtiest region (black spots) lie in the galactic plane.}
\label{masks}
\end{figure}

The entire sky is partitioned in a total of $r=9$ regions
depending upon their foreground contamination. All the $r$ sky parts
are shown in the fig.~\ref{masks}. Individual sky parts are
color-coded according to the number $r$ assigned to them. 
The dirtiest part is labeled
with the maximum index. The procedure of forming these sky
parts are similar to Ref.~\cite{Tegmark} with small variants. We describe
our sky partitioning to identify these $r$ regions in detail in
appendix~\ref{mask_making}.

For each combination, the cleaning procedure is performed in $r$ 
iterations starting from the dirtiest region. As in Ref.~\cite{Tegmark} we 
call the initial foreground contaminated maps as the initial temporary
maps. At the end of each of iteration we obtain a set of partially cleaned 
temporary maps which are used as input in the next iteration.
In $i^{\rm th}$ iteration ($i = 1, 2, 3, ..., r$) we
perform the following three steps~:

\begin{enumerate}
\item{} Using the $i^{\rm th}$ mask we select the $i^{th}$ region of
 the sky from the temporary maps.  Then we obtain power spectrum
 matrix from the $i^{\rm th}$ region only.~\footnote {These power
 spectrum matrix is obtained from the partial sky spherical harmonic
coefficients. In
 practice while using 1 year WMAP data we have not combined K band map
 for $l > $ 603 ($B_l < 0 $ for K band for $l> 603$ and the negative
 values of beam function specified is unphysical.). KA and Q band beam
 functions are available till $l = 850$ and $l =1000$. Therefore we do
 not combine these bands beyond these values. Similar limits
 determined by the condition $B_l > 0 $ were used for 3 year data
 analysis also.}
\item{} Using the weight factors we obtain a full sky cleaned map
using eq.~(\ref{c_map}).
\item{}Next we replace the $i^{\rm th}$ region of the temporary
maps by the corresponding region of the cleaned map. Before
replacement the cleaned map is smoothed to the resolution of the
different frequency channels. After replacement the maps define new
temporary maps for the next iteration.

\end{enumerate}

The entire cleaning procedure is automated to carry the three steps
iterated $r$ times to obtain the final cleaned maps.
 The final cleaned maps have the
resolution of the W band.

In eq.~(\ref{weight}) the weights could have numerical errors if the
$C_l$ matrix becomes ill-conditioned at specific values of $l$. This
can happen if by chance any mode has almost equal contribution combined
from the CMB, foreground and detector noise. Hence, in practice, the numerical
implementation obtains weights using the $C_l$ matrix smoothed over a
range of $\Delta l= 11$ prior to inversion.

\begin{figure}
\includegraphics[scale=.33,angle=90]{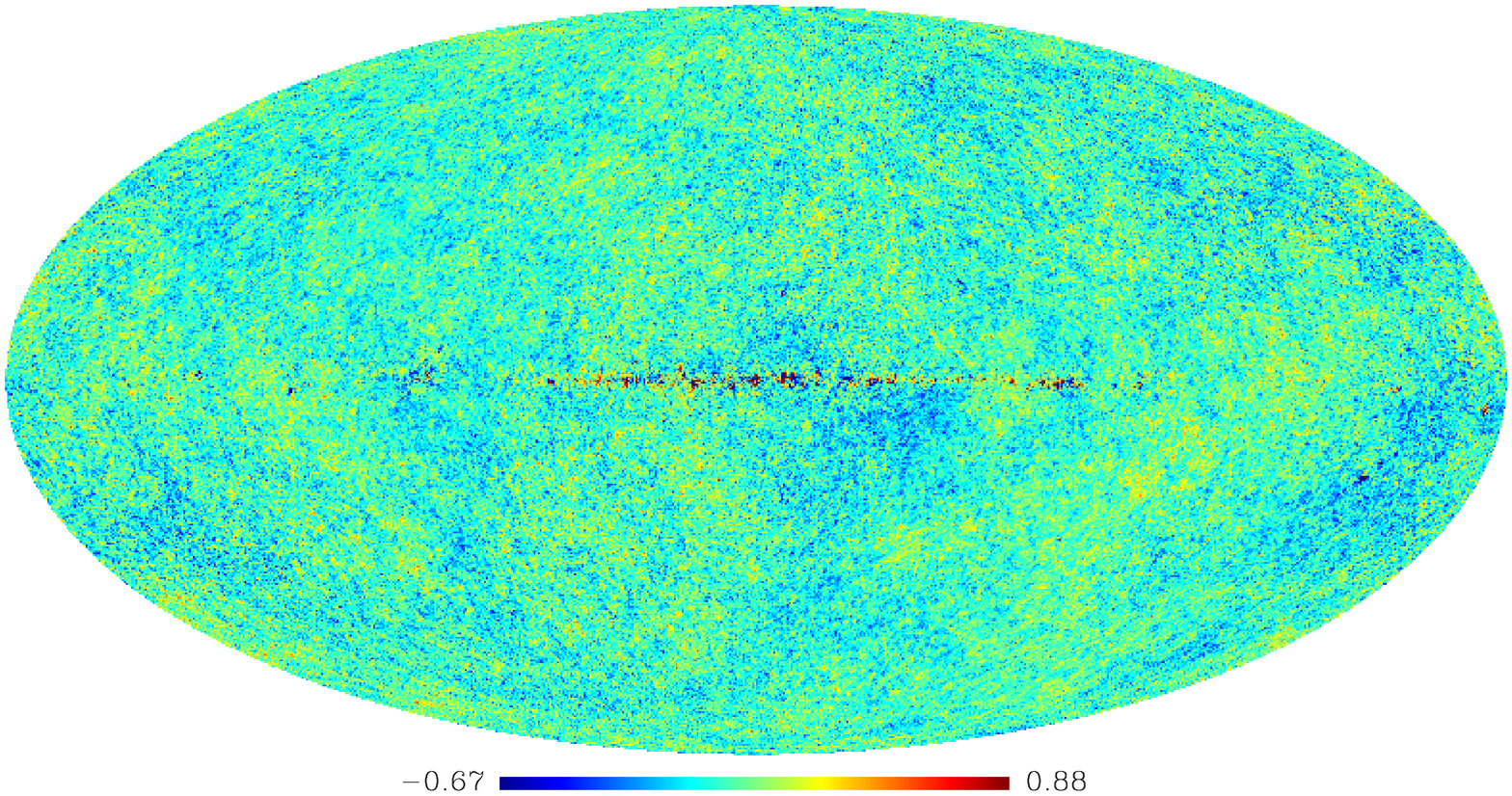}
\includegraphics[scale=.3,angle=00]{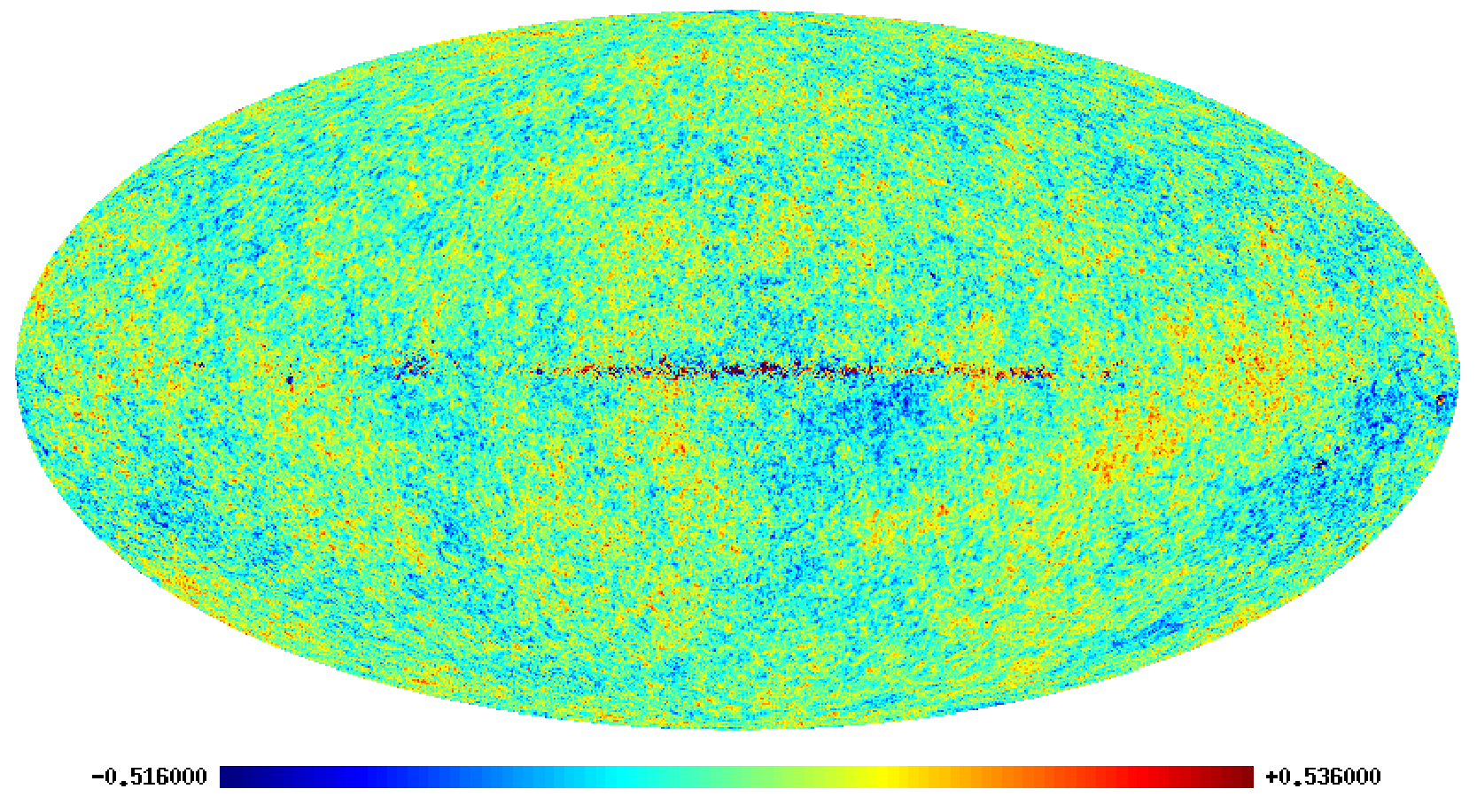}
\caption{ The left panel shows the cleaned map C8
using the 1 year WMAP data. There is some residual foreground
contamination near the galactic regions. 
The right panel shows the CA8 map using the 3
year WMAP data. The temperature scale of both the figures are chosen
from the sky part outside the Kp2 mask.}
\label{w13_ring.ps}
\end{figure}

As shown in table~\ref{tab2}, for the four channel combination
($n_c=4$), a total of $48$ cleaned maps can be obtained using all the
possible combinations of the DA maps for each of the WMAP 1 year and
WMAP 3 year data. The cleaned map C8 for WMAP 1 year data is shown in
the left panel of the figure \ref{w13_ring.ps}. There is some residual
foreground left in the galactic plane as seen in this map. We apply
Kp2 mask,  supplied by the WMAP team, to flag the contaminated pixels
near the galactic plane. Therefore these residuals do not affect our
final estimated power spectrum. A similar map is obtained from the
WMAP 3 year data also. The right panel of the figure \ref{w13_ring.ps} shows CA8
map from WMAP-3 data. Both the maps look similar.

\begin{figure}
\includegraphics[scale=0.35,angle=-90]{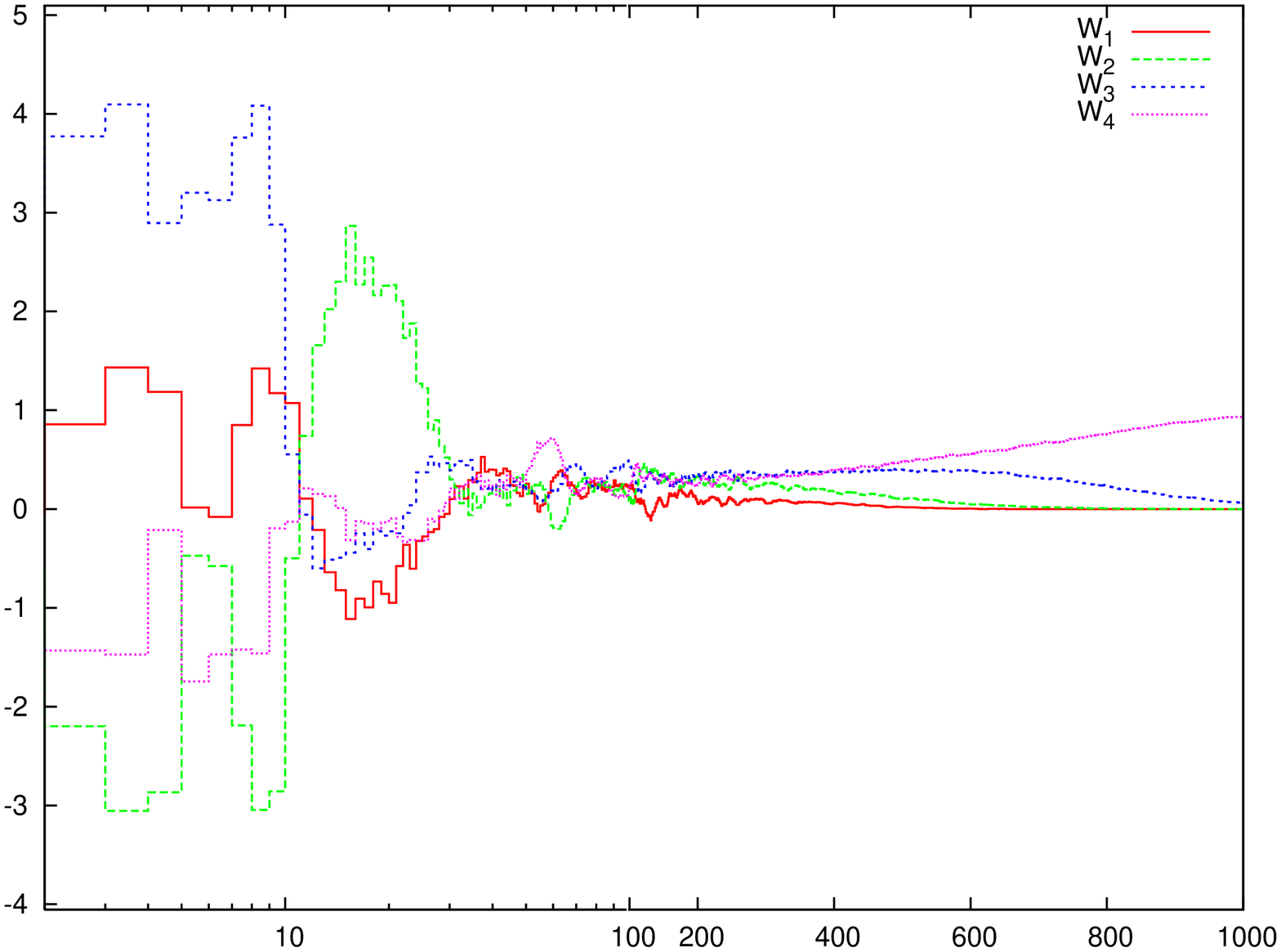}
\includegraphics[scale=0.35,angle=-90]{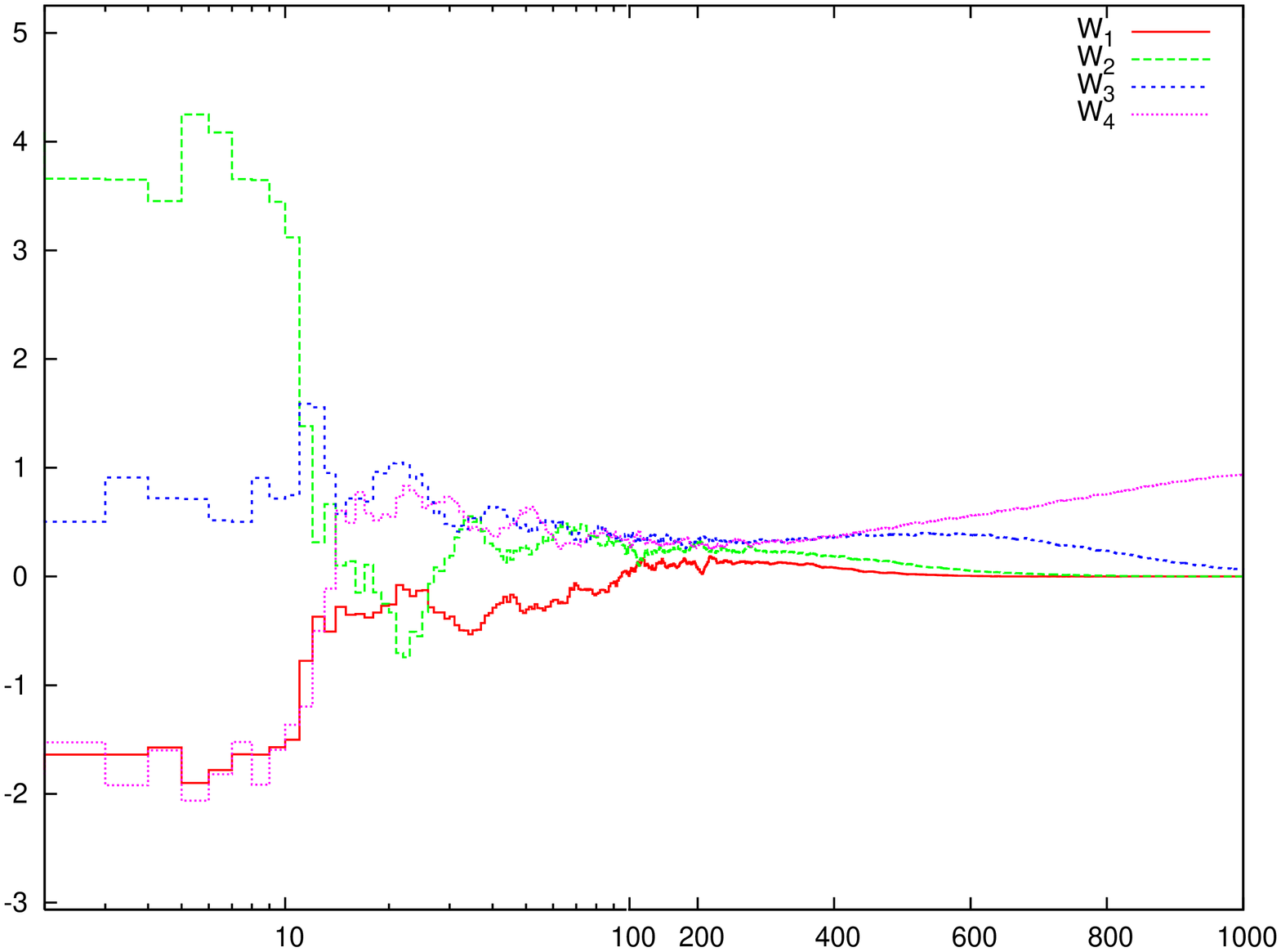}
\caption{The left panel of the figure shows the weights for the
cleanest region for the combination KA, Q1,V1,W12. The right panel
show the weights for the same combination but for the second cleanest
region. At low multipole foregrounds dominate. Therefore, weights take
large positive and negative values to subtract foregrounds.}
\label{wfig}
\end{figure}

The weights $\hat W_l$ are shown in the figures \ref{wfig} for one of the
cleaned map (CA1) for the cleanest and second cleanest
region.
For low $l$ where the diffuse foregrounds are dominant, the weights
take large positive and negative values to subtract the
foregrounds. At large $l$ the maximum weight is given to the W band
channel since it has the highest resolution.

\subsection{Power spectrum estimation}
\label{power_spectrum_estimation}

Our final power spectrum is based upon the MASTER estimate introduced
in Ref.~\cite{Hivon}. Even after performing foreground cleaning the galactic
disk region remains significantly contaminated. Hence one needs to exclude
this region before estimating the cosmological power spectrum. However, 
flagging the contaminated sky region effectively introduces uneven 
weighting of the pixels. 
Those flagged have efectively zero weight
and rest have unit weight. Weighting a map in the pixel space causes
neighboring multipoles to get correlated. The correlation of the multipoles
are described by the coupling matrix which is determined by the geometry of 
the sky coverage. Moreover, the power spectrum of such a map is biased
low because of less sky coverage. The MASTER method debiases the partial sky 
power spectrum on the ensemble average by inverting the coupling matrix.

The MASTER method originally implemented in Ref.~\cite{Hivon}
deals with auto power spectrum only. This can readily be generalized
to the case of cross power spectrum ~\cite{Xspect}. We first exclude the galactic region from all
the $48$ cleaned maps (Table~\ref{tab2},$n_c=4$) using Kp2 
mask as supplied by the WMAP team. Then map pairs C{\bf
i} \& C{\bf j} to be cross correlated are chosen such that they do not have any DA  common between
them. This choice ensures that the detector noise bias does not affect any of the
cross power spectrum. A total of $24$ cross power spectra could be obtained 
which are unaffected by the detector noise bias. Each of these cross spectra
are then debiased from the partial sky effect following Ref.~\cite{Xspect} 
using 
the coupling (bias) matrix corresponding to the Kp2 mask. The small scale 
systematic effects of beam and pixel smoothing were removed using appropriate 
circularized beam transform~\cite{Hivon} and pixel window functions as supplied 
by the HEALPix package. The $24$ cross power spectra are then combined
with equal weights into a single `Uniform average' power 
spectrum.~\footnote{There exists the additional freedom to choose optimal
weights for combining the $24$ cross-power spectra which we do not
discuss in this work.} The final power spectrum is binned in the same
manner as the WMAP's published result for ease of comparison.

Both one and three year power spectra were obtained following this
method. 
We defer a more detailed discussion on the 3 year power spectrum
results to the section \ref{3yr_results}. We also estimate a residual
contamination in the `Uniform average' power spectrum for both WMAP 1
year and WMAP 3 year data from the unresolved point sources. A point
source model used by WMAP team ~\cite{Hinshaw,Hinshaw06} and estimated
entirely from the WMAP data ~\cite{Bennett1} is sufficient for our
estimation. We recover a flat, (approximately $140\mu K^2$ for WMAP 1
year data and $100\mu K^2$ for WMAP 3 year data), residual point source
contamination in the two `Uniform average' power spectra for
$l \gsim 400$. This residual is much less than actual point source
contamination in Q, KA or K band and intermediate between V and W band
point source contamination.

\subsection{Estimation of residual unresolved point source power spectrum}
\label{ps_res}
There have been several studies regarding the 
point source contamination in the CMB maps~\cite{ps1,ps2,ps3}.
We estimate residual contamination due to unresolved point sources in
the `Uniform average' power spectrum following the point source model
constructed by the WMAP team~\cite{Hinshaw,Hinshaw06}. The WMAP point source 
model consists of a point source covariance matrix in thermodynamic temperature 
unit following $C^{ps (ij)}_l = c_{(ij)}A\left(\nu_i/\nu_0\right)^{-2}\left(\nu_j/\nu_0\right)^{-2}$ for two different frequency channel $i$ and $j$. Here $A$ is the amplitude of unresolved
point source spectrum in antenna temperature at a reference frequency $\nu_0$ and $c_{ij}$ is the conversion factor from antenna to thermodynamic temperature. We compute the
residual unresolved point source contamination in each of the $24$
cross power spectra assuming that the residual (extra galactic) point source contamination has a statistically isotropic distribution in the sky. As described 
in section \ref{cleaning}, the foreground cleaning procedure consists 
of $r$ number of iterative steps. The total point source contamination in any of the
cross power spectra consists of contribution from each of the
individual sky parts. The final residual unresolved point source contamination 
is the uniform average over all such $24$
residual point source contamination. Below we describe in detail 
residual point source contamination in a cross combination of two maps $(\bf i,j)$.

As stated in section \ref{power_spectrum_estimation} we apply Kp2 mask 
on the cleaned maps prior to cross power spectrum estimation. Apart from
removing the galactic region, Kp2 mask also removes a circular region of $0.6$
degree radius around each of the resolved point sources. Therefore 
the resolved point sources cannot affect the cross power spectra.
However in the detector noise dominated region, $l\ge 400$, the iterative 
foreground cleaning method cannot remove all power from the 
unresolved point sources. As resolved point 
sources are already masked it is important to estimate only residual unresolved point source contamination in each of the cross power spectrum.  

Let us assume that $a^{\bf i}_{lm} $ is the spherical harmonic
coefficients obtained from cleaned map $\rmCC_{\bf i}$, where ${\bf i}=1,2,
...,48$,
\begin{equation}
a^{\bf i}_{lm}= B^{\bf i}_l a^c_{lm} + B^{\bf i}_l a^{p(\bf i)}_{lm}\,.
\end{equation}
Here $a^{p(\bf i)}_{lm} $ is the residual unresolved
point source contamination. We note that we have not considered any 
detector noise contribution in this equation. However this does not imply any
loss of generality in the discussion of cross power spectrum. The noise
bias does not affect the cross power spectrum. Also we do not include 
any residual diffuse foreground in our study in this 
section. At the large multipole region where 
point source contribution becomes significant diffuse galactic contaminations 
are entirely subdominant and thus they are not a concern.

We assume a statistically isotropic model of the residual unresolved 
point source distribution over the sky.  Point sources are uncorrelated with CMB. On the ensemble average 
a partial cross power spectrum obtained from cleaned maps $\bf i$ and $\bf j$ is related 
to the CMB and point source power spectra as follows
\begin{eqnarray}
\left< \tilde C^{\bf ij}_l\right> = M_{ll'}\left( \left<\hat C^c_{l'} \right> + 
\left<\hat C^{p(\bf ij)}_{l'} \right>\right)B^{\bf i}_{l'}B^{\bf j}_{l'}p^2_{l'}\,.
\end{eqnarray}
Here, $M_{ll'}$ is the coupling matrix, $\hat C^{p(\bf ij)}_{l'}$ is the estimate for the full sky residual unresolved point source power spectrum and $B^{\bf i}_{l'}B^{\bf j}_{l'}p^2_{l'}$ denote combined effect of beam and pixel smoothing. We can recast the above equation as,
\begin{eqnarray}
\left<\hat C^{c}_l \right> = \frac{M^{-1}_{ll'} \left<C^{\bf ij}_{l'}\right>}{B^{\bf i}_{l'}B^{\bf j}_{l}p^2_{l}}-  \left<\hat C^{p(\bf ij)}_{l'}\right>.
\end{eqnarray}
The next task is to obtain the estimates $\hat C^{p(\bf ij)}_{l'} $ themselves
in each cross spectra. To compute them, let us assume that
$f^{\bf i}(\theta,\phi)$ is the residual point source function present in the ${\bf i}^{th}$ cleaned map. Our cleaning method partitions the sky into $9$
parts. The entire sky is then cleaned in a total of $9$ iterations. If
$g_k(\theta,\phi)$ is the point source residual present in the $k
^{th}$ sky part, we have
\begin{eqnarray}
f^{\bf i}(\theta,\phi) = \sum_{k=1}^{k=9}g_k(\theta,\phi) \, .
\end{eqnarray}
After expanding both sides in spherical harmonics we obtain
\begin{eqnarray}
a^{\bf i}_{lm} = \sum_{k=1}^{k=9}\tilde a^k_{lm} \, .
\label{ps1}
\end{eqnarray}
The partial sky unresolved residual point source modes could  be written in terms of the full-sky modes using
\begin{eqnarray}
\tilde a^k_{lm} = \sum_{l'm'}M^k_{lml'm'}a^k_{l'm'} \, .
\label{ps2}
\end{eqnarray}
We note that, $a^k_{l'm'}$ represents the residual point source
contamination in the {\it entire cleaned map} obtained after the $k^{th}$
iteration. This is different from the contamination, $\tilde a^k_{lm}$, that is actually present in the $k^{th}$ sky part. The symbol $ M^k_{lml'm'}$ 
denotes the mode mode
coupling matrix for the spherical harmonic modes for the $k^{th}$ sky
part. We can  rewrite $a^k_{l'm'}$ in terms of the spherical harmonic
coefficients of the temporary maps obtained at $k^{th}$ iteration as,
\begin{eqnarray}
a^k_{l'm'} = \sum_{i=1}^{i=4}a^{ki}_{l'm'}\hat w^{k i}_{l'} \, ,
\label{ps3}
\end{eqnarray}
where $\hat w^{k i}_l$ is the weight for the $i^{th}$ channel at the
$k^{th}$ iteration for multipole $l$. Following eqs.~(\ref{ps1}), (\ref{ps2}) and (\ref{ps3}) we obtain
\begin{eqnarray}
a^{\bf i}_{lm} = \sum_{k=1}^{k=9}\sum_{l'm'}M^k_{lml'm'}\sum_{i=1}^{i=4}a^{k
i}_{l'm'}\hat w^{k i}_{l'} \, .
\end{eqnarray}
The residual unresolved point source spectrum in one cross
combination $(\bf i, j)$ is given by
\begin{eqnarray}
\left<\hat C^{ps (\bf ij)}_l\right> =\sum_{m=-l}^{m=l} \frac{\langle a^{\bf i}_{lm}a^{*\bf
j}_{lm}\rangle}{2l+1} \, .
\end{eqnarray}
Substituting $ a^{\bf i}_{lm}$ and $a^{*\bf j}_{lm}$ we find,
\begin{eqnarray}
\left < \hat C^{ps (\bf ij)}_l\right>=\frac{1}{2l+1}\sum_{m=-l}^{m=l}
\sum_{k,k'=1}^{k,k'=9}\sum_{l',m'}\sum_{l'',m''}M^{k}_{lml'm'}M^{*k'}_{lml''m''}\sum_{i,i'=1}^{i,i'=4}\left<
a^{k i}_{l'm'}a^{*k' i'}_{l''m''}\hat w^{ki}_{l'}\hat w'^{k'i'}_{l''}\right> \, .
\label{pscorrect}
\end{eqnarray}
We note that the two set of weights corresponding to the two cleaned maps which are cross correlated are not strictly identical. Therefore, in the above equation, we have used a prime to distingush between the weights for two cleaned maps. 

We note that the primary foreground contamination at large multipole region comes from (extra-galactic) point sources. Diffuse galactic foregrounds are subdominant at large $l$. The beam deconvolution process leads to 
dominance of the effective noise of the DA maps over the unresolved point source contamination. As a result weights become entirely determined by the noise level of the maps and asymptotically ($l \rightarrow \infty$)
become constant for all realizations. Moreover, small differeneces in weights from realization to realization due to fluctuations in noise from the mean level is not a concern. These fluctuations could be further suppressed by computing binned estimate of the power spectrum. Using simulations of the cleaning procedure with realistic model of point sources and detector noise we have verified
that the weights remain effectively unchanged 
whether we use point sources or not. Since the pairs of cleaned maps that are cross-correlated have uncorrelated
noise, the corresponding weights $\hat w^{ki}$ and $\hat w'^{k'i'}$ could be treated as
uncorrelated with one another. Also, weights are effectively uncorrelated with
unresolved point sources as they are determined by the detector noise
only. In this case the residual unresolved point source contamination
becomes
\begin{eqnarray}
\left<C^{ps \bf ij}_l\right>=\frac{1}{2l+1}\sum_{m=-l}^{m=l} \sum_{k,k'=1}^{k,k'=9}\sum_{l',m'}\sum_{l'',m''}M^{k}_{lml'm'}M^{*k'}_{lml''m''}\sum_{i,i'=1}^{i,i'=4}\left<a^{k i}_{l'm'}a^{*k' i'}_{l''m''}\right>\left <\hat w^{ki}_{l'}\hat w'^{k'i'}_{l''}\right> \, .
\end{eqnarray}
Next we assume that the distribution of unresolved point sources is 
statistically isotropic over the full-sky, i.e.
\begin{eqnarray}
<a^{k i}_{l'm'}a^{*k' i'}_{l''m''}> = C^{ps (ii')}_{l'}\delta_{l'l''}\delta_{m'm''} \, .
\end{eqnarray}
Here $ C^{ps (ij)}_{l'}$ is the point source model as supplied by the WMAP team. This simplifies the expression for the residual unresolved point source contribution,
\begin{eqnarray}
\left<C^{ps \bf ij}_l\right>=\frac{1}{2l+1}\sum_{m=-l}^{m=l} \sum_{k,k'=1}^{k,k'=9}\sum_{l',m'}M^{k}_{lml'm'}M^{*k'}_{lml'm'}\sum_{i,i'=1}^{i,i'=4}C^{ps (ii')}_{l'}\left <\hat w^{ki}_{l'}\hat w'^{k'i'}_{l'}\right> \, .
\end{eqnarray}
The residual unresolved point source contamination then could be written as,
\begin{eqnarray}
\left<\hat C^{ps (\bf ij)}_l\right>=\frac{1}{2l+1}\sum_{m=-l}^{m=l} \sum_{k,k'=1}^{k,k'=9}\sum_{l',m'}M^{k}_{lml'm'}M^{*k'}_{lml'm'}\left<\bf \hat W^{k}_{l'} C^{ps}_{l'}\hat W'^{k'}_{l'}\right> \, .
\end{eqnarray}
Here, $\bf \hat W^k_l$ is the row vector of weights for sky part $k$ as
in eq.~(\ref{weight}). The elements of the point source power
spectrum matrix are given by $ C^{ps (ij)}_{l'}$. We note in passing that
this matrix is not symmetric because we are interested in residual
point source power in the cross combination of two maps obtained from
two cleaned maps which are linear combinations of (K, Q, V, W) and
(KA,Q,V, W) respectively. Explicitly we use the following form of the
matrix 
\begin{eqnarray}
C^{ps (ij)}_{l} =\left(\begin{array}{cccc}
C^{KKA}_l & C^{KQ}_l & C^{KV}_l & C^{KW}_l\\
C^{KAQ}_l & C^{QQ}_l & C^{QV}_l & C^{QW}_l \\
C^{KAV}_l & C^{QV}_l& C^{VV}_l &C^{VW}_l \\
C^{KAW}_l & C^{WQ}_l&  C^{WV}_l & C^{WW}_l
\end{array}
\right) \, .
\label{Clps}
\end{eqnarray}
We note that, in general, point sources may not be perfectly correlated
 in all the frequencies. The point source power spectrum assumed by
 the WMAP team and used in this work assume the existence of a
 single point source template which may not be perfectly
 true. However, our point source correction method could easily
 incorporate the extra information of point source decoherence from
 frequency to frequency in the matrix in eq.~(\ref{Clps}).

\begin{figure}
\includegraphics[scale=0.3,angle=-90]{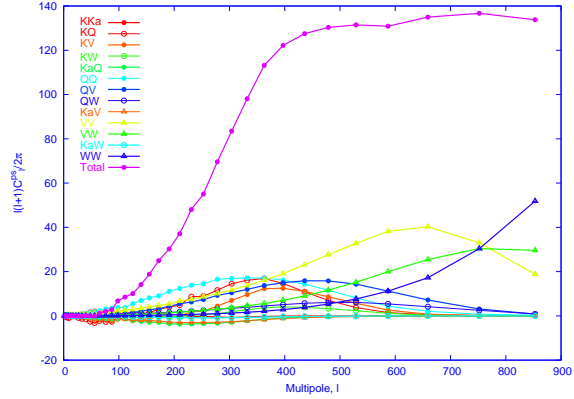}
\caption{The effective contribution from each of the elements of the matrix $ C^{ps (ij)}_{l'}$ in the total residual point source estimation, where $i,j$ are the index representing the 4 DA we use for combinations.}
\label{ps_matrix}
\end{figure}

From a detailed study of a single iteration cleaning
method we verified that a residual point source bias in the auto or cross power spectrum of the 
cleaned maps could be very well approximated by $\bf \hat W C^{ps}_l \hat W^T_l$, without a need to compute $\left <\bf \hat W C^{ps}_l \hat W^T_l\right>$. A detailed discussion of this analysis and the corresponding simulations 
are given in appendix \ref{res_ps_method}. 
Thus we propose $\left<\hat C^{ps ({\bf ij})}_l\right>\approx \hat C^{ps ({\bf ij})}_l$.
This simplifies the expression for the residual unresolved point source contribution
\begin{eqnarray}
\hat C^{ps (\bf ij)}_l=\frac{1}{2l+1}\sum_{m=-l}^{m=l} \sum_{k,k'=1}^{k,k'=9}\sum_{l',m'}M^{k}_{lml'm'}M^{*k'}_{lml'm'}{\bf \hat W^k_{l'} C^{ps}_{l'}\hat W'^{k'^T}_{l'}} \, .
\label{ups}
\end{eqnarray}
 Eq.~\ref{ups} can further be written as
\begin{eqnarray}
\hat C^{ps (\bf ij)}_l=\sum_{k=1}^{k=9}\sum_{l'}M^{k}_{ll'}{\bf \hat  W^k_{l'}C^{ps}_{l'}\hat W'^{k^T}_{l'}} +\frac{1}{2l+1}\left(\sum_{m=-l}^{m=l} \sum_{k,k'(k\ne k')}\sum_{l',m'}M^{k}_{lml'm'}M^{*k'}_{lml'm'}{\bf  \hat W^k_{l'}C^{ps}_{l'}\hat W'^{k'^T}_{l'}}\right) \, .
\label{pscrorrect}
\end{eqnarray}
Here $M_{ll'}$ is the mode mode coupling matrix. The first term of the bracket is obtained in analogous method as shown in Ref.~\cite{Hivon}. The last term can also be further simplified. For this purpose we note that
\begin{eqnarray}
M^{k}_{lml'm'} = \sum_{l''m''}w^k_{l''m''}(-1)^{m'}\left[\frac{(2l+1)(2l'+1)(2l''+1)}{4\pi}\right]^{1/2}\left(
\begin{array}{ccc}
l & l' & l'' \\
0   &  0  &  0
\end{array}
\right)\left(
\begin{array}{ccc}
l & l' & l'' \\
m   &  -m'  &  m''
\end{array}
\right)\, .
\end{eqnarray}
where $w^k_{lm}$ are the spherical harmonic coefficients from the
$k^{th}$ mask.  Hence
\begin{eqnarray}
\nonumber \sum_{mm'}M^{k}_{lml'm'}M^{*k'}_{lml'm'}=\frac{(2l+1)(2l'+1)}{4\pi}\sum_{l''m''}\sum_{l'''m'''}w^k_{l''m''}w^{k'}_{l'''m'''}\left((2l''+1)(2l'''+1)\right)^{1/2}\left(
\begin{array}{ccc}
l & l' & l'' \\
0   &  0  &  0
\end{array}
\right) \\
\times \left(
\begin{array}{ccc}
l & l' & l''' \\
0   &  0  &  0
\end{array}
\right)\sum_{mm'}\left(
\begin{array}{ccc}
l & l' & l'' \\
m   &  -m'  &  m''
\end{array}
\right)\left(
\begin{array}{ccc}
l & l' & l''' \\
m   &  -m'  &  m'''
\end{array}
\right) \, .
\end{eqnarray}
Using the property of the Wigner $3jm$ symbol we have 
\begin{eqnarray}
\nonumber \sum_{mm'}M^{k}_{lml'm'}M^{*k'}_{lml'm'}=\frac{(2l+1)(2l'+1)}{4\pi}\sum_{l''m''}\sum_{l'''m'''}w^k_{l''m''}w^{k'}_{l'''m'''}\left((2l''+1)(2l'''+1)\right)^{1/2}\left(
\begin{array}{ccc}
l & l' & l'' \\
0   &  0  &  0
\end{array}
\right) \\
\times \left(
\begin{array}{ccc}
l & l' & l''' \\
0   &  0  &  0
\end{array}
\right)\delta_{l''l'''}\delta_{m''m'''}\frac{1}{2l''+1} \, .
\end{eqnarray}
Hence
\begin{eqnarray}
\frac{1}{2l+1}\sum_{mm'}M^{k}_{lml'm'}M^{*k'}_{lml'm'}=\frac{(2l'+1)}{4\pi}\sum_{l''m''}\sum_{l'''m'''}\hat w^k_{l''m''}\hat w'^{k'}_{l'''m'''}\left(
\begin{array}{ccc}
l & l' & l'' \\
0   &  0  &  0
\end{array}
\right)^2 \, .
\end{eqnarray}
This is easily written in terms of the cross power spectra $w^{kk'}_{l''}$ of two masks $k, k'$
\begin{eqnarray}
\frac{1}{2l+1}\sum_{mm'}M^{k}_{lml'm'}M^{*k'}_{lml'm'}=\frac{(2l'+1)}{4\pi}\sum_{l''}(2l''+1)w^{kk'}_{l''}\left(
\begin{array}{ccc}
l & l' & l'' \\
0   &  0  &  0
\end{array}
\right)^2= M^{kk'}_{ll'} \, .
\end{eqnarray}
We have defined $M^{kk'}_{l'}$ as the cross coupling matrix of two masks $k, k'$. Finally we can write eq.~(\ref{pscorrect}) as
\begin{eqnarray}
\hat C^{ps (\bf ij)}_l=\sum_{k=1}^{k=9}\sum_{l'}M^{k}_{ll'}{\bf  \hat W^k_{l'}C^{ps}_{l'}\hat W'^{k^T}_{l'}} +\sum_{k,k' k\ne k'}\sum_{l'}M^{kk'}_{ll'}{\bf \hat W^k_{l'} C^{ps}_{l'} \hat W'^{k'^T}_{l'}} \, ,
\end{eqnarray}
which estimates unresolved residual point source contamination in
individual cross power spectrum. After correcting each of the
cross power spectra we form a simple average  to obtain final
point source corrected spectrum.  We note that the second term on the 
right hand side 
of the above equation gives entirely negligible contribution to the total
estimate of the residual point source correction. {\it This is a result
of the fact that weights are effectively determined by the detector noise 
in the large 
$l$ limit. Two different sky parts have uncorrelated noises. So weights 
from two different sky regions are uncorrelated at large $l$ limit.} In
Ref.~\cite{sah06_proc}, we have shown that the residual point source
contamination is significantly smaller than the contamination arising
from K, Ka, Q or V band. In fact point source residual is intermediate
between V and W band point source power. Figure \ref{ps_matrix}
 shows, residual unresolved point source contamination $\sum_{k=1}^{k=9}\sum_{l'}M^{k}_{ll'}
\hat W^{ki}_{l'}C^{ps (ij)}_{l'}\hat W^{kj}_{l'}$ for different values of $i,j$. Here $i,j$ are the index representing the $4$ DA. All possible
combinations $(i,j)$ are explicitly shown in eq.
(\ref{Clps}). The total unresolved point source spectrum is shown as the pink line with
filled circular points. The dominant contributors to the total
unresolved point source spectrum at large $l$ are the WW, VW and VV
combinations. This is expected since the V and W band share most of
the weights at large $l$ because of their higher angular
resolutions. Weights are negligible for K, Ka, Q bands in the large $l$
limit and hence point source contamination from these bands is
heavily suppressed.

\begin{figure}
\begin{center}
\includegraphics[scale=0.3,angle=-90]{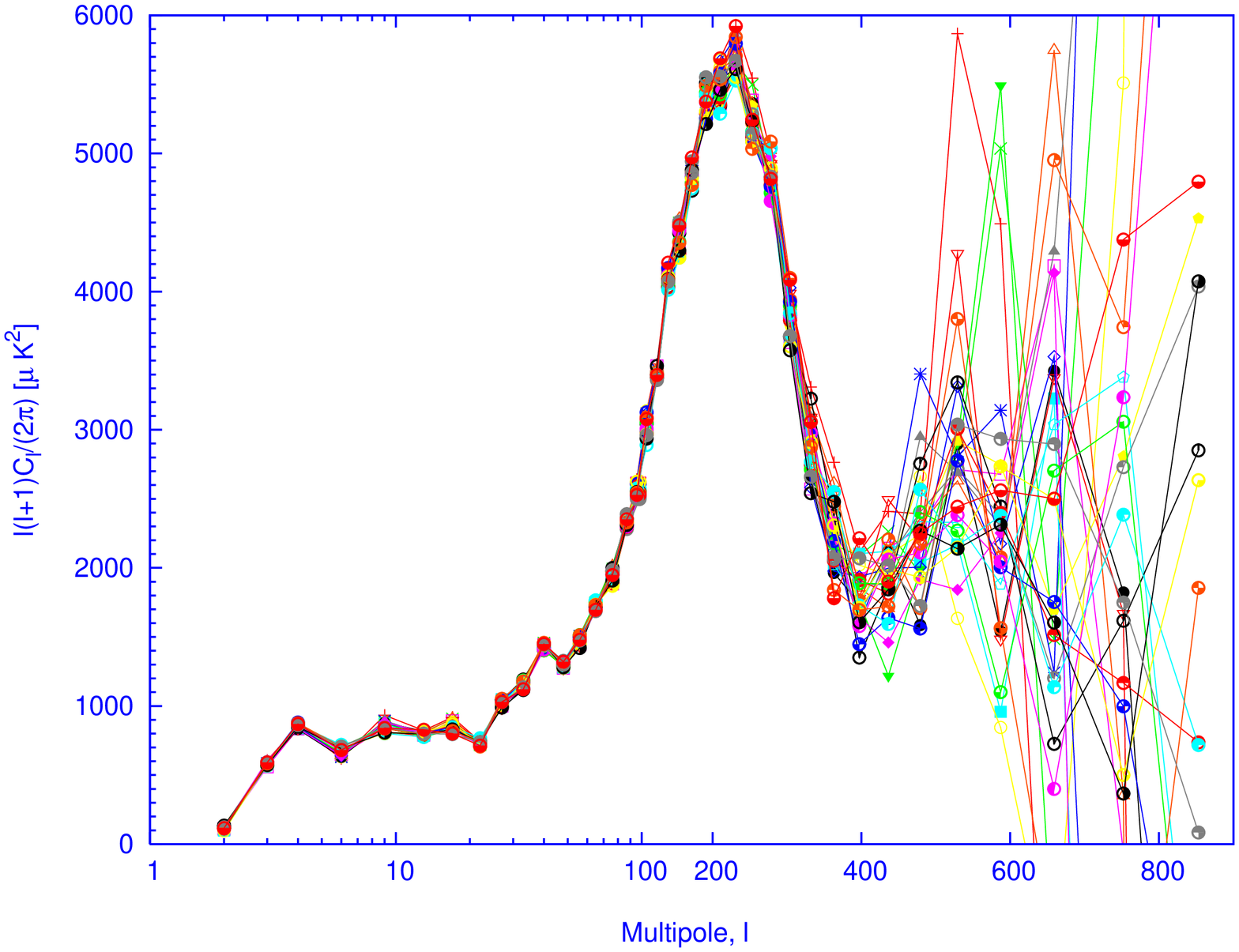}
\includegraphics[scale=0.3,angle=-90]{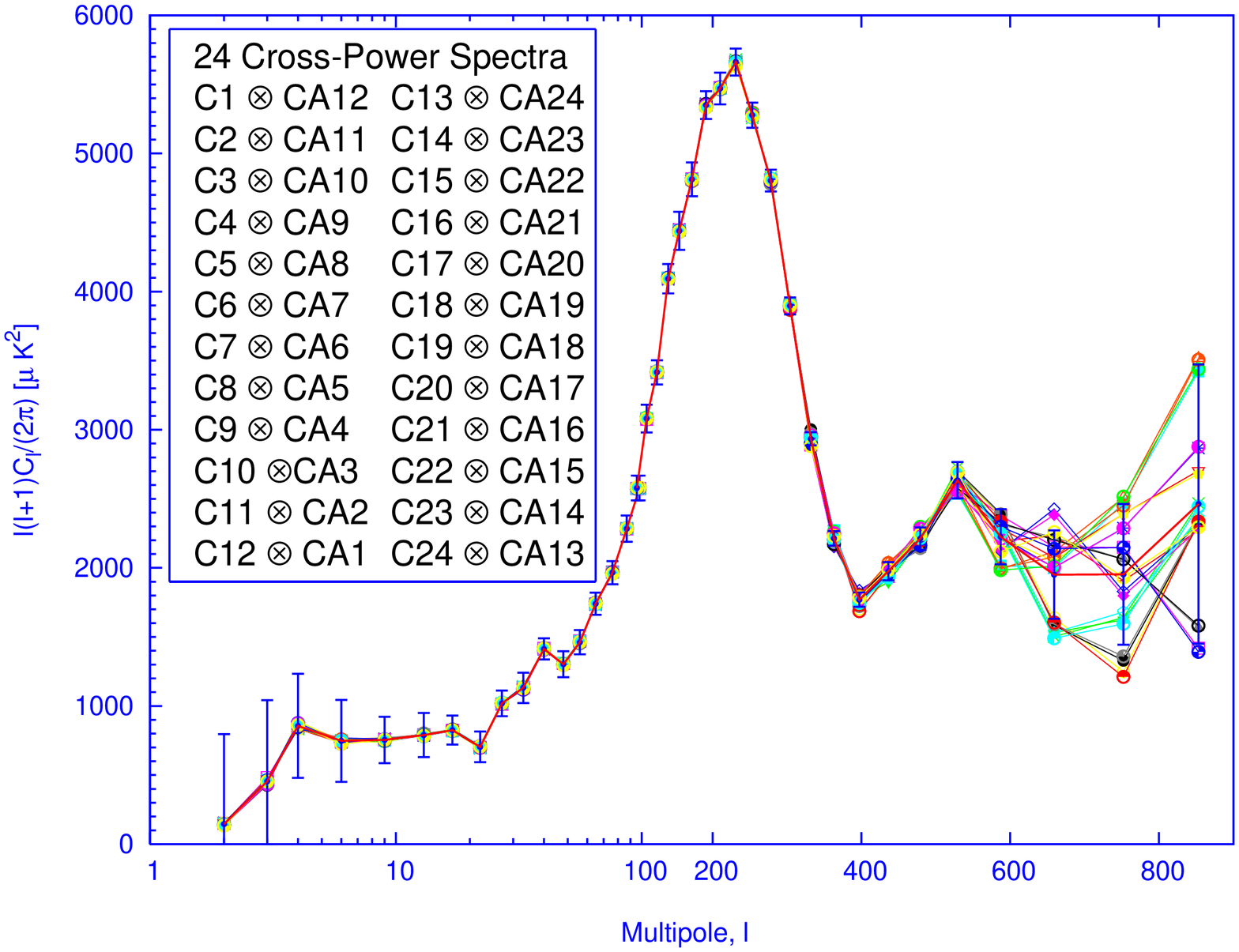}
\caption{ The left panel shows  28 cross power spectra that would 
have been obtained by WMAP  team's  $1$ year analysis without any point source subtraction. 
The right panel shows our $24$ cross spectra based on $1$ year data.}
\label{each_bin}
\end{center}
\end{figure}

\begin{figure}
\begin{center}
\includegraphics[width=6cm,angle=-90]{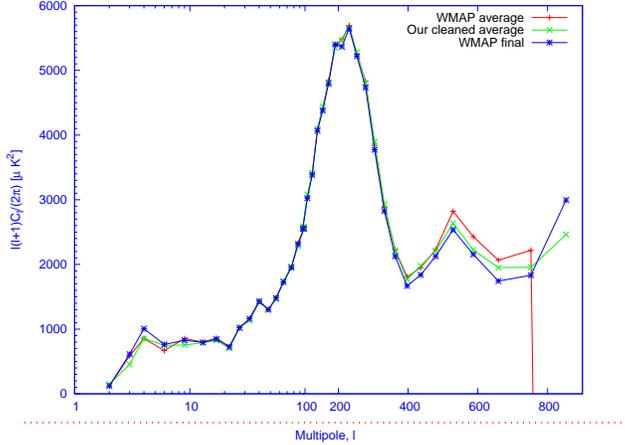}
\caption{Comparison of the WMAP's average  power spectrum (red) with  our binned power spectrum without any point source subtraction ( green). Clearly starting from the first acoustic peak we have less point source contamination. WMAP's final binned power spectrum is also shown in  blue for comparison. Interestingly the notch at l=4 appears to be reduced  in the WMAP's average power spectrum (red) .}
\label{WMAP_final}
\end{center}
\end{figure}

The basis of the WMAP team's 1 year power spectrum are the $28$ cross power
spectra which are available from the LAMBDA website in the unbinned
form. These $28$ cross power spectra are not corrected for the
residual unresolved point sources. Following the WMAP team's 1 year
bins we compare these 28 cross power spectra
in  Fig.~\ref{each_bin} with $24$ cross spectra obtained from our own 1 year
results. These $24$ cross spectra also are not corrected for the
residual unresolved point sources and 
show very little dispersion compared to the WMAP results. 
The `uniform average' power spectrum plotted in green line in fig. \ref{WMAP_final} has less excess power near the second acoustic peak compared to an `uniform average' of the WMAP's 28 cross spectra (red line). This merely shows that we have removed  some amount of point sources at this range of $l$ during our cleaning.

\subsection{Computing error bars}
\label{er_bar}
We rely upon Monte-Carlo simulation to compute the error bars on our
power spectrum. We generate synchrotron, free-free and thermal dust
maps corresponding to different frequencies in a given combination
using the publicly available Planck Sky Model. Each of the random
realizations of CMB and foreground maps are convolved by the
appropriate beam function for each detector. Random noise maps
corresponding to each detector are generated by first sampling a
Gaussian distribution with unit variance. In the final step we
multiply each Gaussian variable by the number $\sigma_0/\sqrt{N_p}$ to
form realistic detector noise maps. Here $\sigma_0$ is the noise per
observation of the detector under consideration and $N_p$ 
the effective number of observations at each pixel. These realistic
maps with detector noise, foreground and CMB signal are then passed through the
cleaning pipeline. The error-bars for our power spectrum correspond
to the standard deviation of the power spectrum obtained from Monte-Carlo simulations .

\begin{figure}
\centering
\includegraphics[scale= 0.4,angle = 0]{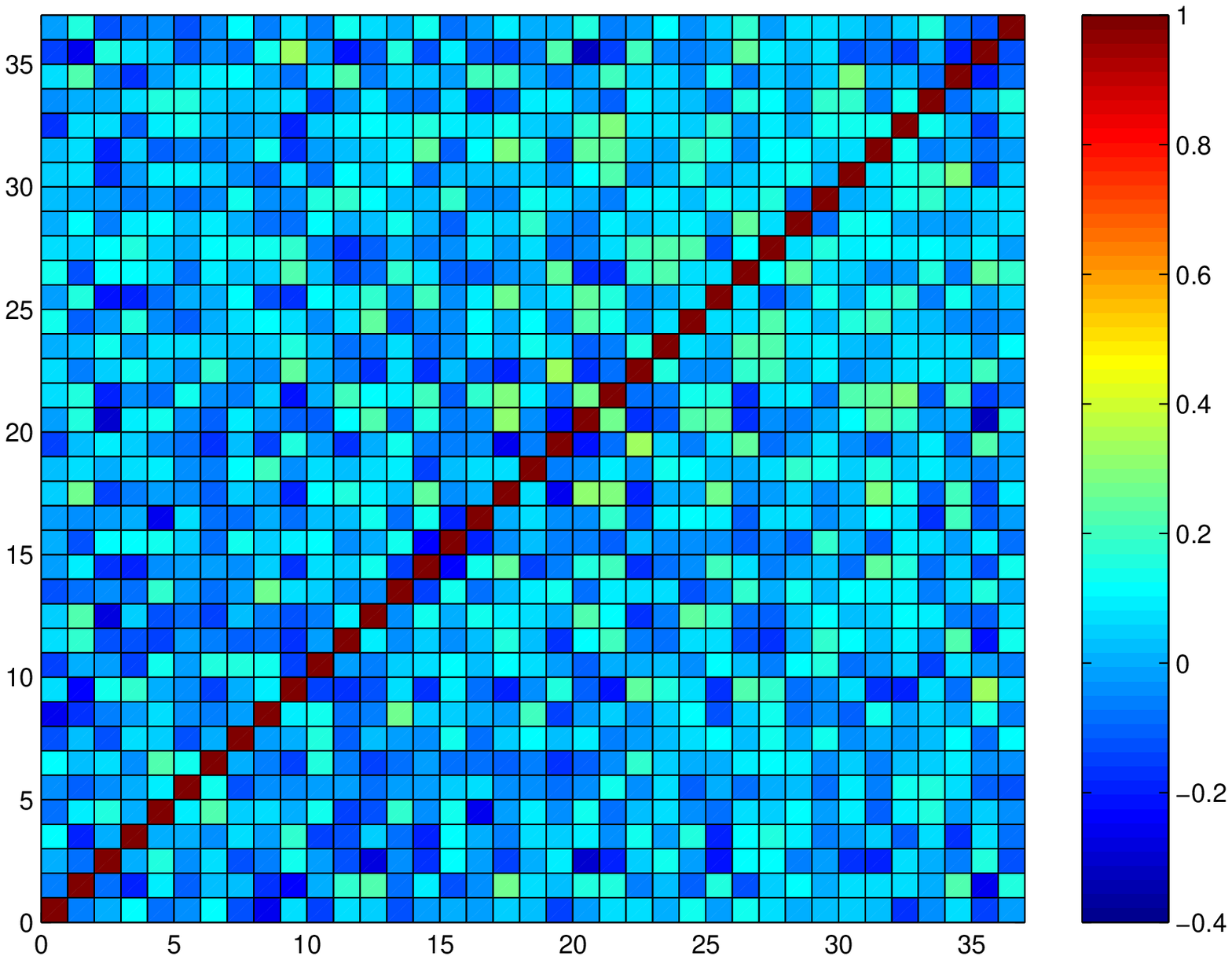}
\includegraphics[scale= 0.4,angle = 0]{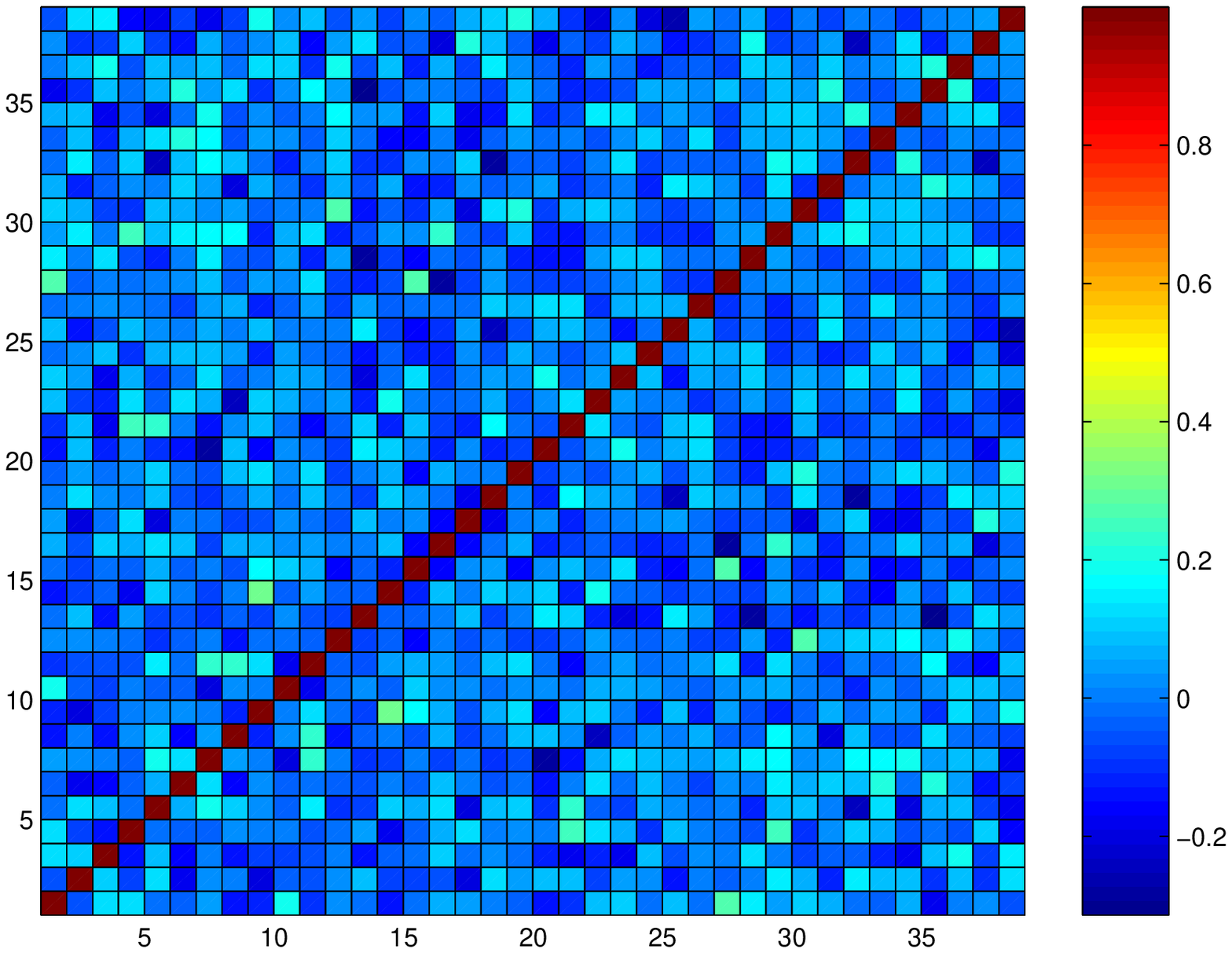}
\caption{Correlation matrix $\frac{<\Delta C_b \Delta C_{b'}>}{\sqrt{<(\Delta{ C_b})^2 ><(\Delta {C_{b'}})^2>}}$ from our simulation plotted with respect to the bin index. As the figure shows the matrix is mostly dominated by the diagonal elements.}
\label{Cov}
\end{figure}

Due to the (Kp2) mask applied to remove potential foreground
contaminated regions, the neighboring multipoles
become coupled. In the presence of detector noise, 
correlation between neighboring $\hat C^c_l$ becomes stronger. The
covariance matrix $\left <\Delta \hat C^c_l \Delta \hat C^c_{l'}\right>= \left<(\hat C^c_l-\left<\hat C^c_l\right>)(\hat C^c_{l'}-\left<\hat C^c_{l'}\right>)\right>$ obtained from 
simulations is therefore expected  to have non-diagonal elements. 
It is convenient to bin the power spectrum in order to
minimize the correlations and errors. 
We have considered a binning identical to that used by the WMAP team
in their analysis. Let $\hat C_b$ denotes the binned power spectrum. Then the covariance matrix of the 
binned spectrum is  obtained as $\left <\Delta \hat C^c_b \Delta \hat C^c_{b'} \right>= \left<(\hat C^c_b-\left<\hat C^c_b\right>)(\hat C^c_{b'}-\left<\hat C^c_{b'}\right>)\right>$. The standard deviation obtained from the diagonal elements of the binned covariance matrix was used as the error-bars on the binned final spectrum extracted from the WMAP data. We also define a normalized covariance matrix of the binned power spectrum following,
\begin{equation}
	C_{bb'}=\frac{\left <\Delta \hat C_b \Delta \hat C_{b'}\right>}{\sqrt{\left<(\Delta \hat C_b)^2 \right> \left<(\Delta \hat C_{b'})^2\right>}}\,,
\end{equation} 
 wherein all the elements of this matrix are bound to lie between $[-1,1]$. This correlation matrix 
represents the actual bin to bin correlation matrix following the cleaning method. In the left panel of the figure~\ref{Cov} we show the correlation matrix for WMAP 1 year simulations.
 The right hand panel of this figure is the corresponding plot for the WMAP 3 year analysis.
 Both these matrices are seen to be sufficiently diagonal dominated.

\section{Results}
\label{results}
Figure \ref{1yr3yr} shows the main result of the power spectrum for CMB anisotropy estimated 
using our analysis of WMAP 1 year and WMAP 3 year data.
The blue line shows the WMAP 3 year power spectrum and the red line shows the  
1 year 
spectrum. All these spectra are corrected for residual unresolved point source contamination.
In the lower panel of this figure we show the residual unresolved point source contamination
for 1 year and 3 year respectively. In what follows we describe the power spectrum obtained by us
for WMAP 1 year data and WMAP 3 year data respectively.

\subsection{WMAP 1 year data}
\label{res1}
\begin{figure}[h]
\includegraphics[scale=0.3,angle=-90]{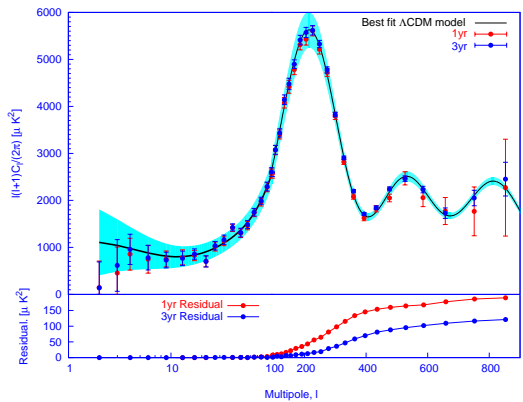}
\caption{Comparison of 1 year 4 channel 
 with that of 3 year 4 channel power spectrum. The best fit WMAP's
power spectrum is shown in black line along with cosmic variance band. The
bottom panel of this figure shows the residual unresolved point source
contamination for both the power spectra. }
\label{1yr3yr}
\end{figure}

Using the $48$ foreground cleaned maps obtained from the WMAP 1 year maps 
we obtain a `Uniform average' power spectrum following the method mentioned in
section~\ref{power_spectrum_estimation}.  The residual 
unresolved point source contamination is removed following ~\ref{ps_res}. The estimated power spectrum
with error bars is plotted in red in figure~\ref{1yr3yr}.

\begin{figure}
\includegraphics[scale= 0.35,angle = -90]{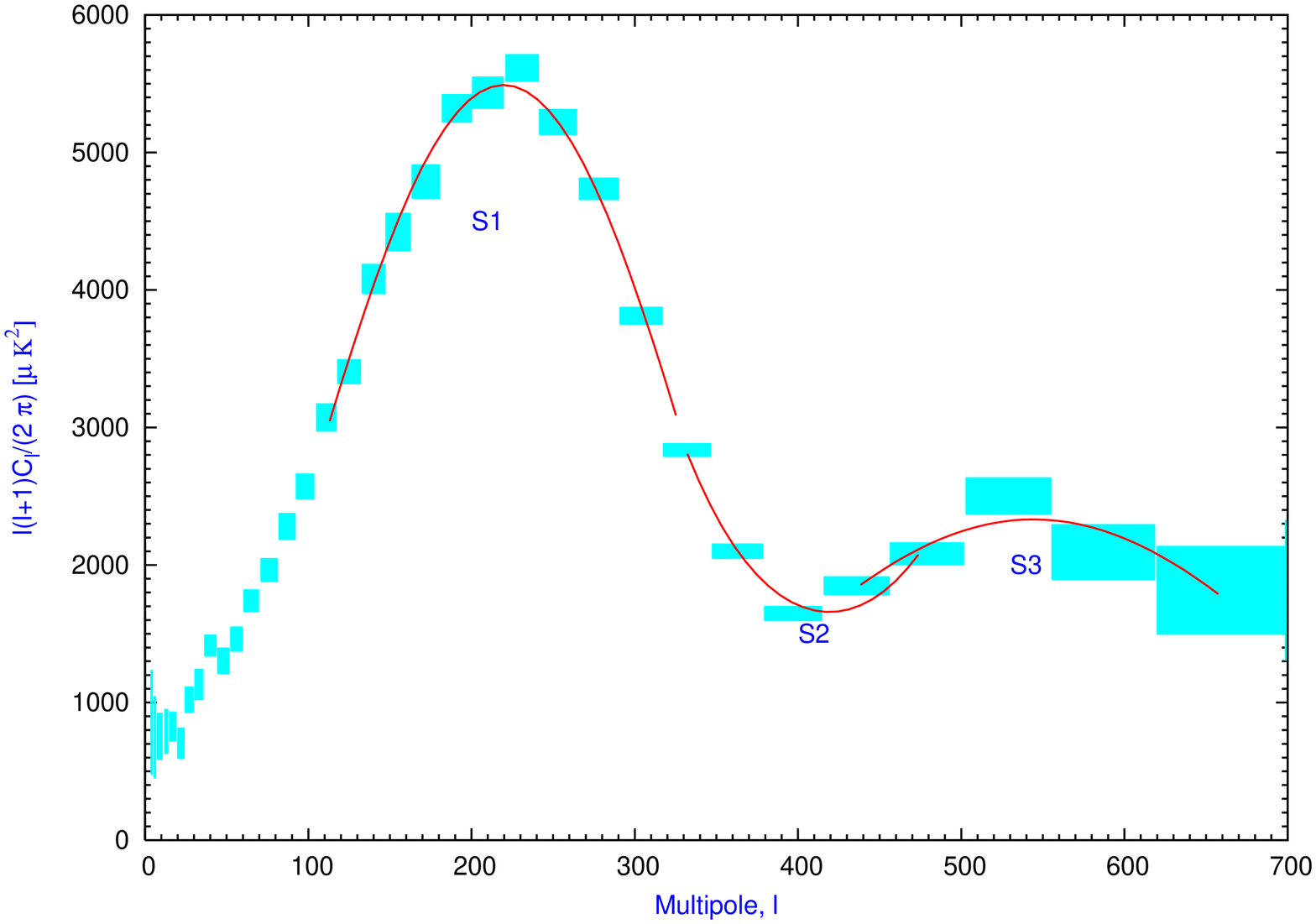}
\includegraphics[scale=0.35,angle=-90]{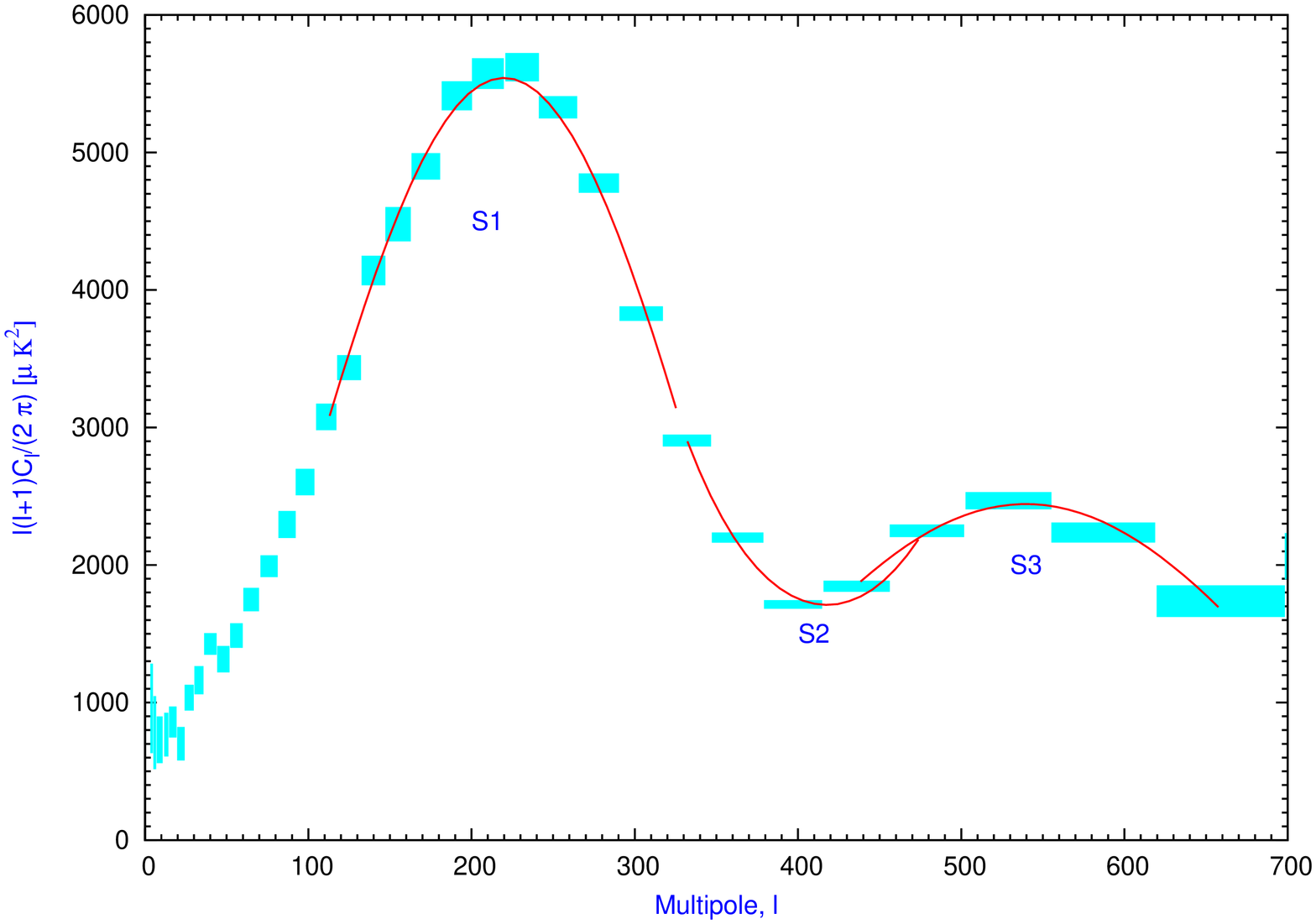}
\caption{The left panel shows the 3 different
multipole ranges used to obtain positions of the first peak, the first
trough and the second peak from our point source subtracted power spectrum
using 1 year WMAP data. Before fitting the 1 year power spectrum was
binned in the same manner as the WMAP's binning. The box error-bars are
used to indicate x and y error-bars. 
The right panel shows the same figure but using the
WMAP 3 year data. A correction due to residual unresolved point
sources was performed prior to fitting.
}
\label{peak_fit}
\end{figure}

We find a suppression of power in the quadrupole and octupole moments
consistent with the results published by the WMAP team. However, our quadrupole moment
($ 146 \mu K^2 $) is little larger than the quadrupole moment
estimated by WMAP team ($ 123 \mu K^2 $) and Octupole ($ 455 \mu K^2$)
is less than the WMAP team result ($ 611 \mu K^2$).  The `Uniform
average' power spectrum does not show the `bite' like feature present at the 
first acoustic peak in the power spectrum reported by
WMAP~\cite{Hinshaw}. We perform a quadratic fit of the form $\Delta T_l 
=\Delta T_{l_0} + \alpha(l-l_0)^2$ to the peaks and
troughs of the binned spectrum similar to WMAP
analysis~\cite{page}. For the residual point source corrected
 power spectrum we obtain the first acoustic peak
at $l = 219.8 \pm 0.8 $ with the peak amplitude $\Delta T_l = 74.1 \pm
0.3 \mu K $, the second acoustic peak at $l = 544 \pm 17 $ with the
peak amplitude $\Delta T_l = 48.3 \pm 1.2 \mu K $ and the first trough
at $l = 419.2 \pm 5.6 \mu K $ with peak amplitude $\Delta T_l = 41.7
\pm 1 \mu K$.  The left panel of figure
\ref{peak_fit} shows the three different ranges of multiples used to
find out peak and trough positions and their corresponding amplitudes
$\Delta T_l$. (A similar plot for the WMAP 3 year data is shown in the right panel 
of this figure. The results for WMAP 3 year analysis are summarized 
in section \ref{3yr_results}.)

As a cross check of the method, we have carried out the analysis with other
possible combinations of the DA maps.

\begin{enumerate}
\item{} The WMAP team also provide foreground cleaned maps
corresponding to Q1 to W4 DA (LAMBDA). The Galactic foreground signal,
consisting of synchrotron, free-free, and dust emission, was removed
using the 3-band, 5-parameter template fitting
method~\cite{Bennett1}. We also include K and Ka band maps which are
not foreground cleaned.
The resulting power spectrum from our analysis
matches closely to the `Uniform average' power spectrum.

\item{}Excluding the K and Ka band from our analysis we get a power
spectrum close to the `Uniform average' results. Notably, we find a
more prominent notch at $l = 4$ similar to WMAP's published results.

\end{enumerate}
 
In case of `Uniform average' a maximum difference of $92~\mu K^2$ is
observed only for octupole. For the large multipole range the
difference is small and for $l=752$ it is approximately $48~\mu
K^2$. This is well within the $1\sigma$ error bar ($510 \mu K^2$)
obtained from the simulation.  This shows that our foreground cleaning
is comparable in efficiency to that obtained by
employing template fitting methods that
rely on a model of foreground emission 
to estimate the contamination at the CMB dominated frequencies.

\begin{figure}
\centering
\includegraphics[scale= 0.3,angle = -90]{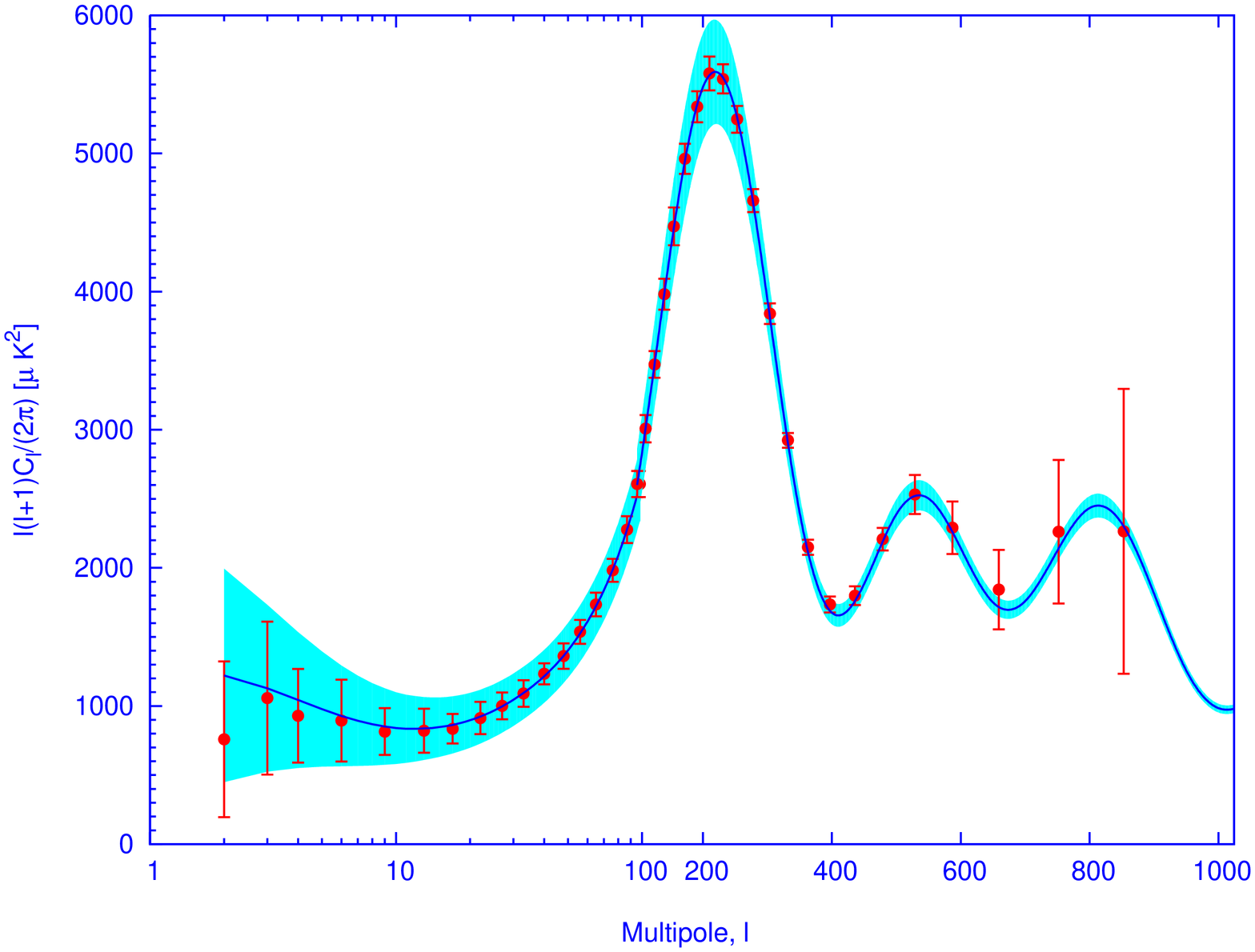}
\includegraphics[scale= 0.3,angle = -90]{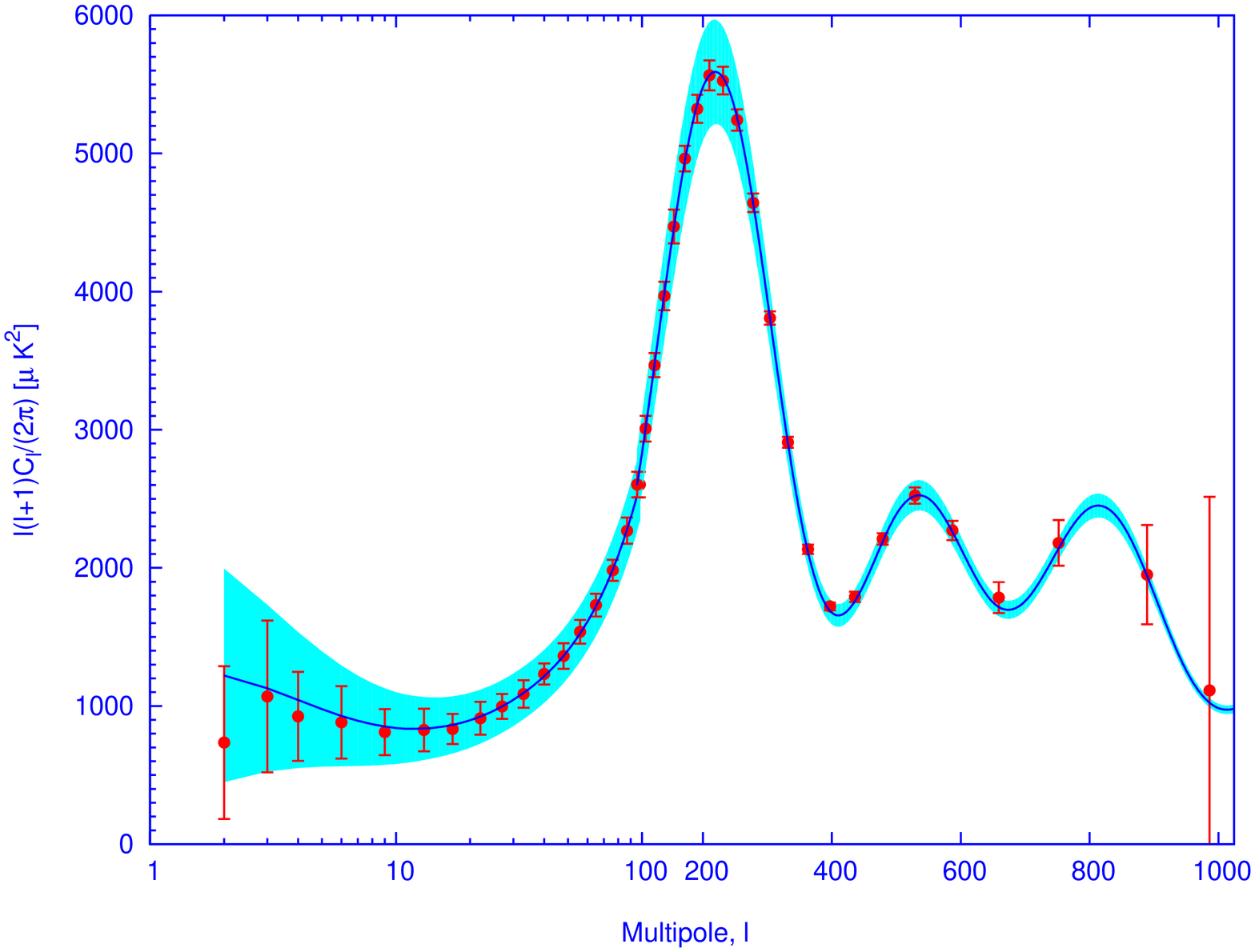}
\caption{The left panel shows (in red points) ensemble
averaged power spectrum from 110 Monte Carlo simulations of our power
spectrum estimation method. The simulations were carried out using the
1 year WMAP detector noise maps available from the LAMBDA website. We use
publicly available Planck Sky Model to generate the diffuse foreground
models. The recovered spectrum is binned in the same manner as WMAP 1
year power spectrum. The input theoretical spectrum is shown in blue
line with cosmic variance. The right panel is same figure but with 3
year noise maps. The 3 year noise maps were generated following the
method described in the text. The spectrum is binned following the
binning scheme of the WMAP team's analysis of $3$ year data.}
\label{log_linear_simulation.ps}
\end{figure}

The Monte Carlo simulations of our cleaning method also reveals the
negative bias in the low $l$ moments. The origin of this negative bias is
explained in
section~\ref{Bias}. For $l=2$ and $l=3$ the bias is $-27.4\%$
and $-13.8\%$ respectively. However this bias become negligible at
higher $l$, e.g. at $l=22$, it is only $-0.8\%$. This bias can be
explained in terms of an anti-correlation of the CMB with the residual
foregrounds in the cleaned map. For further details of the bias we
refer to appendix~\ref{bias_lowl}. The standard deviation obtained
from the diagonal elements of the covariance matrix is used as the
error bars on the $C_l$'s obtained from the data.  The ensemble
average of $110$ cleaned power spectrum is shown in the left panel of
the figure~\ref{log_linear_simulation.ps}. The presence of bias at the
low multipole moments is visible is this figure.

\subsection{WMAP 3 year data}
\label{3yr_results}

\subsubsection{4 channel combinations}

\begin{figure}[h]
\includegraphics[scale=0.4,angle=-90]{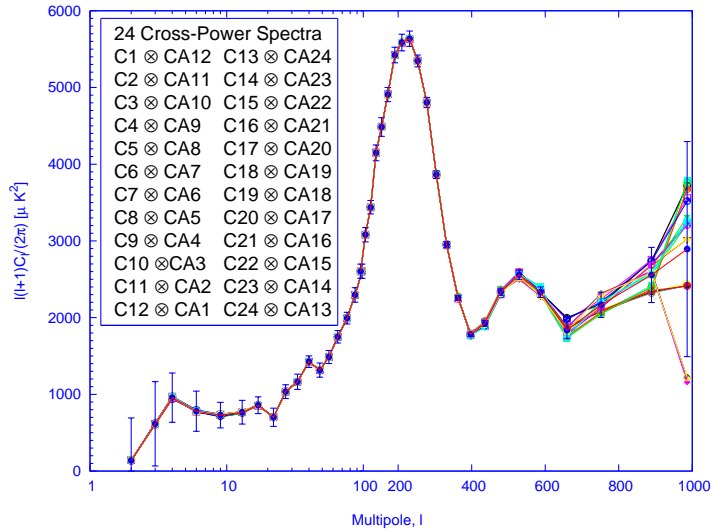}
\caption{The 24 cross power spectra for the 3 year WMAP data
are shown with the detector noise bias removed. 
The red line with blue error-bars is the
`Uniform average' power spectrum. In the inset of this figure we show
all the possible cross correlations of the cleaned maps that give rise
to the 24 cross spectra.}
\label{24_bin3yr}
\end{figure}

We analyse the 3 year WMAP data by using a procedure identical to that used
for the 1 year data. The details are given in section~\ref{tab2}. 
In figure~\ref{24_bin3yr} we show each of the $24$ cross power. 
The `Uniform average' of these 24 cross
power spectra is also shown in this figure in red line with blue
error-bar. This figure is similar to the $24$ individual cross power
spectra obtained for WMAP 1 year data~\cite{sah06,sah06_proc}. For the 3
year data all the $24$ cross power spectra show very little dispersion
till the second trough. In comparison the $1$ year data 
shows small dispersion
only till the second acoustic peak~\cite{sah06}. 
This may be explained due to the 
effectively lower detector noise
in the $3$ year data compared to the $1$ year data.

We perform a peak fitting to the peaks and troughs of the 3 year
power spectrum as well. A parabolic function of the form $\Delta T_l =
\Delta T_{l_0} + \alpha(l-l_0)^2$ is fitted to the peaks and troughs
of the power spectrum amplitude. We use the 3 year binned data for the
fitting purpose.  For the point source corrected power spectra, 
the first acoustic peak has an amplitude $\Delta T_{l_0} = 74.4
\pm 0.3 \mu K$ at the multipole position $l =219.9 \pm 0.8$. The first
trough is located at $l=417.7 \pm 3.2$ with an amplitude $\Delta
T_{l_0} = 41.4\pm 0.6 \mu K $. The amplitude and position of the
second acoustic peak are given by $\Delta T_{l_0}= 49.4 \pm 0.4 \mu K, l = 539.5 \pm
3.7$. In the right panel of figure \ref{peak_fit} we show the different multipole
ranges used to obtain the position of the peaks and troughs and their
amplitudes.

\subsubsection{3 channel combinations}
We also follow an alternative approach in which we use only the Q,V
and W band DA maps. This is similar to the 4 channel
cleaning, however now we only get $24$ cleaned maps.
  We form a total of $12$ cross power spectra from these
cleaned maps after applying 3 year Kp2 mask. After debiasing all the
power spectra by the coupling matrix and removing beam and pixel
effect we obtain an `3 channel uniform average' power spectrum. 
We find that the 3 channel spectrum matches  well with the  4 channel spectrum. 

\subsection{Comparison of 1 year and 3 year power spectra}
	
Fig.~\ref{1yr3yr} compares the  power spectra obtained from 
the $1$ year and $3$ WMAP data binned identically using the 
WMAP team's $1$ year binning method.
They match closely with one another and
also with the WMAP's best fit power spectrum available from the LAMBDA
website.  For the residual point source corrected 
power spectrum we obtain for WMAP 1 year (WMAP 3 year) data the first acoustic
peak at $l = 219.8 \pm 0.8 $ $(219.9 \pm 0.8 )$ with the peak
amplitude $\Delta T_l = 74.1 \pm 0.3 \mu K $ $ (74.4 \pm 0.3 \mu K)$,
the second acoustic peak at $l = 544 \pm 17 $ $ (539.5 \pm 3.7 )$ with
the peak amplitude $\Delta T_l = 48.3 \pm 1.2 \mu K $ $(49.4 \pm 0.4
\mu K)$ and the first trough at $l = 419.2 \pm 5.6 \mu K $ $(417.7 \pm
3.2 )$ with peak amplitude $\Delta T_l = 41.7 \pm 1 $ $ \mu K (41.4
\pm 0.6 \mu K)$.

We note that our cleaning method significantly removes
unresolved point source contamination. The original $l^2$ dependence
of the unresolved point source power spectrum present in the
foreground contaminated maps (as well present in template cleaned
maps) is significantly reduced and becomes independent of $l$ at
large $l$. We also note that the unresolved residual point source
contamination is less by about $\approx 50 \mu K^2 $ in 3 year power
spectrum than the 1 year power spectrum. This is expected. The WMAP
supplied 3 year Kp2 mask removes more point sources than the 1 year Kp2
mask.

 \section{Conclusion}
\label{conclu}
   The rapid improvement in the sensitivity and resolution of the CMB
experiments has posed increasingly stringent requirements on the level
of separation and removal of the foreground contaminants. We carry out
an estimation of the CMB power spectrum from the WMAP data
that is independent of foreground model. The method does not rely upon
any foreground template and employs the lack of noise correlation
between independent channels. This paper is a detailed
description of the first estimate of the CMB angular power spectrum
solely based upon the WMAP data. In this paper, we present an indepth
study of the biases that arise in the foreground cleaning. In particular, 
we provide an understanding and correction for
the negative bias at low multipoles reported in our earlier
work~\cite{sah06}.

Usual approaches to foreground removal, usually incorporate the extra
information about the foregrounds available at other frequencies, the
spatial structure and distribution in constructing a foreground
template at the frequencies of the CMB measurements. These approaches may
be susceptible to uncertainties and inadequacies of modeling
involved in extrapolating from the frequency of observation to CMB
observations that a blind approach, such as presented here,
evades. The understanding of polarized foreground for CMB polarization
maps is rather scarce. Hence modeling uncertainties could dominate the
systematics error budget of conventional foreground cleaning.  The
blind approach extended to estimating polarization spectra after
cleaning CMB polarization maps could prove to be particularly
advantageous.

\acknowledgments 

The analysis pipeline as well as the entire
simulation pipeline is based on primitives from the HEALPix
package.~\footnote {The HEALPix distribution is publicly available from
the website http://healpix.jpl.nasa.gov.} We acknowledge the use of
version 1.1 of the Planck reference sky model, prepared by the members
of Working Group 2 and available at www.planck.fr/heading79.html
The entire analysis procedure was carried out on the IUCAA HPC
facility as well as on the computing facilities at IAP.
RS acknowledges the Indo-French Sandwich
Fellowship granted by the French Embassy in India and EGIDE in Paris,
France. RS thanks IAP for hosting his visit. RS thanks Francois Bouchet, Christophe
Pichon, Karim Benabed, Pawel Bielewicz and Planck group members at IAP
for useful and illuminating discussions. We are grateful to Lyman
Page, Olivier Dore, Charles Lawrence, Kris Gorski,
Hans Kristian Eriksen and Max Tegmark for thoughtful comments and
suggestions. We acknowledge a private communication with Garry Hinshaw
on the unresolved point source model. We thank Amir Hajian, Subharthi
Ray and Sanjit Mitra in IUCAA for helpful discussions. 
 We thank the WMAP team for producing excellent quality CMB
maps and making them publicly available.

\appendix

\section{Analytic derivation of Weights and Cleaned Power spectrum}
\label{Weight_eqn}
The main idea behind the blind foreground cleaning method used here is
entirely based upon the minimization of total power in the cleaned map
in multipole space~\cite{Tegmark}. Weights for different channels are
obtained minimizing the total power $\hat C_l^{Clean}$ of the
cleaned map. However, we ensure that the CMB angular power spectrum is conserved
during cleaning by imposing the constraint ${\bf \hat W_le_0=e^T_0\hat
W^T}=1$ on the weights. The solution for the weights that satisfy these
conditions is the point in weight space where normals to the functions
$ f(\bf\hat W_l)= \hat W_l\hat C_l\bf\hat W_l^T $ and $g(\bf\hat W_l)=\hat W_le_0$ are parallel to one another.  Following Lagrange's
multiplier method, this is cast to an equivalent problem of minimizing
\begin{equation}
   \bf \hat W_l\hat C_l\bf\hat W_l^T - \lambda \hat W_le_0 \, .
\label{lagvar}
\end{equation}
Here $\lambda $ is the unknown Lagrange multiplier parameter which could
be determined from variational principle. At the extrema, the expression
in eq.~(\ref{lagvar}) is unchanged under small variations in $\bf \hat
W_l$ leading to
\begin{equation}
    {\bf \Delta \hat W_l\hat C_l\bf\hat W_l^T +\hat W_l\hat C_l\bf\Delta \hat W_l^T-\lambda \Delta \hat W_le_0}=0 \, .
\end{equation}
Since the power spectrum matrix, $\bf \hat C_l $ is a symmetric
matrix, the first two terms of the left hand side
are equal to one another. Hence we obtain,
\begin{eqnarray}
    {\bf \Delta \hat W_l\left[ 2 \hat C_l\bf\hat
      W_l^T-\lambda e_0\right]}=0 \, .
\end{eqnarray}
Since this relation is true for any arbitrary variation $\bf \Delta
\hat W_l $, we obtain
\begin{eqnarray}
    { \bf  \left [2 \hat C_l\bf\hat W_l^T-\lambda e_0\right]}= 0 \label{w1}\, .
\end{eqnarray}
We introduce a (non zero) square matrix $\bf \hat G_l$. Later we will
 identify $\bf \hat G_l$ as the {\em Moore Penrose Generalized
 Inverse} (MPGI) of the covariance matrix $\bf \hat C_l$. After
 multiplication from left by this matrix we can rewrite the above
 equation as
  \begin{eqnarray}
    {\bf  2 \hat G_l \hat C_l \hat W^T_l  -\lambda \hat G_l e_0} = 0 \, .
 \end{eqnarray}
Hence we obtain
   \begin{eqnarray}
    \lambda=  \bf \frac{2 e^T_0 \hat G_l \hat C_l \hat W^T_l}{e^T_0 \hat G_l e_0} \label{w2} \, .
    \end{eqnarray}
Now we use the constraint ${\bf \hat W_l e_0}=1$. Assuming $ \bf \hat W_l \ne 0$ we multiply
eq.~(\ref{w1}) from left by $\bf \hat W_l$ and to obtain
\begin{equation}
 {\bf  2 \hat W_l \hat C_l\bf\hat W_l^T=\lambda \hat W_le_0} =\lambda\,.
 \label{w3}
\end{equation}
Using eq.~(\ref{w2}) and eq.~(\ref{w3})
we obtain
\begin{eqnarray}
 2 \bf \hat W_l \hat C_l\hat W_l^T = 2 \frac{e^T_0 \hat G_l \hat C_l
 \hat W^T_l}{{e^T_0 \hat G_le_0}}\,,
\end{eqnarray}
or,
\begin{eqnarray}
{\left( \bf  {\hat W_l}  -\frac{e^T_0 \hat G_l }{{{e^T_0 \hat G_l e_0}}}\right) \bf \hat C_l\hat W_l^T }=  0\,.
\end{eqnarray}
If we neglect solutions which belongs to null space of $\bf \hat
C_l$,~\footnote{There is a physical justification behind neglecting
this solution. The weights satisfying $\bf \hat C_l\hat W_l^T = 0$
do not preserve CMB power while minimizing total power in the
cleaned map. These solution for the weights merely sets the total power
in the cleaned map $ \bf \hat W_l\hat C_l \hat W^T_l =0$.  Thus we are
not interested in the space of solutions to $\bf \hat C_l\hat W_l^T=0$.} i.e., assuming $\bf \hat C_l\hat W_l^T \ne 0$, we obtain
\begin{eqnarray}
\bf \hat W_l=\frac{e^T_0\hat G_l}{e^T_0 \hat G_le_0} \,.
\end{eqnarray}
We impose the restriction on $\bf \hat G_l$ that it is symmetric
(since the MPGI of a symmetric matrix is symmetric) to obtain
 \begin{eqnarray}
\bf \hat W^T_l=\frac{\hat G_l e_0}{e^T_0 \hat G_le_0} \, .
\end{eqnarray}
The corresponding power spectrum of the cleaned map is given by
\begin{eqnarray}
\bf \hat W_l \hat C_l \hat W^T_l = \frac{e^T_0\hat G_l}{e^T_0 \hat G_le_0} \hat C_l \frac{\hat G_l e_0}{e^T_0 \hat G_l e_0} =\frac{1}{e^T_0 \hat G_l e_0}\,,\label{GCL}
\end{eqnarray}
where we have imposed the condition $\bf \hat G_l \hat C_l = \hat C_l\hat G_l$. It is easy to note that the choice of symmetric $\bf \hat
G_l $ helps to obtain a simplified expression for the cleaned power
spectrum. Also one can verify that, if $\bf \hat G_l \hat C_l \hat
G_l= \hat G_l$ is satisfied and $\bf \hat G_l$ is symmetric then $\bf \hat
G_l$ satisfies all the defining conditions of MPGI of $\bf \hat
C_l$. Hence, we can identify $\bf \hat G_l$ as the Moore-Penrose
Generalized Generalized Inverse (MPGI) of $\bf \hat C_l$ (and vice versa).

Eq.~(\ref{GCL}) has an interesting property. It remains valid
even when the full covariance matrix $\bf \hat C_l$ is singular. A
singular full covariance matrix is encountered in noiseless case (or
numerically very low noise case) when the foreground components follow
a rigid frequency scaling and total number of components (all
foregrounds and CMB) become less than the number of available channels.
If $\bf \hat C_l$ is non singular we can replace $\bf \hat
G_l$ everywhere by $\bf \hat C^{-1}_l$. This is because MPGI of a
non-singular matrix is its inverse.

\section{Partitioning the sky}
\label{mask_making}
An important advantage of the foreground cleaning analyzed 
here is that we can allow the weights to vary with sky positions as well
as with the multipole moment. To allow the weights to vary with sky
positions we can partition the sky into several regions depending upon
the level of foreground contamination. (Alternatively the partition
could be done directly using 
the knowledge of expected spectral index dependence on the 
sky.) We followed the procedure of
Ref.~\cite{Tegmark} to partition the sky. Each of these
partitions are identified with a sky masks. The mask takes non zero
value at all pixels contained within the sky partition represented and
is zero outside. In this section we describe the procedure of
constructing these masks.

There are 2 Difference Assemblies for Q band, 2 for V band and 4 for W
band. To make the masks we first averaged all the DA maps for a given
frequency band.  Correspondingly we averaged the beam functions for
each frequency band. For the K and KA bands there are only one
difference assembly in each case. Therefore for these bands no
averaging was done. We smoothed all the five maps (e.g. K, Ka, Q, V, W) 
by the resolution
function of the K band, which has the lowest resolution. We
obtained four difference maps W-V, V-Q, Q-K, K-Ka, out of these 5
smoothed maps. The smoothing was performed first obtaining 
the $a_{lm} $ coefficients. As each averaged map was effectively
smoothed by the averaged beam function (corresponding to each channel)
during the observations, we decided to first remove the beam effect.
 Then we smoothed each map by the common resolution of the
lowest frequency band. Mathematically, the difference maps were
obtained as follows:
\begin{equation}
a_{lm}^{W-V}=a_{lm}^W\frac{B_{l}^K}{B_{l}^W}-a_{lm}^V\frac{B_{l}^K}{B_{l}^V} \, ,
\end{equation}
\begin{equation}
a_{lm}^{V-Q}=a_{lm}^V\frac{B_{l}^K}{B_{l}^V}-a_{lm}^Q\frac{B_{l}^K}{B_{l}^Q} \, ,
\end{equation}
\begin{equation}
a_{lm}^{Q-K}=a_{lm}^Q\frac{B_{l}^K}{B_{l}^Q}-a_{lm}^K \,,
\end{equation}
\begin{equation}
a_{lm}^{K-KA}=a_{lm}^K-a_{lm}^{KA}\frac{B_{l}^K}{B_{l}^{KA}} \, .
\end{equation}
The $a_{lm}$ were converted to difference maps using HEALPix supplied
subroutine alm2map.

Next we construct a junk map out of these four difference maps
assigning at each pixel the absolute maximum value among the four
difference maps.  We down-sample the junk map using the HEALPix
supplied program udgrade to a resolution of $\nside = 64$. We identify
$7$ different sky mask partitions from this low resolution junk map
after applying cutoff corresponding to the following temperature thresholds
(in $\mu K$) $T > 30000$, $30000 \ge T> 10000$, $10000 \ge T > 3000$,
$3000 \ge T >1000$, $1000 \ge T >300$, $300 \ge T > 100$ and $T <
100$. 
The partition $ T > 30000$ is maximally contaminated by 
foreground emission. The second dirtiest partition is disjoint on the sky and we
use $3$ separate masks corresponding this partition. The resulting $9$
masks are then converted back to the HEALPix resolution $\nside=512$ using
udgrade routine of HEALPix. Next we smooth each mask using a Gaussian
beam of FWHM $30^\prime$ and redefine smoother mask boundaries at the
threshold of $0.5$.~\footnote{We found that if we smooth them by a
Gaussian function of FWHM 2 degree (as mentioned in
Ref.~\cite{Tegmark}) and then define the boundaries at the threshold of
$0.5$ the resulting 9 masks cannot cover the entire sky. In that case
there are some regions near the galactic plane which do not belong to
any of our 9 masks. Obviously if such regions are not covered by any
of the masks and remain present in our final cleaned map then final
power spectrum will be contaminated by the foreground. Therefore we
chose to use a Gaussian function of lower FWHM of $30^\prime$ for
smoothing. The reason why in our case, masks smoothed by a 2 degree
Gaussian function cannot cover the entire sky is clear. This is
actually dependent on the common beam function by which we are
smoothing each map before forming the difference maps. We found that
the masks near the galactic regions contain a few small isolated
regions. Therefore smoothing by a Gaussian with FWHM as large as 2
degree, gives this isolated small regions maximum value far less than
unity, in fact maximum value becomes quite near $0.5$, the cutoff
value, after smoothing.  Consequently applying a cut of $0.5$ removes
most of the part of these isolated regions, leading to some part of
the sky, uncovered by the masks.}  In this case we found that almost
the entire sky is covered by the 9 masks (except for a few pixels in the sky).

To convert the second dirtiest region into 3 mask files we upgraded
the $\nside =64$ resolution map to the $\nside =512$. Then using IDL
task mollcursor, we found out the extension in galactic $\theta, \phi$
coordinates of the $3$ different parts of the second dirtiest region. We
then implemented a method which determines the pixel index in `ring'
format for these $3$ regions. Finally we converted them to $3$ different
masks at $\nside =512$. The masks were then smoothed by Gaussian
function of FWHM =$30^\prime$. We applied a cutoff of 0.5 to each of
them. The $8^{th}$, $7^{th}$ and $6^{th}$ masks were numbered according to the
descending order of maximum pixel value in the junk map at resolution
$\nside =512$.

Figure \ref{masks} shows our $9$ different masks that partition the sky
based on estimated level of foreground contamination. These regions
are similar to what is shown in Ref.~\cite{Tegmark}. As the figure
shows, one side of the band near the galactic plane is more severely
foreground contaminated.

\section{Combining cross power spectra}
\label{app_c}

The basis of our final power spectrum are a set of 24 cross power
spectra where the detector noise bias has been removed. An uniform weighting of
cross spectra is used to obtain the final power spectrum. In this
section we describe the procedure to obtain the 24 cross spectra and their 
combination to form the final spectrum.

Let us assume that $\Delta T ^i(\theta , \phi) $ represents one of the
48 final cleaned map, where ($i$ = 1, 2, 3, ..., 48). To remove the
residual foreground contamination near the galactic plane (figure
\ref{w13_ring.ps}) we apply the Kp2 mask supplied by the WMAP team on each
of these 48 maps. The masked map can be represented as 
\begin{equation}
\Delta {T'}^i(\hat n) = W(\hat n ) \Delta T ^i(\hat n) \, .
\label{weigh_map}
\end{equation}
Here $W(\theta,\phi)$ represents the Kp2 mask. In the next step, we
obtain a cross power spectrum between pairs of foreground cleaned maps
that have uncorrelated noise (recall that noise in different
DAs are uncorrelated). If $\tilde a^i_{lm} $ and $\tilde a^j_{lm} $
are the spherical harmonic coefficients obtained from two such
maps
\begin{equation}
\tilde a^{i,j}_{lm}= \int W(\hat n) \Delta T ^{i,j}(\hat n) d \Omega \, . 
\end{equation}
The cross power spectrum is obtained using, 
\begin{eqnarray}
\tilde C^{ij}_l = \sum_{m=-l}^{m=l}\frac {\tilde a^i_{lm} \tilde a^{j*}_{lm}}{2l+1} \, .
\end{eqnarray}
Here the superscript $ij$ represents the cross power spectrum
obtained from the $i^{th}$ and $j^{th}$ foreground cleaned
maps. Following Ref.~\cite{Hivon}, the ensemble average of the $C_l$
estimated from the partial sky is related to the $C_l$ from the full
sky as
\begin{eqnarray}
\left< {\tilde C_l}\right> = \sum_{l'}M_{ll'}\left<\hat C^{c}_{l'}\right> \, ,
\label{Coupling}
\end{eqnarray}
where $M_{ll'}$ is a coupling matrix. This matrix represents the fact
that when we multiply our map by the weight function (Kp2 mask)
 in the pixel space we are
effectively performing a smoothing operation and neighboring spherical
harmonic coefficients get coupled. An analytic expression for this
coupling matrix is given in Ref.~\cite{Hivon},
\begin{equation}
M_{l_1l_2}= \frac{2l_2+1}{4 \pi}\sum_{l_3=|l_1-l_2|}^{l_3=l_1+l_2}(2l_3+1)W_{l_3}
\left(
\begin{array}{ccc}
l_1 & l_2 & l_3 \\
0   &  0  &  0
\end{array}
\right)^2 \, ,
\end{equation}
where the last term is the Wigner-3j symbol and $W_{l}$ is the power
spectrum of the mask under consideration.  Although in our case we are
interested in the partial sky power spectrum which has been obtained
by cross correlating two different maps, it is easy to show that in
this case also eq.~(\ref{Coupling}) remains valid. The rank of the
coupling matrix is limited to $l_{max}=1024$. We use numerical
routines from the {\sc netlib } package~\cite{netlib} to compute the Wigner -3j 
symbol.

Eq.~(\ref{Coupling}) is true only for an ideal observation with
infinite angular resolution. In practice all instruments have finite
angular resolution
given by the beam function $B(\theta,\phi)$ of the
instrument. Mathematically, we may write the observed temperature anisotropy,
\begin{eqnarray}
\Delta T(\hat n ) = \int \Delta T'(\hat n') B(\hat n, \hat n')
d\Omega_{n'} \, ,
\label{DeltaT}
\end{eqnarray}
where $\Delta T'(\hat n')$ is now the full sky map in the absence of
the beam. In most of the CMB experiments the
beam function is circularly symmetric to a good approximation, i.e., depends only on the angle
$\theta '' = \cos^{-1}(\hat n\cdot \hat n')$ between two directions $\hat n $ and $\hat n' $, 
\begin{eqnarray}
B(\hat n, \hat n') \equiv  B(\hat n \cdot \hat n') \, .
\end{eqnarray}
We may expand this function in terms of Legendre polynomials,
\begin{eqnarray}
B(\hat n \cdot \hat n')=\sum_{l=0}^{l=\infty} \frac{2l+1}{4\pi} B_l P_l(\hat n \cdot \hat n') \, .
\end{eqnarray} 
Here, $B_l$'s are Legendre transform of the beam function.
Substituting this in eq.~(\ref{DeltaT}) and expanding $\Delta T$ and
$\Delta T'$ in spherical harmonics, we obtain 
\begin{eqnarray}
\sum_{l=0}^{l=\infty}\sum_{m=-l}^{m=l} a_{lm}Y_{lm}(\hat n ) = \sum_{l'=0}^{l'=\infty}\sum_{m'=-l'}^{m'=l'} a'_{l'm'}\int Y_{l'm'}(\hat n' )\sum_{l''=0}^{l''=\infty}\frac{2l''+1}{4\pi} B_{l''} P_{l''}(\hat n \cdot \hat n')d\Omega_{\hat n '} \, .
\end{eqnarray}
Using the addition formula 
\begin{eqnarray}
P_{l''}(\hat n \cdot \hat n') =\frac{4\pi}{2l''+1}\sum_{m''=-l''}^{m''=l''}Y_{l''m''}(\hat n )Y^*_{l''m''}(\hat n' ) \, 
\end{eqnarray}
and the orthonormality property of spherical harmonics
\begin{eqnarray}
\int Y_{l'm'}(\hat n' )Y^*_{l''m''}(\hat n' ) = \delta_{l'l''}\delta_{m'm''}\,,
\end{eqnarray}
we obtain, 
\begin{eqnarray}
\sum_{l=0}^{l=\infty}\sum_{m=-l}^{m=l} a_{lm}Y_{lm}(\hat n ) = \sum_{l'=0}^{l'=\infty}\sum_{m'=-l'}^{m'=l'} a'_{l'm'}Y_{l''m''}(\hat n ) B_{l'} \, .
\end{eqnarray}
We again use orthonormality of spherical harmonics to obtain
\begin{eqnarray}
 a_{lm} = a'_{lm} B_{l} \, .
\end{eqnarray}
This relation shows that due to the finite resolution of the instrument
spherical harmonic coefficients get multiplied by the Legendre
transform of the beam function. Hence it is easier to account for the
effect of a circular beam in the spherical harmonic space than
deconvolving the map by the beam function in the pixel space.

The effect of finite pixel size of the map has similar effect on the
recovered spherical harmonic coefficients. The pixel window functions
$p_l$ have been supplied with the HEALPix distribution for different
resolutions. Taking into account finite pixel size we have
\begin{eqnarray}
 a_{lm} = a'_{lm} B_{l}p_l \, .
\end{eqnarray}
Hence the recovered power spectrum
is related to the actual CMB sky power spectrum as
\begin{eqnarray}
 \hat C_{l} = \hat C'_{l} B^2_{l}p^2_l \, .
\end{eqnarray}
 In presence of both beam and finite pixel effects,
eq.~(\ref{Coupling}) obtained for the ideal case~\footnote{In
Ref.~\cite{Hivon} the authors followed the convention that $B_{l}$
represents combined smoothing effect due to pixel as well as beam. Our
notation is different from them.} is now modified to
\begin{eqnarray}
\left <\tilde C_l\right> = \sum_{l'}M_{ll'}\left<\hat C^c_{l'}\right>B^2_{l'}p^2_{l'} \, .
\end{eqnarray}
A generalization of this expression in case of cross power spectrum
has been also reported in Ref.~\cite{Xspect}. The partial sky cross
power spectrum is related to the full sky power spectrum in the
following manner
\begin{eqnarray}
\left<\tilde C^{ij}_l \right>= \sum_{l'}M_{ll'}\left<\hat C^c_{l'}\right>B^i_{l'}B^j_{l'}p^2_{l'} \, .
\end{eqnarray}
Here we note that the noise terms drop out in the cross-correlation
procedure.

The final estimate of the full-sky spectrum is obtained by simply
inverting the coupling matrix. We have used a singular value
decomposition technique to invert our matrix. We checked that this matrix
gives unbiased estimates of the full sky power spectrum using Monte
Carlo simulations of CMB maps and using Kp2 cut.

The final full sky estimate of the power spectrum is obtained from
\begin{eqnarray}
 C^{\bf i}_l = \sum_{l'}(M^{-1})_{ll'}\tilde C^{ij}_{l'}/B^i_{l}B^j_{l}p^2_{l} \, .
\end{eqnarray}
with ${\bf i}\equiv (i,j)$, ${\bf i}= (1, 2, 3, 4, ..., 24)$ as there
are $24$ cross-correlations possible that do not have detector noise
bias. Using the $24$ cross-power spectra we form an
averaged power spectrum following
\begin{equation}
\bar C_l=\sum_{i=1}^{i=24}C_l^i N_{cross} 
\end{equation}
where  $N_{cross}={1}/{24}$.
Now we bin them in the same manner as the WMAP. The binned power spectrum is defined as 
\begin{eqnarray}
C_b=\frac{1}{\Delta l}\sum_{l=l_{min}}^{l=l_{max}}\frac{l(l+1)}{2 \pi}\bar C_l \, ,
\label{Binned_power}
\end{eqnarray}
with
$\Delta l=l_{max}+1-l_{min}\label{deltal}$.
Eqn.~(\ref{Binned_power}) defines our final power spectrum. 
 
\section{Estimation of residual unresolved point source contamination}
\label{res_ps_method}
In this appendix we study  
the point source bias in the auto power
spectrum of a cleaned map and in the cross power
of two cleaned maps. 
From a detailed study of a single iteration cleaning
method we verify that a residual point source bias in the auto or cross power spectrum of the 
cleaned maps could be very well approximated by $\bf \hat W C^{ps}_l \hat W^T_l$, without a need to compute $\left <\bf \hat W C^{ps}_l \hat W^T_l\right>$. 
For the auto power spectrum we optain an analytic point source bias in terms
of the noise covariance matrix and the point source covariance matrix.
 We note that, in practice when we estimate
cosmological power spectrum from the WMAP data we do not use the auto
power spectrum of a cleaned map. Nevertheless the auto power spectrum
is important to study along with the cross power spectrum for a deeper
understanding of the  point source bias. We note that in all the Monte
Carlo simulations in this section we treat point sources as fixed templates. The
sources of randomness come from CMB and detector noises.

\subsection{Auto Power Spectrum}

Following eq.~(\ref{bias_fg_nse}) the cleaned auto power spectrum obeys,
\begin{eqnarray}
\left < \hat C^{Clean}_l \right >= \left< \hat C^c_l \right>+ {\left< \frac{1}{\bf e^T_0{(\hat C^{f+N}_l)}^{-1}e_0} \right>} + (1-n_c)\frac{\left<\hat C^{c}_l\right>}{2l+1} \, .
\label{ps_b}
\end{eqnarray}
Clearly the positive bias is given by ${ \left< 1/{\bf e^T_0{(\hat C^{f+N}_l)}^{-1}e_0} \right>}$ which is caused by the detector noise and foreground
covariance matrix of the map. In this appendix we are basically interested
in studying point source
bias and since diffuse foreground is not a concern at large $l$, 
we do not include any diffuse foreground component in this study. Considering the second term on the right hand side of eq.~{(\ref{ps_b}) we note that
  it is not possible to identify a bias which solely comes from point sources. Another 
  point to note is that we know only the mean noise covariance matrix, $\bf C^N_l$, not the empirical noise covariance matrix, $\bf \hat C^N_l$, 
  where $\bf \hat C^N_l = C^N_l + \delta \hat C^N_l$ and 
  $\bf \delta \hat C^N_l$ denotes noise fluctuation
  on the true noise level. Therefore we express 
 ${ \left< 1/{\bf e^T_0{(\hat C^{f+N}_l)}^{-1}e_0} \right>}$ in terms of 
 a mean noise covariance matrix and the foreground covariance matrix. As we will
see such a simplification helps us to obtain an analytical expression for
the point source bias. Keeping
in mind that WMAP detector noise level is much larger than the point 
source spectrum, we simplify ${\left< 1/{\bf e^T_0{(\hat C^{f+N}_l)}^{-1}e_0} \right>}$ analytically 
by making expanding to first order~{\footnote {We note that we cannot use Sherman-Morrison formula to decouple the foreground and detector noise bias from the term ${\bf \left< \frac{1}{e^T_0{(\hat C^{f+N}_l)}^{-1}e_0} \right>}$
in an useful form.}}, 
in terms of $\bf C^{ps}_l{(\hat C^{N}_l)}^{-1}$,
\begin{eqnarray}
\label{ex1}
\bf ({\hat C^{f+N}_l)}^{-1} = \hat {(C^{N}_l)}^{-1} -\hat {(C^{N}_l)}^{-1}C^{ps}_l \hat {(C^{N}_l)}^{-1}\,,
\end{eqnarray}
Due to the low point source contamination relative to the WMAP noise level 
it is reasonable to assume that ${(\bf e^T_0{(\hat C^{N}_l)}^{-1} C^{ps}_l {(\hat C^{N}_l)}^{-1}e_0)}/{(\bf e^T_0 {(\hat C^{N}_l)}^{-1}e_0)} \ll 1$. 
Expanding to first order we obtain,
\begin{eqnarray}
\left < \frac{1}{\bf e^T_0 {(\hat C^{f+N}_l)}^{-1}e_0}\right> = \left < \frac{1}{\bf e^T_0 {(\hat C^{N}_l)}^{-1}e_0}\right> + \bf \left<\frac{ e^T_0 {(\hat C^{N}_l)}^{-1}}{e^T_0 {(\hat C^{N}_l)}^{-1}e_0}C^{ps}_l\frac{ {(\hat C^{N}_l)}^{-1}e_0}{e^T_0 {(\hat C^{N}_l)}^{-1}e_0}\right> \,.
\label{auto_bias}
\end{eqnarray}
We interprete the first term on the right hand side
as the detector noise bias. The second term is the point source 
bias. In the following subsections
we recast these results in terms of the mean
noise covariance matrix. We also verify our analytic expressions 
using Monte Carlo simulations.

\subsubsection{Noise induced  bias}
We first turn our attention to the detector noise bias. We 
estimate this term upto second order in 
$\bf \delta \hat C^N_l {(C^{N}_l)}^{-1}$,
\begin{eqnarray}
\nonumber \biggl< \frac{1}{\bf e^T_0 {(\hat C^{N}_l)}^{-1}e_0}\biggr> = \biggl<
\frac{1}{ \bf e^T_0 {(C^{N}_l)}^{-1} e_0} \biggl(1+ \frac{\bf e^T_0 {(C^{N}_l)}^{-1}\delta \hat
C^N_l {(C^{N}_l)}^{-1}e_0}{\bf e^T_0 {(C^{N}_l)}^{-1}e_0} -\frac{\bf e^T_0 {(C^{N}_l)}^{-1}\delta \hat
C^N_l {(C^{N}_l)}^{-1} \delta \hat C^N_l {(C^{N}_l)}^{-1}e_0}{\bf e^T_0 {(C^{N}_l)}^{-1}e_0}  \\
+\frac{(\bf e^T_0 {(C^{N}_l)}^{-1}\delta \hat C^N_l {(C^{N}_l)}^{-1}e_0)^2}{(\bf e^T_0 {( C^{N}_l)}^{-1}e_0)^2}\biggr) \biggr>\,.
 \label{ns_b1}
\end{eqnarray}
The noise fluctuations are assumed to have zero mean, $\left< \bf \delta \hat C^N_l\right> =0$. Hence the second term in the
bracket vanishes on the ensemble average. 
Only the numerators of the third and fourth terms in the 
bracket on the right hand side of eq.~(\ref{ns_b1}) are stochastic variables. 
On the ensemble average
the numerator of the third term becomes,
\begin{eqnarray}
\left < \bf e^T_0 {(C^{N}_l)}^{-1}\delta \hat C^N_l {(C^{N}_l)}^{-1} \delta \hat C^N_l {(C^{N}_l)}^{-1}e_0\right> = \left <\sum_{i,j,k}C^{N i}_l \delta \hat
C^{N(ij)}_lC^{N j}_l\delta \hat C^{N(jk)}_lC^{N k}_l \right> = \sum_{i,j,k}C^{N i}_l C^{N j}_l C^{N k}_l \left < \delta \hat
C^{N(ij)}_l\delta \hat C^{N(jk)}_l\right>\,, \nonumber
\end{eqnarray}
where we have assumed that the noise covariance matrix of WMAP is
diagonal. We simplify the ensemble averaged quantities using the
relation eq. ($28$) of Ref.~\cite{Xspect}. If we consider a general 
term of the form $\left <\delta \hat C^{N(ij)}_l\delta \hat C^{N(kp)}_l\right>$
then for uncorrelated detector
noises, the ensemble averaged quantities will survive only for the
same pair, i.e. when $(i,j) = (k,p)$. We further
note that,
\begin{equation}
\left<\delta \hat C^{{N(ij)}^2}_l\right> =\Bigg\lbrace 
\begin{matrix}\frac{1}{2l+1}C^{Ni}_lC^{Nj}_l,\ \  i\ne j \\
 \frac{2}{2l+1}C^{Ni}_lC^{Ni}_l,\ \ i= j\end{matrix}  \ .
\end{equation}
Using above relations the third and fourth term in the bracket of eq.~(\ref{ns_b1}) become
\begin{eqnarray}
\left<\frac{\bf e^T_0{(C^{N}_l)}^{-1}\delta \hat C^N_l {(C^{N}_l)}^{-1}\delta \hat C^N_l {( C^{N}_l)}^{-1}e_0}{\bf e^T_0 {(C^{N}_l)}^{-1}e_0} \right> = \frac{(1+n_c)}{2l+1} \,,
\label{T2}
\end{eqnarray}
Following a similar method we also simplify the fourth term in eq.~(\ref{ns_b1}). The 
final result is
\begin{eqnarray}
\left<\frac{(\bf e^T_0 {(C^{N}_l)}^{-1}\delta \hat C^N_l {(C^{N}_l)}^{-1}e_0)^2}{(\bf e^T_0 {( C^{N}_l)}^{-1}e_0)^2}\right> = \frac{2}{2l+1} \,.
\label{T3}
\end{eqnarray}
Using eqs.~(\ref{ns_b1}), (\ref{T2}), (\ref{T3}) we obtain,
\begin{eqnarray}
 \left < \frac{1}{\bf e^T_0 {(\hat C^{N}_l)}^{-1} e_0}\right> = \left <\frac{1}{ \bf e^T_0 {( C^{N}_l)}^{-1}e_0} \left(1 - \frac{1+n}{2l+1}+\frac{2}{2l+1}\right) \right>=\frac{1}{ \bf e^T_0 {(C^{N}_l)}^{-1}e_0}\frac{2l+2-n_c}{2l+1}\,.
 \label{nb}
\end{eqnarray}
This enables us to
compute the noise bias in terms of the theoretical noise models.
We verified this equation by
Monte-Carlo simulations of $1000$ noise covariance matrices
corresponding to WMAP V and W bands. The noise maps from which 
these empirical noise covariance matrices were computed 
were formed following a method similar to that mentioned in 
section \ref{er_bar}. In Fig.~\ref{ns_b1_fig} 
we show the difference between the empirical
noise bias $\left < 1/{\bf
e^T_0 {(\hat C^{N}_l)}^{-1} e_0}\right>$ and the second order 
analytical expression for noise bias,
given by the right hand side of eq.~(\ref{nb}).
The difference is consistent with zero. The large fluctuations at large
$l$ is caused by the beam deconvolution effect in the
effective noise covariance matrix.
\begin{figure}
\includegraphics[scale=0.4,angle=-90]{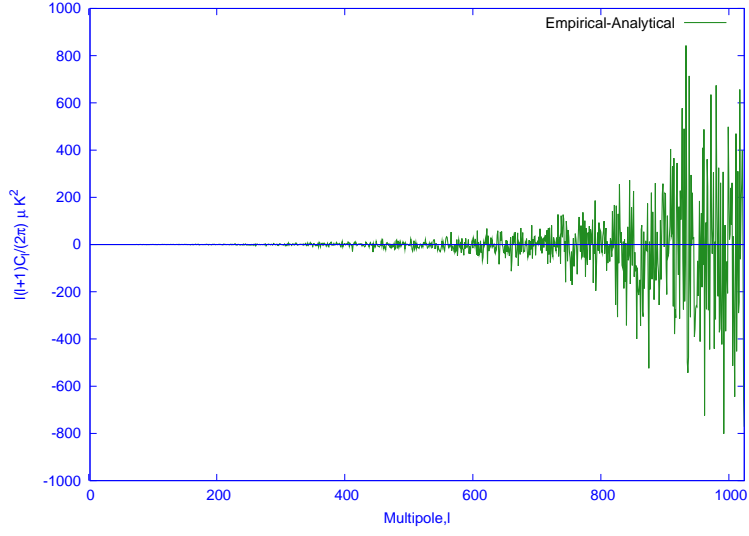}
\caption{The difference between the empirical noise bias and the theoretical
estimate of noise bias given in eq.~\ref{nb} for V and W bands.  
}
\label{ns_b1_fig}
\end{figure}

\subsubsection{Point source bias}
Here we study the point source bias.
We perform Monte-Carlo simulations in a single 
iteration cleaning with CMB, point sources 
and detector noise having WMAP's noise level. We compute ensemble averaged cleaned power
spectrum $\left< \hat C^{Clean}_l\right>$ from $1000$ such simulations. The validity of eq.~(\ref{nb}) allows us to compute noise bias in terms of 
the theoretical noise covariance matrix and subtract this analytical
result from $\left< \hat C^{Clean}_l\right>$. After correcting 
for noise bias we find that $\left< \hat C^{Clean}_l\right> - \left(1/{ \bf e^T_0{(C^{N}_l)}^{-1}e_0}\right)\left({(2l+2-n_c)}/{(2l+1)} \right)$ 
still has a residual positive bias at multipole range $l >400$. This excess is a
clear demonstration of the presence of point source bias. The brown line 
in Fig.~\ref{Ps_b1} is the power spectrum, corrected for the noise and the 
negative CMB bias of the form $-C^c_l/(2l+1)$. The excess is caused by the point source bias. We analytically compute the point source bias in terms of theoretical noise covariance matrix similar to
eq~({\ref{nb}). Following a first order expansion in $\bf C^{ps}_l {(\hat C^{N}_l)}^{-1}$ and a second order expansion in terms of $\bf \delta \hat C^N_l {(C^{N}_l)}^{-1}$ we
obtain,
\begin{eqnarray}
{\bf \left<\frac{ e^T_0 {(\hat C^{N}_l)}^{-1}}{e^T_0 {(\hat C^{N}_l)}^{-1}e_0}C^{ps}_l\frac{ {(\hat C^{N}_l)}^{-1}e_0}{e^T_0 {(\hat C^{N}_l)}^{-1}e_0}\right> =  \frac{ e^T_0 {(C^{N}_l)}^{-1}}{e^T_0 {( C^{N}_l)}^{-1}e_0}C^{ps}_l\frac{{(C^{N}_l)}^{-1}e_0}{e^T_0 {(C^{N}_l)}^{-1}e_0}} \frac{2l}{2l+1} + \frac{tr(\bf C^{ps}_l{(C^{N}_l)}^{-1})}{(2l+1)(\bf e^T_0 {(C^{N}_l)}^{-1}e_0)} \,.
\label{ps_1st_ord_ana}
\end{eqnarray}
\begin{figure}
\includegraphics[scale=0.35,angle=-90]{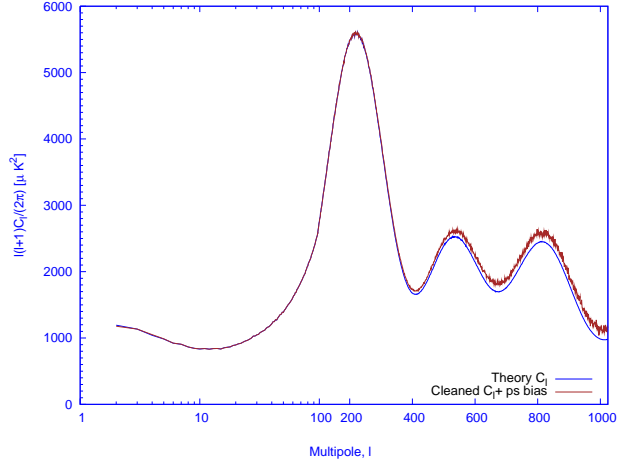}
\caption{The auto power
spectrum without correcting for the residual unresolved point source
bias is shown in brown line. The blue line shows the theoretical CMB power
spectrum.}
\label{Ps_b1}
\end{figure}
Again we verify the validity of the above expression following Monte Carlo simulations. We 
compute the point source bias from $1000$ cleaning simulations following 
$\left< \hat C^{Clean}_l\right> - C^c_l - C^c_l/(2l+1) - \left(1/{ \bf e^T_0 {( C^{N}_l)}^{-1}e_0}\right)\left((2l+2-n_c)/(2l+1)\right)$ where $C^c_l$ denotes the theoretical CMB 
power spectrum. In Fig. \ref{ps_b2} the green line shows the binned estimate of the point source 
bias from simulations. The error-bars plotted in this figure are valid for
the mean of 
$1000$ simulations of random samples.
The black line with filled points shows the analytical estimate 
of the point source bias computed from the right hand side of the eq.~(\ref{ps_1st_ord_ana}).
These two results match closely except at the last two points. However, one can easily
 conclude from the error-bars  plotted in this figure that differences between 
the two lines at these two points are not significant. We interprete these deviations as noise 
 fluctuations. 

Given the large noise level of WMAP maps, one 
might expect that $\left <\bf \hat W_l C^{ps}_l \hat W^T_l \right>$
would be a good approximation of the point source bias that is given by
the second term of eq.~(\ref{auto_bias}). In the limit ${\bf \hat C_l} \rightarrow {\bf \hat C^N_l}$, applicable when noise dominates at large $l$ and 
point source bias is also significant, these two terms are identical. Using the analytical form of weights as in eq.~(\ref{Weight}), we find that up to the first order in $\bf C^{ps}_l\hat C^{N-1}_l$,
\begin{eqnarray}
\nonumber {\bf \left < \hat W_l C^{ps}_l \hat W^T_l \right >} = {\bf \left<\frac{ e^T_0{(\hat C^{N}_l)}^{-1}}{e^T_0 {(\hat C^{N}_l)}^{-1}e_0}C^{ps}_l\frac{ e^T {(\hat C^{N}_l)}^{-1}}{e^T_0 {(\hat C^{N}_l)}^{-1}e_0}\right>} - \frac{C^c_l}{2l+1} {\bf \left< \frac{e^T_0 {(\hat C^{N}_l)}^{-1}C^{ps}_l{(\hat C^{N}_l)}^{-1} e_0}{e^T_0 {(\hat C^{N}_l)}^{-1}e_0} \right>} \\
+ \frac{C^c_l}{2l+1}tr(\bf C^{ps}_l\left<{(\hat C^{N}_l)}^{-1}\right>) \,.
\label{ps_b_wt}
\end{eqnarray}
Here, $C^c_l$ is the theoretical CMB power spectrum. However, the last
two terms in the above equations are negligibly small compared to the
first term on the right hand side. From simulations we 
find that these two terms contribute less than $0.1 \mu K^2$ to the point source 
residual when multiplied by the prefactor $l(l+1)/(2\pi)$. The reason why these terms
  become negligible is that, apart from a multiplicative first order term,  $\bf C^{ps}_l {(\hat C^{N}_l)}^{-1}$, the first term on the right hand side go 
as $\bf C^N_l$, whereas the last two terms goes   as $C^c_l/(2l+1)$. Indeed, the noise becomes much larger than CMB power spectrum at large $l$ due to beam deconvolution, making the two last two terms insignificant compared to the first 
term. In  figure~\ref{Ps_b2} we show magnitude of the individual contributions arising from the second  and third term following $1000$ Monte Carlo simulations of the cleaning method. We use V and W bands  in the simulations using realistic WMAP noise levels and residual unresolved point source model. The red line shows the difference  between the two terms which is  less than  $0.01 \mu K^2$ or even smaller at $l>400$. Hence it is justified to assume,
\begin{eqnarray}
{\bf \left < \hat W_l C^{ps}_l \hat W^T_l \right >} = {\bf \left<\frac{ e^T_0 {(\hat C^{N}_l)}^{-1}}{e^T_0 {(\hat C^{N}_l)}^{-1}e}C^{ps}_l\frac{ e^T_0 {(\hat C^{N}_l)}^{-1}}{e^T_0{(\hat C^{N}_l)}^{-1}e_0}\right>} \,.
\label{ps_b3}
\end{eqnarray}
In Fig.~\ref{Ps_b1} we show the point source bias computed following${\bf \left < \hat W_l C^{ps}_l \hat W^T_l \right >}$ in blue line. This matches well
with the 
black line with filled circle, justifying eq.~(\ref{ps_b3}).



\begin{figure}
\includegraphics[scale=0.35,angle=-90]{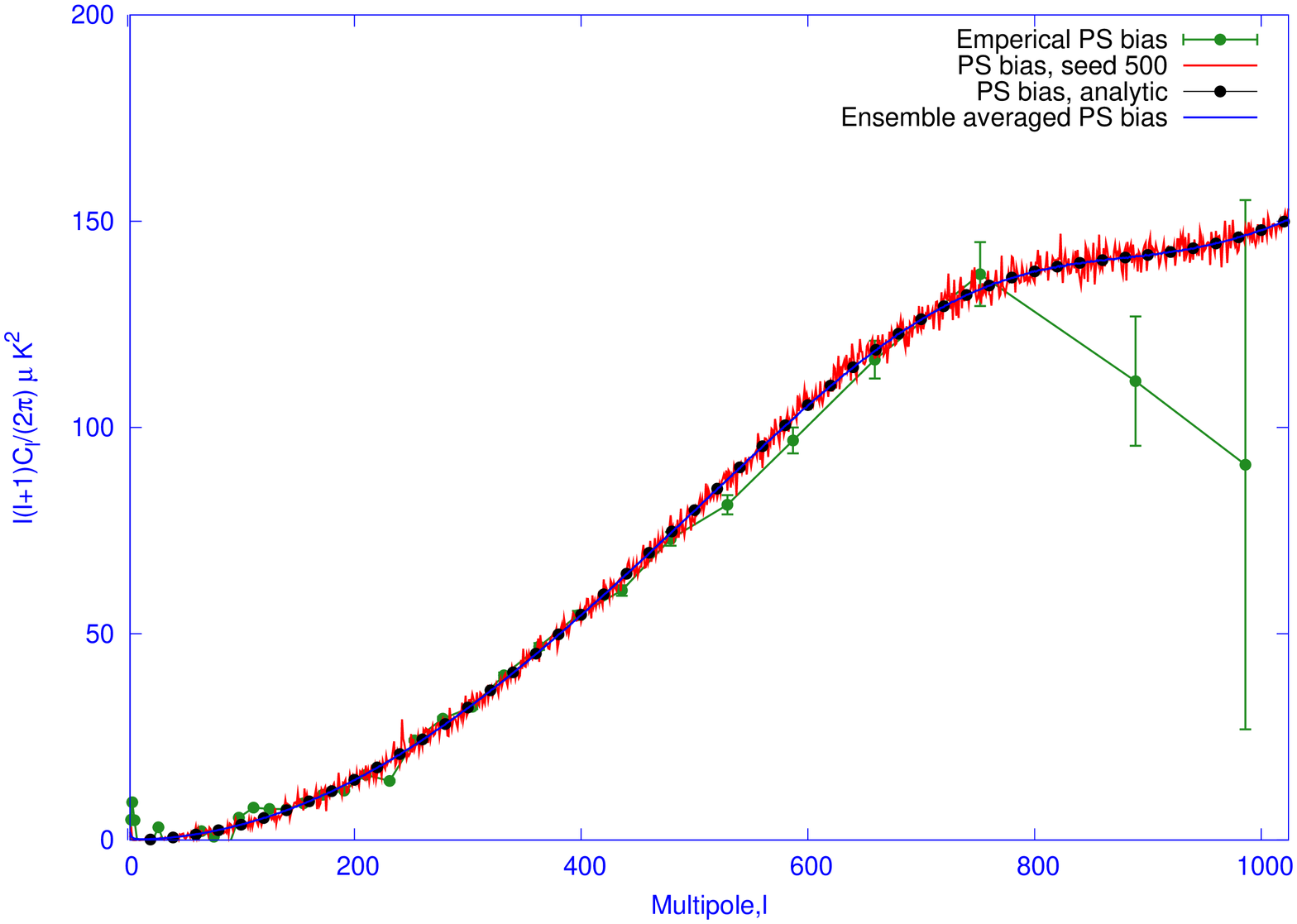}
\caption{
The point source bias, $\bf \left< \hat W_l C^{ps}_l\hat
W^T_l\right>$ (blue line), and the approximation $\bf \hat
W_l C^{ps}_l\hat W^T_l$ (red line) for the case of 
auto power spectrum of the cleaned maps. The empirical point
source bias computed by using $\left< \hat C^{Clean}_l\right> - C^c_l - C^c_l/(2l+1) - \left(1/{ \bf e^T_0 {( C^{N}_l)}^{-1}e_0}\right)\left((2l+2-n_c)/(2l+1)\right)$ is shown in
green line. The error bars are computed only for the term $\left <
1/{\bf e^T_0 {(\hat C^{N}_l)}^{-1}e_0}\right>$ without including the cosmic
variance. All the data points are binned following the WMAP's 3 year
binning method. The analytical point source bias computed following
eq.~(\ref{ps_1st_ord_ana}) is also shown in black line with
black dots.}
\label{ps_b2}
\end{figure}

\begin{figure}
\includegraphics[scale=0.35,angle=-90]{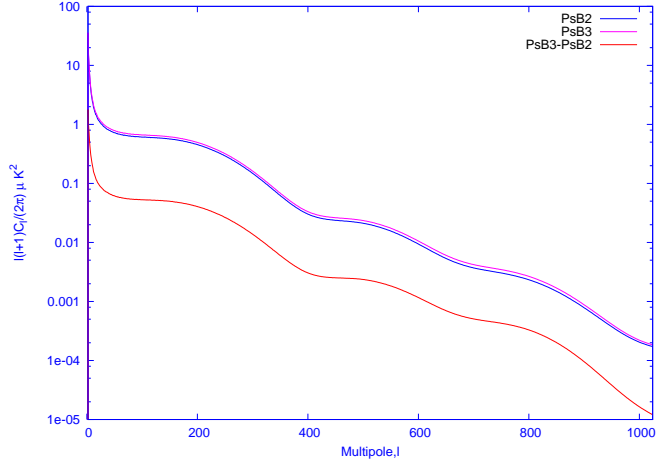}
\caption{ The contributions arising from the third 
(blue line) and fourth (pinkline) terms of eq.~(\ref{ps_b_wt}) and their
difference. } 
\label{Ps_b2}
\end{figure}

The unresolved point source bias is then $\left<{\bf \hat
W_lC^{ps}_l\hat W^T_l}\right>$. Now we propose that, the point source
bias could also be taken into account following ${\bf \hat
W_lC^{ps}_l\hat W^T_l} \sim \left<{\bf \hat W_lC^{ps}_l\hat
W^T_l}\right>$. The reason is that, at large $l$ where detector noise
dominates. The weights are entirely determined by the mean noise level. 
Therefore weights become approximately constant from realization to realization.
Small fluctuations in weights due to noise fluctuations are not important
compared to the magnitude of total point source bias $\left<{\bf \hat
W_lC^{ps}_l\hat W^T_l}\right>$. The advantage of this method is that,
we can estimate point source bias in terms of the weights which are the 
results of a cleaning. In the figure \ref{ps_b2} we show how
well ${\bf \hat W_lC^{ps}_l\hat W^T_l}$ (red line) matches with $\left<{\bf \hat
W_lC^{ps}_l\hat W^T_l}\right>$. This result is obtained from $1000$
Monte Carlo simulations of our cleaning method using V and W bands. 
 
\subsection{Cross Power Spectrum}
We now consider point source bias in case of cross power spectrum of
two cleaned maps. In this case the cleaned power spectrum is
given by,
 \begin{eqnarray}
\left <\hat C^{Clean}_l \right > = C^c_l + 2 (1-n_c) \frac{C^c_l}{2l+1} + \bf <\hat W^{1}_l C^{ps}_l \hat W^{2^T}_l> + <\hat W^{1}_l \hat C^{N(12)}_l \hat W^{2^T}_l> \,.
 \end{eqnarray}
Here $\bf \hat W^{1}_l$ and $\bf \hat W^{2}_l$ are the weight vectors for 
the two cleaned maps and $\bf \hat C^{N(12)}_l$ denotes  a  chance noise correlation in the two sets of maps used to obtain two cleaned maps. From Monte-Carlo simulations involving cross spectrum, $v1w12\otimes v2w34,v1w13\otimes v2w24,v1w14\otimes v2w23, v1w24\otimes v2w13,v1w34\otimes v2w12 $ we recover a point source bias at the large multipole range. In the left panel of the figure~\ref{ps_cross} we show the point source bias from $1000$ such simulations. 
 \begin{figure}
 \includegraphics[scale=0.35,angle=-90]{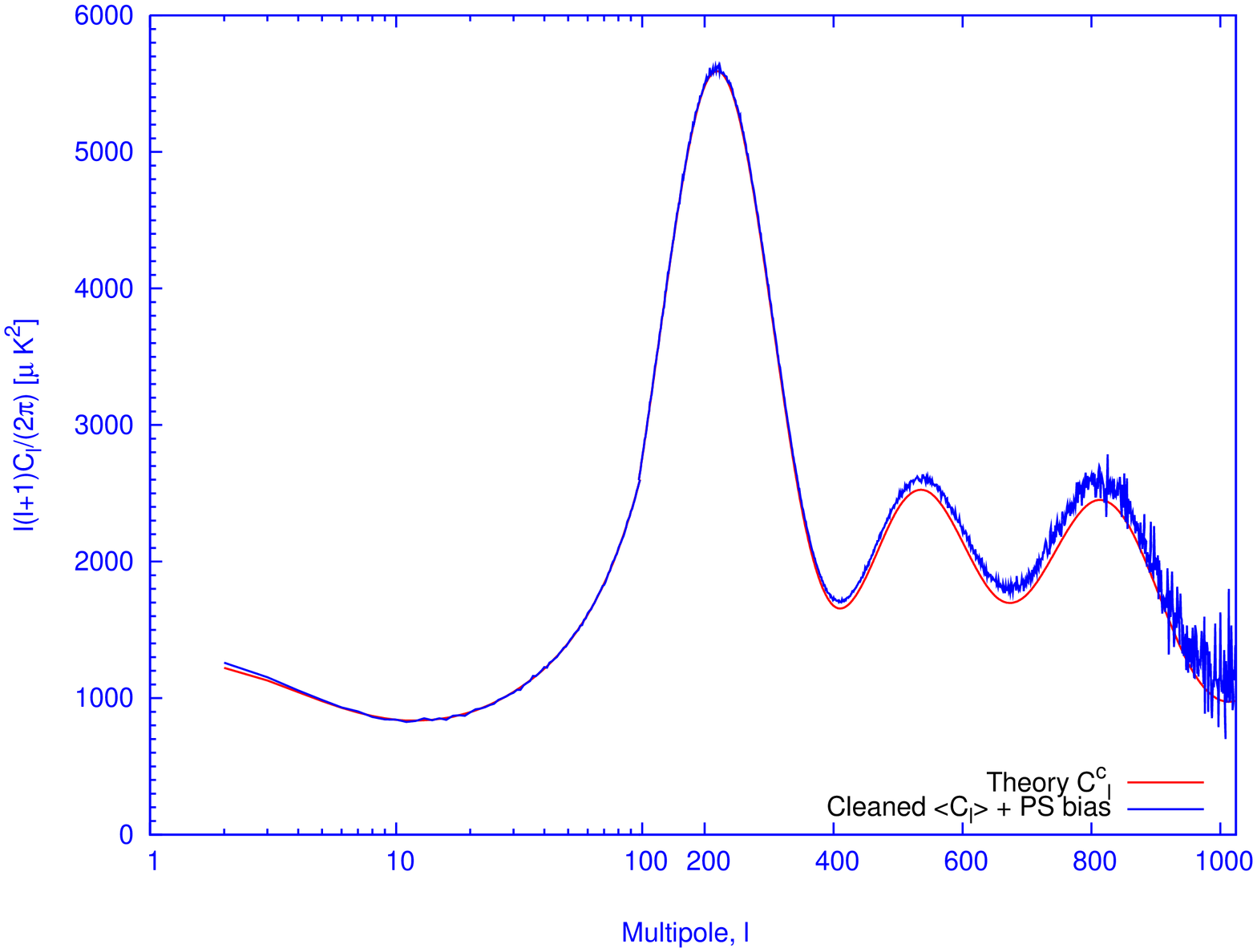}
 \includegraphics[scale=0.35,angle=-90]{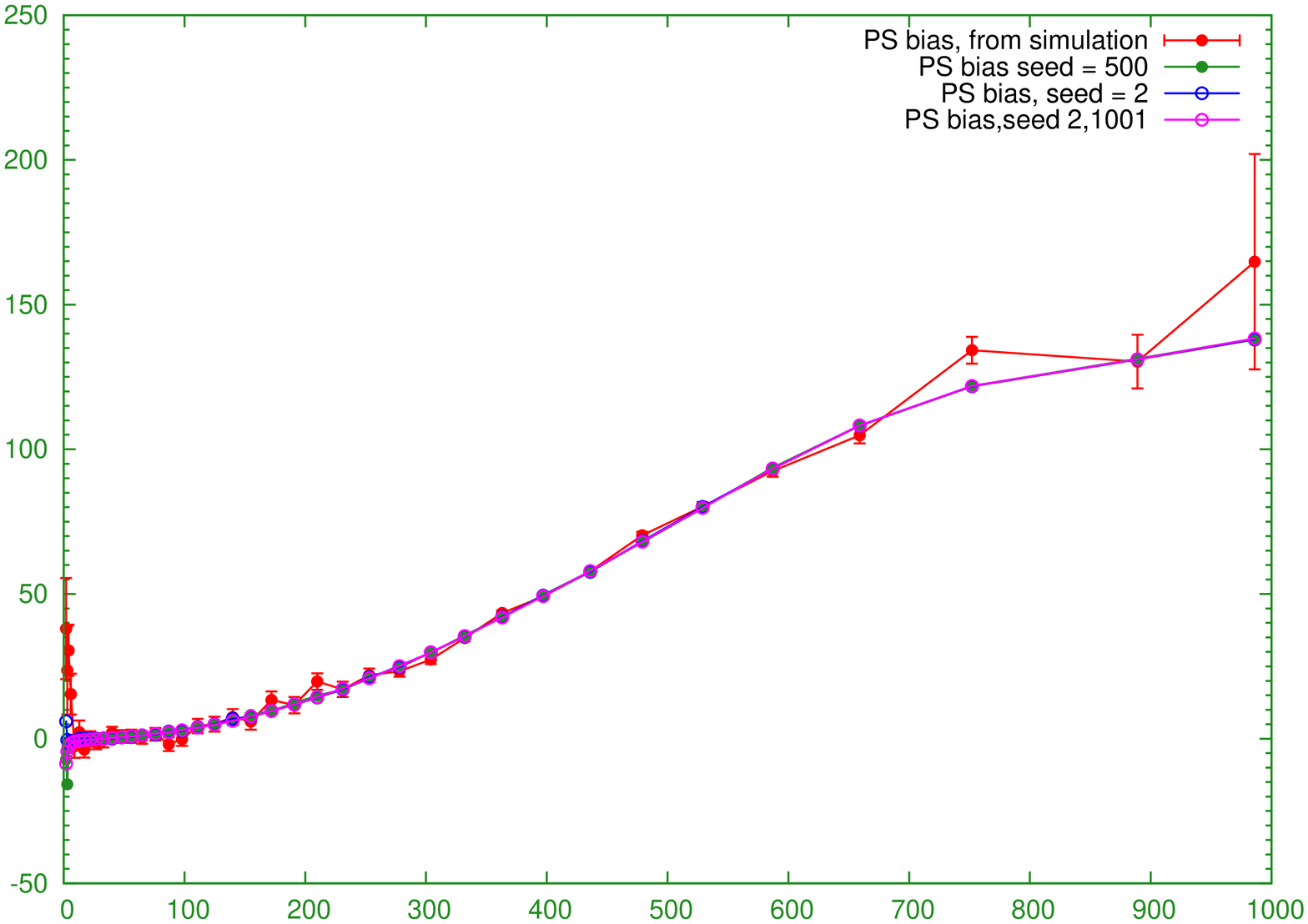}
 \caption{{\it Left:} The point source bias at large multipole range
 obtained from $1000$ Monte Carlo simulations of the cleaning method
 involving V and W bands is shown in blue line. The CMB bias
 $\frac{1-n_c}{2l+1}C^c_l$ has been removed before plotting the blue
 line to show the bias coming from point source only. The red line
 shows the theoretical power spectrum from which random realization of
 CMB were drawn.{\it Right:} Point source bias computed from $\left
 <\hat C^{Clean}_l \right > - C^c_l - 2 (1-n_c) \frac{C^c_l}{2l+1}$
 is shown in red line. The magenta line shows point source bias computed
 from $\left< \bf \hat W^{1}_l C^{ps}_l \hat W^{2^T}_l\right>$ which
 agrees excellently with $\bf \hat W^{1}_l C^{ps}_l \hat W^{2^T}_l$
 computed from two arbitrary realizations shown in green and blue
 lines. }
 \label{ps_cross}
 \end{figure}
Like the auto power case, here too the point source bias could be
obtained as $\bf \hat W^{1}_l C^{ps}_l \hat W^{1^T}_l$.  We show this
in the right panel of the figure~\ref{ps_cross}. The red line shows
bias computed from simulations, i.e. $\left <\hat C^{Clean}_l \right >
- C^c_l - 2 (1-n_c) \frac{C^c_l}{2l+1}$.  We compute $\bf \hat W^{1T}_l
C^{ps}_l \hat W^2_l$ for any two arbitrary realizations. These are
shown in blue and green colors respectively. The magenta colored line
shows $\left < \bf \hat W^{1}_l C^{ps}_l \hat W^{2^T}_l\right>$ which is
in excellent agreement with blue and green lines. This justifies the 
approximation
$\left < \bf \hat W^{1}_l C^{ps}_l \hat W^{2^T}_l\right> \sim \bf \hat
W^{1}_l C^{ps}_l \hat W^{2^T}_l$, so that point source bias could as well
be computed from the weights from a single realization without any need of
ensemble average. We have followed this approach of point source removal
in the multi-region iterative cleaning method described in section
\ref{ps_res}.

\section{Bias at low-$l$}
\label{bias_lowl}

Assume that there are $n_f$ number of foreground components. Each of
these follow rigid frequency scaling. The number of channels is $n_c$,
where $n_c \ge n_f+1$. The rank of full covariance matrix is determined
by the number of independent components. Therefore, for $n_f$ number of
independent foreground components and an additional CMB component the
rank of the full covariance matrix is  $n_f+1$. In many cases we
choose total number of channels $n_c$ to be equal to the rank of the
matrix, $n_f+1$. This ensures that the full covariance matrix is of full
rank, (order=rank) so that it is invertible.

However it is also interesting to explore the freedom of using a
singular covariance matrix $\bf \hat C_l$. This may arise, for
example, if we consider $n_f$ number of foreground components (each with
rigid frequency scaling) but with $n_c$ number of channels with $n_c >
n_f+1$. The motivation of this work lies in the fact we have shown that
the cleaned power spectrum could still be defined in terms of Moore
Penrose Inversion (in eq.~(\ref{Clean_mpi})) in case we encounter a
singular covariance matrix. So it is worth investigating whether there
is a further simplified form of eq.~(\ref{Clean_mpi}).

The calculations proceed as follows:
\begin{itemize}
\item{} Obtain an analytic expression for the full covariance
matrix. We find that the full
covariance matrix can be obtained as three successive rank-one updates
of three different matrices.
\item{}Identify the appropriate cases of the generalized versions of
Sherman Morrison formula which applies to each of three rank-one
updates that have to be carried out.
\item{} Apply Sherman Morrison formula successively for the three rank
one updates to obtain analytical expression for the bias.
\end{itemize}

Let the $p^{th}$ foreground component for channel $i$ be denoted by
$F^p_0(\theta,\phi)f^i_p$. Here, $F^p_0(\theta, \phi)$ is the $p^{th}$
foreground template based on frequency $\nu_0$, so that $f^i_p=1$, for
frequency $\nu_0$. The full signal map at $i^{th}$ frequency channel
is given by
\begin{eqnarray}
S^i(\theta, \phi) = C(\theta, \phi) + \sum_{p=1}^{n_f}F^p_0(\theta,\phi)f^i_p \,.
\end{eqnarray}
Alternatively, in the spherical harmonic space,
\begin{eqnarray}
a^i_{lm}= a^c_{lm} + \sum_{p=1}^{n_f}f^i_pa^{p0}_{lm} \,.
\end{eqnarray}
The auto power spectrum of the $i^{th}$ channel
\begin{eqnarray}
\hat C^i_l = \hat C^c_l + 2\sum_{p=1}^{n_f}f^i_p\hat C^{cf(p)0}_l + \sum_{p,p'}^{n_f}f^i_pf^i_{p'} C^{(pp')0}_l\,,
\end{eqnarray}
where $C^{(pp')0}_l $ is the correlation between any two foreground
components $p,p'$ and $\hat C^{cf(p)0}_l$ is the chance correlation
between CMB and $p^{th}$ foreground component.  The cross power
spectrum between two channels $i,j$ are given by
\begin{eqnarray}
\hat C^{ij}_l = \hat C^c_l + \sum_{p=1}^{n_f}f^i_p\hat C^{cf(p)0}_l +
\sum_{p=1}^{n_f}f^j_p\hat C^{cf(p)0}_l+ \sum_{p,p'}^{n_f}f^i_pf^j_{p'}
C^{(pp')0}_l \,,
\end{eqnarray}
or
\begin{eqnarray}
\hat C^{ij}_l = \hat C^c_l + \sum_{p=1}^{n_f}f^i_p\hat C^{cf(p)0}_l +
\sum_{p=1}^{n_f}f^j_p\hat C^{cf(p)0}_l+ F^{ij}_l\,.
\end{eqnarray}
Introducing explicit matrix notations for the equation, we write 
\begin{eqnarray}
\hat {\bf  C_l} = \hat C^c_l \left(\begin{array}{cccc}
1 & 1 & ... & 1 \\
1 & 1 & ... & 1 \\
. & . & ... & . \\
1 & . & ... & 1
\end{array}
\right)_{(n_c\times n_c)}
+ \sum_{p=1}^{n_f}\hat C^{cf(p)0}_l\left(\begin{array}{cccc}
f^1_p & f^1_p  & ... & f^1_p  \\
f^2_p & f^2_p  & ... & f^2_p  \\
.     & .      & ... & . .   \\
f^{n_c}_p & f^{n_c}_p  & ... & f^{n_c}_p 
\end{array}
\right)_{(n_c\times n_c)} \nonumber \\
+ \sum_{p=1}^{n_f}\hat C^{cf(p)0}_l\left(\begin{array}{cccc}
f^1_p & f^2_p  & ... & f^{n_c}_p  \\
f^1_p & f^2_p  & ... & f^{n_c}_p  \\
.     & .      & ... & . .   \\
f^1_p & f^2_p  & ... & f^{n_c}_p 
\end{array}
\right)_{(n_c\times n_c)}
+ \left(\begin{array}{cccc}
F^{11} & F^{12}  & ... & F^{1n_c}  \\
F^{21} & F^{22}  & ... & F^{2n_c}  \\
.     & .      & ... & . .   \\
F^{n_c1} & F^{n_c2}  & ... & F^{n_cn_c} 
\end{array}
\right)_{(n_c\times n_c)} \,.
\end{eqnarray}
We define
\begin{eqnarray}
 {\bf  \hat f^{p0}_{l}}= \hat C^{cf(p)0}_l\left(\begin{array}{c}
f^1_p  \\
f^2_p  \\
.     \\
f^{n_c}_p 
\end{array}\right) \,,
\end{eqnarray}
and 
\begin{eqnarray}
 {\bf  e_{0}}= \left(\begin{array}{c}
1  \\
1 \\
.     \\
1 
\end{array}\right)\,.
\end{eqnarray}
Clearly, ${\bf  \hat f^{p0}_{l},  e_{0}} \in  \mathbb{R}_{n_c,1}$, where $\mathbb{R}_{n_f,n_c}$ denotes set of  real $n_f \times n_c$ matrices.
Full covariance matrix  can then be written as,
\begin{eqnarray}
{\bf \hat C_l} = \hat C^c_l{\bf e_0e^T_0} + \left(\sum_{p=1}^{n_f}{\bf \hat f^{p0}_l} \right){\bf e^T_0} + {\bf e_0}\left(\sum_{p=1}^{n_f}{\bf \hat f^{p0}_l} \right)^T + {\bf A_3} \,.
\end{eqnarray}
Define
\begin{equation}
{\bf \hat f^{0}_l}= \sum_{p=1}^{n_f}{\bf \hat f^{p0}_l} \,,
\end{equation}
 then, it is  possible to rewrite
\begin{eqnarray}
{\bf \hat C_l} = \hat C^c_l{\bf e_0e^T_0} + \underbrace{{\bf \hat f^{0}_l} {\bf e^T_0} + \underbrace{{\bf e_0}{\bf \hat f^{0T}_l} + {\bf A_3}}_{\bf A_2}}_{\bf A_1}  \,, 
\end{eqnarray} 
with ${\bf  \hat f^{0}_{l}} \in  \mathbb{R}_{n_c,1}$.

\subsection{Generalized Sherman Morrison Formula for rank one updates}

This section discusses some mathematical theorems that
will be useful to us. The results are mainly based upon the two papers
\cite{CDM,JKB}. In Ref.~\cite{CDM} analytic expressions for MPGI of rank one modified
matrices of the form $\bf M= A+bc^*$, (notations changed) are
reported. Here $\bf M, A$ are any $m \times n$ matrices, i.e. ${\bf M,
A }\in \mathbb{C}_{m,n}$,  ${\bf b} \in \mathbb{C}_{m,1}$ and ${\bf c}
\in \mathbb{C}_{n,1}$. $\mathbb{C}_{m,n}$ is the set of all $m \times
n$ complex matrices, (*) denotes conjugate transpose. The motivation
of the work in Ref.~\cite{CDM} was to generalize the Sherman Morrison
formula
\begin{eqnarray}
{\bf M^{-1} = A^{-1}} -\frac{1}{\lambda}\bf A^{-1}bc^*A^{-1}
\end{eqnarray}    
where $\lambda = 1 +c^*A^{-1}b$ in case of any arbitrary $m \times n$
matrix $\bf A$. The Sherman Morrison formula given by the above
equation is valid only for square and nonsingular matrix.

The main results of the Ref.~\cite{CDM} are a set of formulas for
the MPGI depending upon the different set of conditions, namely

\begin{itemize}

\item[(i)] ${\bf b} \notin \mathcal C ({\bf A})$ and $ {\bf c} \notin \mathcal C (\bf A^*)$

\item[(ii)] ${\bf b} \in \mathcal C ({\bf A})$ and $ {\bf c} \notin \mathcal C (\bf A^*)$ and $\lambda=0$ 

\item[(iii)] ${\bf b} \in \mathcal C ({\bf A})$ and $ {\bf c}$ arbitrary and $\lambda \ne 0$ 

\item[(iv)]${\bf b} \notin \mathcal C ({\bf A})$ and $ {\bf c} \in \mathcal C (\bf A^*)$ and $\lambda=0$ 

\item[(v)] ${\bf b}$ arbitrary and $ {\bf c} \in \mathcal C (\bf A^*) $  and $\lambda \ne 0$ 

\item[(vi)] ${\bf b} \in \mathcal C ({\bf A})$ and $ {\bf c} \in
\mathcal C (\bf A^*)$ and $\lambda=0$
\end{itemize}
 where $\mathcal C ({\bf A})$ represents the column space of a matrix
$\bf A $.

Ref. \cite{JKB} shows that it is sufficient to consider $5$
independent cases only. All these $5$ different cases are listed below
along with an useful theorem proved regarding the rank modification
that occurs during the rank one update of the matrix $\bf A$.

\subsection{Theorem}
\label{Theorem}
{\it For given ${\bf A} \in \mathbb{C}_{m,n}$ and nonzero ${\bf b} \in
\mathbb{C}_{m,1}$ and ${\bf c} \in \mathbb{C}_{n,1}$, let $\bf M$ be
the modifications of $\bf A$ of the form $\bf M=A+bc^*$ and let
$\lambda= 1+\bf c^*A^\dagger b$. Then} \\
\begin{equation}
\nonumber r({\bf M}) = r({\bf A}) -1 \Leftrightarrow {\bf b} \in \mathcal C ({\bf A}),{\bf c} \in \mathcal C ({\bf A^*}), \lambda=0
\end{equation}
\begin{equation*}
r({\bf M}) = r({\bf A})  \Leftrightarrow \begin{cases} {\bf b} \in \mathcal C ({\bf A}),{\bf c} \in \mathcal C ({\bf A^*}), \lambda \ne 0 \\
{\bf b} \in \mathcal C ({\bf A}),{\bf c} \notin \mathcal C ({\bf A^*})\\
{\bf b} \notin \mathcal C ({\bf A}),{\bf c} \in \mathcal C ({\bf A^*})\\
\end{cases}
\end{equation*}
\begin{equation}
r({\bf M}) = r({\bf A}) + 1 \Leftrightarrow {\bf b} \notin \mathcal C ({\bf A}),{\bf c} \notin \mathcal C ({\bf A^*}), \lambda=0. \nonumber
\end{equation}
Analytical expressions of the MPGI $\bf M^\dagger$ is given in
Ref.~\cite{CDM} and in Ref.~\cite{JKB} depending upon the $6$ or $5$
conditions that they find sufficient. The expressions are of the form
\begin{eqnarray}
\bf M^\dagger = A^\dagger + G  \, ,
\end{eqnarray}
where $\bf G$ is a matrix obtained from only sums and products of $\bf A, A^\dagger, b,c$ and their conjugate transposes. We do not mention the explicit form of all the cases here. Instead  we give expressions for $\bf M^{\dagger}$  for those cases which will be useful to us. 

\subsubsection{Case~1} 
If ${\bf b} \in \mathcal C ({\bf A}),{\bf c} \in \mathcal C ({\bf A^*}), \lambda \ne 0 $ then 
\begin{eqnarray}
\bf M^\dagger = A^\dagger-\lambda^{-1}de^* \, ,
\end{eqnarray}
where 
\begin{eqnarray}
\bf d = A^\dagger b, e = (A^\dagger)^*c.
\end{eqnarray}

\subsubsection{Case~2} 
If ${\bf b} \notin \mathcal C ({\bf A}),{\bf c} \notin \mathcal C ({\bf A^*}) $ then 
\begin{eqnarray}
\bf M^\dagger = A^\dagger- ku^\dagger -v^\dagger h + \lambda v^\dagger u ^\dagger \, ,
\end{eqnarray}
where $\bf k=A^\dagger b, u = (I-AA^\dagger)b, v = c^*(I-A^\dagger A),
h=c^*A^\dagger$.

\subsubsection{Case~3} 
In Ref.~\cite{JKB}, a relation between
the unique projectors between on the column spaces of $\bf M, A$ is reported. If ${\bf b} \notin \mathcal C ({\bf A}),{\bf c} \in \mathcal C ({\bf A^*}) $ then 
\begin{eqnarray}
\bf MM^\dagger = AA^\dagger-\eta ^{-1}{\bf ee^T}+\eta ^{-1}\nu
^{-1}\bf qq^T\, ,
\end{eqnarray}
where $\bf e = (A^\dagger)^*c$, $\eta = \bf e^*e$,$\nu = \lambda \lambda^*+\eta \phi$, $\phi = \bf f^T f$, $\bf f= (I-AA^\dagger)b$.

\subsection{Analytic computation of the bias}

The analytic compution of bias employs the Generalized Sherman-Morrison
formula and the relation between orthogonal projectors mentioned in the above three cases. Consider the previously stated expression for the covariance matrix,
\begin{eqnarray}
{\bf \hat C_l} = \hat C^c_l{\bf e_0e^T_0} + \underbrace{{\bf \hat f^{0}_l} {\bf e^T_0} + \underbrace{{\bf e_0}{\bf \hat f^{0T}_l} + {\bf A_3}}_{\bf A_2}}_{\bf A_1} \,.  
\end{eqnarray} 
First note that $\bf e_0 \notin \mathcal C ({\bf A_3}),{\bf f^0_l} \in
\mathcal C ({\bf A^*_3}) $, so that this is identical to the
conditions for case~3 and $rank({\bf A_3})= rank({\bf A_2}) = n_f$, the number of foreground components.  But when we consider matrix $\bf A_2$ we see that $\bf f^0_l \notin \mathcal C ({\bf A_2}),{\bf e_0} \notin \mathcal C ({\bf A^*_2}) $. So
here conditions of case~2 apply. Here $rank({\bf A_1})= rank({\bf
A_2})+1 = n_f+1$. On the other hand, $\bf e_0 \in \mathcal C ({\bf
A_1}),{\bf e_0} \in \mathcal C ({\bf A^*_1}) $. Hence $rank({\bf \hat
C_1})= rank({\bf A_1}) = n_f+1$. Now one can carry out the analytic simplification in three steps.

\subsubsection{Step~1 : First application of GSM }
We note that ${\bf e_0} \in \mathcal C (\bf A_1)$. Therefore we are in a situation where conditions  of case~1 are applicable. Our aim is to express $\hat C^{Clean}_l =1/{\bf e^T_0C^\dagger_le_0}$ in terms of $\bf A_1$. For this purpose we simply need to express $\bf C^\dagger_l$ in terms of $\bf A_1$ following GSM formula appropriate for this case. At the next step we compute ${\bf e^T_0C^\dagger_1 e_0}$. The analysis is relatively simple and we do not show the detailed calculation here. Instead we give a similar calculation for a more complicated expression at the next subsection. The result of this section is,
\begin{eqnarray}
\hat C^{Clean}_l = \hat C^c_l + \frac{1}{\bf e^T_0A^\dagger_1 e_0}\,. 
\end{eqnarray}

\subsubsection{Step~2 : Second application of GSM}
At this step we shall simplify $1/\bf(e^T_0 A^\dagger_1e_0)$ further in terms of some function of $\bf A^\dagger_2$ where $\bf A_1 = f^0_le^T_0 + A_2$. As mentioned in the previous subsection, this simplification method proceeds in two steps. First, we express $\bf A^\dagger_1$ in terms of $\bf A^\dagger_2$ following GSM. At the next stage we would compute $1/\bf(e^T_0 A^\dagger_1e_0)$. We easily see that, $\bf f^0_l\notin \mathcal C(\bf A_2)$ and $\bf e_0 \notin \mathcal C(\bf A^T_2)$. Therefore GSM formula corresponding to case~2 would be useful for us, 
\begin{eqnarray}
\bf A^\dagger_1 = A^\dagger_2 -ku^\dagger -v^\dagger h+\lambda v^\dagger u^\dagger \,.
\label{sm_st2}
\end{eqnarray}
Each of the individual quantities on the right hand side are explained in the section \ref{Theorem}. All the  terms starting from the second on the right of above equation could be written in terms of the foreground and CMB shape vectors, $\bf f^0_l$ and $\bf e_0$. The vectors $\bf k$ and $\bf h$ are given by $\bf k=A^\dagger_2f^0_l $ and $ \bf h = e^T_0A^\dagger_2$ respectively. Moore Penrose Inverse of the other two vectors $\bf u$ and $\bf v$ could be computed following the definition $\bf u^\dagger = u^*/\lVert u\rVert$, where $\lVert \bf u\rVert$ denotes vector norm. Thus we have, 
\begin{eqnarray}
\bf u^\dagger= \frac{f^{0T}_l(I-A_2A^\dagger_2)}{f^{0T}_l(I-A_2A^\dagger_2)f^0_l} \, ,
\bf v^\dagger  =\frac{(I-A^\dagger_2A_2)e_0}{e^T_0(I-A^\dagger_2A_2)e_0} \,.
\end{eqnarray}
All the vectors computed above could be used in eq \ref{sm_st2} to express $\bf A^\dagger_1$ in terms of $\bf A^\dagger_2$,
\begin{eqnarray}
\bf A^\dagger_1 = A^\dagger_2 -\frac{A^\dagger_2f^0_l f^{0T}_l(I-A_2A^\dagger_2)}{f^{0T}_l(I-A_2A^\dagger_2)f^0_l}-\frac{(I-A^\dagger_2A_2)e_0e^T_0A^\dagger_2}{e^T_0(I-A^\dagger_2A_2)e_0}
+(1+{\bf e^T_0A^\dagger_2f^0_l})\bf \frac{(I-A^\dagger_2A_2)e_0}{e^T_0(I-A^\dagger_2A_2)e_0}\frac{f^{0T}_l(I-A_2A^\dagger_2)}{f^{0T}_l(I-A_2A^\dagger_2)f^0_l} \,.
\end{eqnarray}
Now we can easily compute $\bf e^T_0A^\dagger_1e_0$. This involves contraction of both the indices of $A^\dagger_{1(ij)}$. After some algebra all the terms above simplifies a lot and we are left with, 
\begin{eqnarray}
\frac{1}{{\bf e^T_0A^\dagger_1e_0}}=\bf \frac{f^{0T}_l(I-A_2A^\dagger_2)f^0_l}{f^{0T}_l(I-A_2A^\dagger_2)e_0} \,.
\label{stage2}
\end{eqnarray}
\subsubsection{Step~3 : Application of a relation of orthogonal projectors}
Both the numerator and denominator of eq.~\ref{stage2} contains orthogonal projector $\bf A_2A^\dagger_2$ on the column space of $\bf A_2$. At this stage we only need to reexpress this term in terms of $\bf A_3A^\dagger_3$ which is also an orthogonal projector on the column space of $\bf A_3$. We recall that $\bf A_2 = A_3 + e_0f^T_0$. The foreground shape vector $\bf f^0_l$ lies on the column space of the foreground covariance matrix, $\bf A_3$. However the CMB shape vector is linearly independent on the foreground templates. Therefore the shape vectors follow, ${\bf e_0} \notin \mathcal C ({\bf A_3}),{\bf f^0_l} \in \mathcal C ({\bf A^*_3})$. Thus the conditions of case~3  are applicable. Following the notations of the Ref. ~\cite{JKB} we obtain,
\begin{eqnarray}
{\bf A_2A^\dagger_2= A_3A^\dagger_3} -\eta ^{-1}{\bf ee^T}+\eta ^{-1}\nu ^{-1}\bf qq^T  
\label{jkb_proj}
\end{eqnarray}
To proceed further we need to express  $\bf ee^T$ and $\bf qq^T $ in terms of $\bf f^0_l$ and $\bf e_0$. We note that, $\bf e=A^\dagger_3f^0_l$ and ${\bf q}=\lambda {\bf e} +\eta {\bf f} $. Following notations of Ref.~\cite{JKB} we see that $\bf f$ is the component of $\bf e_0$ on a plane orthogonal to the column space of $\bf A_3$, i.e. $\bf f=(I-A_3A^\dagger_3)e_0$. The matrix $\bf ee^T$ could immediately be identified as $\bf ee^T=A^\dagger_3f^0_lf^{0T}_lA^\dagger_3$. However the other matrix ${\bf qq^T }$ contains several terms, 
\begin{eqnarray}
 {\bf qq^T }= \lambda^2 {\bf ee^T} + \eta^2 (\bf I-A_3A^\dagger_3)e_0e^T_0 (\bf I-A_3A^\dagger_3)+ \lambda \eta \left( \bf A^\dagger_3f^0_le^T_0 ( I-A_3A^\dagger_3)+( I-A_3A^\dagger_3)e_0f^{0T}_lA^\dagger_3 \right)\,.
\label{eq}
\end{eqnarray}
It is easy to see that, the numerator of eq.~(\ref{stage2}) involves computation of the term $\bf f^{0T}_lA_2A^\dagger_2f^{0}_l$. At this point we have all the necessary expressions to compute this term. First we note that, following eqs.~(\ref{jkb_proj}), (\ref{eq}) the projector on the column space of $\bf A_2$ could be simplified as, 
\begin{eqnarray}
{\bf A_2A^\dagger_2= A_3A^\dagger_3} -\eta ^{-1}{\bf A^\dagger_3f^0_lf^{0T}_lA^\dagger_3}+ 
\eta ^{-1}\nu ^{-1}(\lambda^2 {\bf A^\dagger_3f^0_lf^{0T}_lA^\dagger_3} + \eta^2 (\bf I-A_3A^\dagger_3)e_0e^T_0 (\bf I-A_3A^\dagger_3)+ \nonumber \\
\lambda \eta \left( \bf A^\dagger_3f^0_le^T_0 ( I-A_3A^\dagger_3)+( I-A_3A^\dagger_3)e_0f^{0T}_lA^\dagger_3 \right))\,.
\label{proj}
\end{eqnarray} 
Now we can easily get an expression for $\bf f^{0T}_lA_2A^\dagger_2f^{0}_l$. Using eq.~(\ref{proj}) we obtain,
 \begin{eqnarray}
{\bf f^{0T}_lA_2A^\dagger_2f^{0}_l= f^{0T}_lA_3A^\dagger_3f^{0}_l} -\eta ^{-1}({\bf f^{0T}_lA^\dagger_3f^0_l})^2+ \eta ^{-1}\nu ^{-1}(\lambda^2 ({\bf f^{0T}_lA^\dagger_3f^0_l})^2 + \eta^2 (\bf f^{0T}_l( I-A_3A^\dagger_3)e_0)^2+ \nonumber \\
2\lambda \eta \left( \bf f^{0T}_lA^\dagger_3f^0_lf^{0T}_l ( I-A_3A^\dagger_3)e_0 \right))\,.
\label{n1}
\end{eqnarray}
Though there are several terms on the right hand side of the above equation as we will see some of them drop. We  note that, $\bf f^0_l \in \mathcal C(A_3)$, so that $\bf (A_3A^\dagger_3)f^{0}_l=f^{0}_l$. Hence $\bf f^{0T}_l ( I-A_3A^\dagger_3)e_0=  e^T_0( I-A_3A^\dagger_3)f^{0}_l=0$. If we assume $X= \bf f^{0T}_lA_3A^\dagger_3f^{0}_l $ and 
$Y = \bf f^{0T}_lA^\dagger_3f^0_l$ then using eq.~(\ref{n1}) and $\nu = \lambda^2 +\eta \phi $, as stated earlier, we may easily obtain,
\begin{eqnarray}
{\bf f^{0T}_lA_2A^\dagger_2f^{0}_l}= X+ \frac{Y^2 \phi}{\nu}\,.
\label{num}
\end{eqnarray} 
We now identify $X= \bf f^{0T}_lA_3A^\dagger_3f^{0}_l$ and obtain numerator of the eq.~(\ref{stage2}) as 
\begin{eqnarray}
{\bf f^{0T}_l(I-A_2A^\dagger_2)f^0_l} = -\frac{Y^2 \phi}{\nu} \,.
\end{eqnarray}
The denominator of eq.~(\ref{stage2}) can be computed in a similar fashion. 
We do not elaborate the mathematical details any more for this part. The final result is
\begin{eqnarray}
{\bf f^{0T}_l(I-A_2A^\dagger_2)e_0} =  \frac{1}{\nu}Y\phi \,.
\label{din}
\end{eqnarray}
Thus using eqs.~(\ref{stage2}), (\ref{num}), (\ref{din}) we obtain 
\begin{eqnarray}
\frac{1}{\bf e^T_0A^\dagger_1e_0}= -Y=-\bf f^{0T}_lA^\dagger_3f^0_l\,.
\end{eqnarray} 
The quantity $\left < \frac{1}{\bf e^T_0A^\dagger_1e_0}\right > -\left < \bf f^{0T}_lA^\dagger_3f^0_l\right >$ constitutes the
negative bias in the cleaned power spectrum. Below we further simply
this term in terms of the rank of the foreground covariance matrix $\bf A_3$.

\subsubsection{Simplification of the bias expression}

We note that the elements of the foreground covariance matrix are
given by
\begin{eqnarray}
A^{ij}_3 = \sum_{pp'}^{n_f}f^i_pf^j_{p'}C^{(pp')0}_l\,.
\end{eqnarray}
Also the elements of the chance correlation vector between CMB and all the foregrounds
\begin{eqnarray}
{\bf f^{0}_l}^i=f^{0i}_l= \sum_{p}^{n_f}\hat C^{(cp)0}_lf^i_p \,.
\end{eqnarray}
We may rewrite the magnitude of the negative bias in terms of the above matrix elements and components of the foreground shape vector. After a little algebra we get,
\begin{eqnarray}
\left<{\bf f^{0T}_lA^\dagger_3f^0_l}\right> = \sum_{ij}A^{\dagger ij}_3\left<\sum_{pp'}\hat C^{(cp)0}_l\hat C^{(cp')0}_l\right>f^i_p f^j_{p'}\,.
\label{r1}
\end{eqnarray}
Using $\hat C^{(cp)0}_l= \sum_m(a^c_{lm}a^{p0*}_{lm})/(2l+1)$ and the fact that CMB anisotropies are statistically isotropic, i.e. $\left<a_{lm}a_{l'm'}\right>=C^c_l\delta_{ll'}\delta_{mm'}$ we may obtain,
\begin{eqnarray}
\left<\sum_{pp'}\hat C^{(cp)0}_l\hat C^{(cp')0}_l\right>=\frac{C^c_l}{2l+1}\sum_{pp'}C^{(pp')0}_l \,.
\label{r2}
\end{eqnarray}
Using eq.~(\ref{r2}) and $A^{ij}_3 = \sum_{pp'}^{n_f}f^i_pf^j_{p'}C^{(pp')0}_l $ we can  rewrite eq.~(\ref{r1}) in terms of following expression consisting of CMB and foreground covariance matrices,
\begin{eqnarray}
\left< {\bf f^{0T}_lA^\dagger_3f^0_l}\right> =\frac{C^c_l}{2l+1}\sum_{ij}A^{\dagger ij}_3A^{ij}_3 \, .
\end{eqnarray}
However we observe that, $\sum_{ij}A^{\dagger ij}_3A^{ij}_3= tr({\bf A^\dagger_3  A_3)} =rank({\bf A_3)}=n_f$.
Thus we have 
\begin{eqnarray}
 \left<\hat C^{Clean}_l\right> =  \left<\hat C^c_l \right>-n_f\frac{\left<\hat C^c_l\right>}{2l+1} \, .
\end{eqnarray}
which is eq.~(\ref{B1}).
The other formulae given by eq.~(\ref{an_bias}) and
eq.~(\ref{bias_fg_nse}) can be obtained similarly but need much less
algebra.

\end{document}